\documentclass[12pt]{article}
\newcommand{\mez}{\hspace{+0.5cm}}
\newcommand{\mz}{\hspace{+0.25cm}}
\newcommand{\m}{\hspace*{-0.30mm}}
\newcommand{\n}{\hspace*{-0.20mm}}
\newcommand{\be}{\begin{equation}}
\newcommand{\ee}{\end{equation}}
\newcommand{\la}{\langle}
\newcommand{\ra}{\rangle}

\usepackage{bm,amssymb,color,graphicx}
\definecolor{orange}{rgb}{1,0.5,0}

\begin{document}

\begin{centering}
\vspace*{+4.00cm}
{\Large \bf {\color{blue} An Introduction to Scattering Theory}}\\
\vspace*{+0.50cm}
{\bf Lecture notes for the $\bm 1^{\bf st}$ International Summer School on Advanced Quantum Mechanics}\\ {\bf (Prague, September 09-19, 2019)}\\
\vspace*{+1.00cm}
{\color{blue} {\sl Milan \v{S}indelka}}\\
\vspace*{+0.30cm}
{\it Laser Plasma Department, Institute of Plasma Physics of the Czech Academy of Sciences,\\
U Slovanky 2525/1a, 18200 Prague 8, Czech Republic\\ {\tt sindelka@ipp.cas.cz} }\\
\vspace*{+1.00cm}
{\it (\today)}\\
\vspace*{+1.50cm}
\end{centering}

\mez The purpose of these lectures is to give an accessible and self contained introduction to quantum scattering theory in one
dimension. Part A defines the theoretical playground, and develops basic concepts of scattering theory in the time domain
(Asymptotic Condition, in- and out- states, scattering operator $\hat{S}$). The aim of Part B is then to build up, in a
step-by-step fashion, the time independent scattering theory in energy domain. This amounts to introduce the
Lippmann-Schwinger equation for the stationary scattering states (denoted as $| \psi_{E(\pm 1)}^\pm \ra$), to discuss fundamental
properties of $| \psi_{E(\pm 1)}^\pm \ra$, and subsequently to construct $\hat{S}$ and $\hat{T}$ operators in terms of
$| \psi_{E(\pm 1)}^\pm \ra$. Physical contents of the $\hat{S}$ and $\hat{T}$ operators is then illuminated by deriving explicit
formulas for the probability of transmission/reflection of our quantum particle through/from the interaction region of
the potential. An illustrative numerical example is given, which also highlights an existence of scattering
resonances. Finally, Part C elaborates the nonhermitian scattering theory (Siegert pseudostate formalism), which offers an
extremely powerful tool suitable for clear cut understanding of the resonance phenomena.

\newpage

{\color{blue}
\dotfill\\
\begin{centering}
{\sl Lecture \#1}\\
\vspace*{-0.20cm}
\end{centering}
\dotfill\\}

{\bf A. Basic theoretical setup, scattering theory in time domain}\\

{\sl Based upon Refs.~\cite{Taylor,Roman}.}

\vspace*{+0.20cm}
\begin{itemize}
\item \underline{\bf A.1 Our playground}\\
      Nonrelativistic quantum mechanics\\ of a single particle moving in 1D along a coordinate $x \in (-\infty,+\infty)$.\\
      The Hamiltonian is assumed to take the form
      \be \label{hat-sf-H-H0-V}
         \hat{\sf H} \; = \; \hat{\sf H}_0 \; + \; \hat{\sf V} \mez ;
      \ee
      with $\hat{\sf H}_0$ being the kinetic energy operator and $\hat{\sf V}$ the potential energy.\\      
      Explicitly one has
      \be
         \hat{\sf H}_0 \; = \; \frac{\hat{\sf p}^2}{2\,m} \mez , \mez \hat{\sf V} \; = \; V\m(\hat{\sf x}) \mez ;
      \ee
      where as usual $[\hat{\sf x},\hat{\sf p}]=i\hbar\,\hat{\sf 1}$.\\
      The potential function $V\m(x)$ is assumed to be finite ranged, i.e.,
      \be
         V\m(x) \; = \; 0 \mez {\rm whenever} \mez x \leq a \mez {\rm or} \mez x \geq b \mez ;
      \ee
      where $a$ and $b$ are given parameters ($a<b$).\\
      Dynamical time evolution of the state vector $| \psi(t) \ra$ is governed by the Schr\"{o}dinger equation
      \be \label{TDSCHE}
         i\hbar \, \partial_t \, | \psi(t) \ra \; = \; \hat{\sf H} \, | \psi(t) \ra \mez .
      \ee
      Recall that the Schr\"{o}dinger equation (\ref{TDSCHE}) implies probability conservation.\\
      This is expressed mathematically by the continuity equation which is derived in {\sl Appendix I}.\\
      An elementary treatment of the eigenvalue problem of $\hat{\sf H}$ is given in {\sl Appendix II}.
\item \underline{\bf A.2 Time evolution operators}\\
      Let $\hat{\sf U}\n(t)$ be the full evolution operator, defined through an initial value problem
      \be
         i\hbar \; \partial_t \, \hat{\sf U}\n(t) \; = \; \hat{\sf H} \, \hat{\sf U}\n(t) \mez , \mez
         \hat{\sf U}\n(0) \; = \; \hat{\sf 1} \mez .
      \ee
      Clearly, one has
      \be \label{hat-sf-U-explicit}
         \hat{\sf U}\n(t) \; = \; e^{-(i/\hbar)\hat{\sf H}t} \mez .
      \ee
      Dynamical state of our system is then expressible in the form
      \be
         | \psi(t) \ra \; = \; \hat{\sf U}\n(t) \, | \psi \ra \mez ;
      \ee
      of course with $| \psi \ra = | \psi(0) \ra$.\\
      Let $\hat{\sf U}_0(t)$ be the free evolution operator, corresponding to a situation when the potential is absent.\\
      This entity is defined through an initial value problem
      \be
         i\hbar \; \partial_t \, \hat{\sf U}_0(t) \; = \; \hat{\sf H}_0 \, \hat{\sf U}_0(t) \mez , \mez
         \hat{\sf U}_0(0) \; = \; \hat{\sf 1} \mez .
      \ee
      Clearly, one has
      \be
         \hat{\sf U}_0(t) \; = \; e^{-(i/\hbar)\hat{\sf H}_0t} \mez .
      \ee
      Freely evolving quantum state vectors are then expressible as
      \be
         | \phi(t) \ra \; = \; \hat{\sf U}_0(t) \, | \phi \ra \mez ;
      \ee
      of course with $| \phi \ra = | \phi(0) \ra$.
\end{itemize}
      
\newpage
      
{\color{blue}
\dotfill\\
\begin{centering}
{\sl Lecture \#1}\\
\vspace*{-0.20cm}
\end{centering}
\dotfill}
\begin{itemize}
\item \underline{\bf A.3 The asymptotic condition: A qualitative look}\\
      Suppose that we have prepared an initial quantum state $| \psi \ra$ at $t=0$.\\
      We may view $|\psi\ra$ as a square integrable wavepacket $\psi(0,x) = \la x | \psi \ra$.\\
      Assume that this wavepacket does not populate any bound states of the studied problem.\\
      According to the central assumption of the scattering theory,\\
      our wavepacket behaves for $t \neq 0$ as follows:\\
      As $t$ increases from $0$ to $+\infty$, the wavepacket $\psi(t,x) = \la x | \psi(t) \ra$\\
      tends to leave the interaction region (= region of $x \in (a,b)$ with nonzero potential)\\
      and becomes thus freely evolving at very large positive times.\\
      $[\,$Note that some portion of $\psi(0,x)$ would generally escape to the right ($x \to +\infty$)\\
      \phantom{$[\,$}and some portion to the left ($x \to -\infty$).$\,]$\\
      Similarly, as $t$ decreases from $0$ to $-\infty$, the wavepacket $\psi(t,x) = \la x | \psi(t) \ra$\\
      tends to leave the interaction region and becomes thus freely evolving at very large negative times.\\
      What we just described is the so called {\color{red} {\sl Asymptotic Condition} (AC)}.\\
      It has crucial importance for all our subsequent elaborations.
\item \underline{\bf A.4 Formal statement of the Asymptotic Condition}\\
      Let us formulate now the AC in a more formal mathematical language.\\
      The AC says that, for each prescribed physical state $| \psi \ra = | \psi(0) \ra$\\
      (which does not populate any eventual bound states of the problem),\\
      there exists an unique incoming state $| \phi_{\rm in} \ra$ such that
      \be
         \hat{\sf U}\n(t) \, | \psi \ra \; \to \; \hat{\sf U}_0(t) \, | \phi_{\rm in} \ra \mez {\rm as} \mez
         t \to -\infty \mez .
      \ee
      Similarly, one assumes that there exists an unique outgoing state $| \phi_{\rm out} \ra$ such that
      \be
         \hat{\sf U}\n(t) \, | \psi \ra \; \to \; \hat{\sf U}_0(t) \, | \phi_{\rm out} \ra \mez {\rm as} \mez
         t \to +\infty \mez .
      \ee
      Written equivalently,\\ the AC asserts that the in-state $| \phi_{\rm in} \ra$ and the out-state $| \phi_{\rm out} \ra$
      are related to $| \psi \ra$ as follows:
      \be \label{psi-from-phi-in}
         | \psi \ra \; = \; \lim_{t \to -\infty} \, \hat{\sf U}^\dagger\n(t) \, \hat{\sf U}_0(t) \, | \phi_{\rm in} \ra
         \mez ;
      \ee
      \be \label{psi-from-phi-out}
         | \psi \ra \; = \; \lim_{t \to +\infty} \, \hat{\sf U}^\dagger\n(t) \, \hat{\sf U}_0(t) \, | \phi_{\rm out} \ra
         \mez .
      \ee
      Alternatively, one may state that
      \be
         | \phi_{\rm in} \ra \; = \; \lim_{t \to -\infty} \, \hat{\sf U}_0^\dagger\n(t) \, \hat{\sf U}(t) \, | \psi \ra \mez ;
      \ee
      \be
         | \phi_{\rm out} \ra \; = \; \lim_{t \to +\infty} \, \hat{\sf U}_0^\dagger\n(t) \, \hat{\sf U}(t) \, | \psi \ra \mez .
      \ee
      So far, we have just declared the AC to hold without proving it.\\
      A simple and explicit {\color{red} proof of the AC} is worked out in {\sl Appendix III}.\\
      It is based upon the stationary phase method.
\end{itemize}
      
\newpage
      
{\color{blue}
\dotfill\\
\begin{centering}
{\sl Lecture \#1}\\
\vspace*{-0.20cm}
\end{centering}
\dotfill}
\begin{itemize}
\item \underline{\bf A.5 The Scattering Operator: Definition}\\
      It should be obvious by now\\ that there exists a well defined unitary mapping of the in-states into the out-states.\\
      Indeed, any given dynamical in-state $\hat{\sf U}_0(t) \, | \phi_{\rm in} \ra$ $[t \to -\infty]$\\
      is converted by an interaction (occurring at finite times)\\
      into some uniquely determined dynamical out-state $\hat{\sf U}_0(t) \, | \phi_{\rm out} \ra$ $[t \to +\infty]$.\\
      Such that
      \be
         | \phi_{\rm out} \ra \; = \; \lim_{T \to +\infty} \, \hat{\sf U}_0^\dagger\n(+T) \; \hat{\sf U}(+T) \;
         \hat{\sf U}^\dagger\n(-T) \; \hat{\sf U}_0(-T) \; | \phi_{\rm in} \ra \mez .
      \ee
      We shall formally write
      \be \label{hat-S-formal}
         | \phi_{\rm out} \ra \; = \; \hat{S} \, | \phi_{\rm in} \ra \mez ;
      \ee
      and call
      \be \label{hat-S-def}
         \hat{S} \; = \; \lim_{T \to +\infty} \, \hat{\sf U}_0^\dagger\n(+T) \; \hat{\sf U}(+T) \;
         \hat{\sf U}^\dagger\n(-T) \; \hat{\sf U}_0(-T)
      \ee
      as the {\color{red} Scattering Operator}.\\
      The just defined scattering operator $\hat{S}$ takes an input\\
      (= an incoming in-state wavepacket set up to collide with our potential)\\
      and converts it into an appropriate output\\
      (= the outgoing out-state wavepacket departing away from the interaction region to $x \to \pm\infty$).\\
      Thus the knowledge of $\hat{S}$ enables us to determine\\
      an overall output of the studied scattering process from any given input.\\
      This is why {\sl {\color{red} $\hat{S}$ contains all the information about scattering}.}\\
      Remark: Since we are interested in scattering,\\ we have tacitly restricted the domain of $\hat{S}=(\ref{hat-S-def})$\\
      just to the scattering sector of our quantum state space.\\ In other words, the above established definition (\ref{hat-S-formal})-(\ref{hat-S-def}) of $\hat{S}$\\
      is valid only for such vectors $ | \phi_{\rm in} \ra$ which do not populate eventual bound states of the problem.
\item \underline{\bf A.6 The Scattering Operator: Physical meaning, matrix elements}\\
      Suppose that in the infinite past (before the scattering takes place)\\
      our system has been prepared in a given freely evolving quantum state $\hat{\sf U}_0(t) \, | \phi_{\rm in} \ra$.\\
      We wish to look into the infinite future (after the scattering has taken place)\\
      and determine the probability of finding our system in some freely evolving quantum state
      $\hat{\sf U}_0(t) \, | \chi_{\rm out} \ra$.\\ Clearly, the sought probability equals to
      \be
         \wp \; = \; \Bigl| {\cal S} \Bigr|^2 \mez ;
      \ee
      where
      \be \label{S-matels}
         {\cal S} \; = \; \la \chi_{\rm out} | \, \hat{\sf U}_0^\dagger(t) \; \hat{\sf U}_0(t) \, | \phi_{\rm out} \ra \; = \;
         \la \chi_{\rm out} | \phi_{\rm out} \ra \; = \; \la \chi_{\rm out} | \hat{S} | \phi_{\rm in} \ra \mez .
      \ee
      Redisplayed once again in final form, one has
      \be \label{wp-out-in-def}
         \wp_{\rm out/in} \; = \; \Bigl| \, \la \chi_{\rm out} | \hat{S} | \phi_{\rm in} \ra \, \Bigr|^2 \mez .
      \ee
      This formula gives {\color{red} physical meaning to the scattering matrix elements}.
\end{itemize}

\newpage
      
{\color{blue}
\dotfill\\
\begin{centering}
{\sl Lecture \#1}\\
\vspace*{-0.20cm}
\end{centering}
\dotfill}

\vspace*{+0.30cm}

{\bf B. Passage from the time dependent picture of scattering\\ \phantom{\bf B.} into the time independent scattering theory}\\

{\sl Based upon Refs.~\cite{Taylor,Roman}.}

\begin{itemize}
\item \underline{\bf B.1 Motivation}\\
      The above described time dependent picture of scattering (based upon wavepackets)\\
      is physically intuitive yet still rather formal.\\
      It does not lend itself well to any further theoretical investigation of $\hat{S}$.\\
      Yet we wish to analyze the properties of $\hat{S}$,\\
      in order to gain deep physics insights (and encounter surprises)\\
      and thus better understand the nature of scattering processes.\\
      Further progress is facilitated via switching from the time domain into the energy domain.
\item \underline{\bf B.2 Mathematical developments \#1}\\
      Our forthcoming task is to relate $| \psi \ra$ and $| \phi_{\rm in} \ra$, $| \phi_{\rm out} \ra$ more explicitly.\\
      It is a simple matter to verify an identity
      \be
         \hat{\sf U}^\dagger\n(t) \, \hat{\sf U}_0(t) \, | \phi_{\rm in} \ra \; = \; | \phi_{\rm in} \ra \; - \; \int_t^0 \m
         \left\{ \, \hat{\sf U}^\dagger\n(t'\n) \, \hat{\sf U}_0(t') \, | \phi_{\rm in} \ra \, \right\}^{\m\m\bm\cdot} {\rm d}t' \mez ;
      \ee
      here the upper dot superscript denotes time differentiation. Subsequently one has
      \be
         | \psi \ra \; = \; | \phi_{\rm in} \ra \; - \; \int_{-\infty}^0 \m
         \left\{ \, \hat{\sf U}^\dagger\n(t'\n) \, \hat{\sf U}_0(t') \, \right\}^{\m\m\bm\cdot} | \phi_{\rm in} \ra \; {\rm d}t' \mez .
      \ee
      Similarly,
      \be
         \hat{\sf U}^\dagger\n(t) \, \hat{\sf U}_0(t) \, | \phi_{\rm out} \ra \; = \; | \phi_{\rm out} \ra \; + \; \int_0^t \m
         \left\{ \, \hat{\sf U}^\dagger\n(t'\n) \, \hat{\sf U}_0(t') \, | \phi_{\rm out} \ra \, \right\}^{\m\m\bm\cdot} {\rm d}t' \mez ;
      \ee
      and hence
      \be
         | \psi \ra \; = \; | \phi_{\rm out} \ra \; + \; \int_0^{+\infty} \m\m
         \left\{ \, \hat{\sf U}^\dagger\n(t'\n) \, \hat{\sf U}_0(t') \, \right\}^{\m\m\bm\cdot} | \phi_{\rm out} \ra \; {\rm d}t' \mez .
      \ee
      Direct calculation yields the required time derivative,
      \begin{eqnarray}
         \left\{ \, \hat{\sf U}^\dagger\n(t'\n) \, \hat{\sf U}_0(t') \, \right\}^{\m\m\bm\cdot} & = &
         \hat{\sf U}^{\dagger\bm\cdot}\m(t'\n) \, \hat{\sf U}_0(t') \; + \; \hat{\sf U}^\dagger\n(t'\n) \, \hat{\sf U}_0^{\bm\cdot}(t') \; = \nonumber\\
         & = & \frac{i}{\hbar} \; \hat{\sf U}^\dagger\n(t'\n) \, \left( \hat{\sf H} \, - \, \hat{\sf H}_0 \right) \, \hat{\sf U}_0(t') \; = \;
         \frac{i}{\hbar} \; \hat{\sf U}^\dagger\n(t'\n) \, \hat{\sf V} \, \hat{\sf U}_0(t') \mez .
      \end{eqnarray}
      Correspondingly, one obtains
      \be \label{LS-take-1-in}
         | \psi \ra \; = \; | \phi_{\rm in} \ra \; - \; \frac{i}{\hbar} \, \int_{-\infty}^0 \m
         \hat{\sf U}^\dagger\n(t'\n) \, \hat{\sf V} \, \hat{\sf U}_0(t') \, | \phi_{\rm in} \ra \; {\rm d}t' \mez .
      \ee
      \be \label{LS-take-1-out}
         | \psi \ra \; = \; | \phi_{\rm out} \ra \; + \; \frac{i}{\hbar} \, \int_0^{+\infty} \m\m
         \hat{\sf U}^\dagger\n(t'\n) \, \hat{\sf V} \, \hat{\sf U}_0(t') \, | \phi_{\rm out} \ra \; {\rm d}t' \mez .
      \ee
      This is the sought more explicit relationship between $| \psi \ra$ and $| \phi_{\rm in} \ra$, $| \phi_{\rm out} \ra$.\\
      Formulas (\ref{LS-take-1-in})-(\ref{LS-take-1-out}) include integration over time and still do not offer much insight.\\
      Hence we need to pursue additional developments as detailed below.
\end{itemize}
      
\newpage
      
{\color{blue}
\dotfill\\
\begin{centering}
{\sl Lecture \#1}\\
\vspace*{-0.20cm}
\end{centering}
\dotfill}

\vspace*{+1.00cm}

\begin{itemize}
\item \underline{\bf B.2 Mathematical developments \#2}\\
      Further elaborations based upon equations (\ref{LS-take-1-in})-(\ref{LS-take-1-out}) are possible.\\
      An eigenproblem of $\hat{\sf H}_0$ can be formally written as follows. One has
      \be \label{hat-sf-H-0-eigenproblem}
         \hat{\sf H}_0 \, | \phi_{E\eta} \ra \; = \; E \, | \phi_{E\eta} \ra \mez ; \mez [\,E>0\,,\,\eta\in\{-1,+1\}\,]
      \ee
      with the usual orthonormality and closure relations
      \be \label{hat-sf-H-0-oncl}
         \la \phi_{E\eta} | \phi_{E'\n\eta'\n} \ra \; = \; \delta(E-E'\n) \; \delta_{\eta\eta'} \mez , \mez
         \int_{0}^{+\infty} \m {\rm d}E \, \sum_\eta \, | \phi_{E\eta} \ra \la \phi_{E\eta} | \; = \; \hat{\sf 1} \mez .
      \ee
      It is an elementary task (a simple exercise for you) to verify that\\
      an explicit solution of the eigenproblem (\ref{hat-sf-H-0-eigenproblem}) satisfying also relations (\ref{hat-sf-H-0-oncl})\\
      is given by the prescription
      \be \label{phi-E-eta-def}
         \la x | \phi_{E\eta} \ra \; = \; \phi_{E\eta}\n(x) \; = \; \sqrt{\frac{m}{2\,\pi\,\hbar^2 K}} \;\, e^{+i \eta Kx} \mez ;
      \ee
      where $\hbar K = \sqrt{2\,m\,E}$.\\
      Clearly, vector $| \phi_{\rm in} \ra$ can be uniquely expanded in the $| \phi_{E\eta} \ra$ basis, such that
      \be \label{phi-in-expansion}
         | \phi_{\rm in} \ra \; = \; \int_{0}^{+\infty} \m {\rm d}E \, \sum_\eta \; c_{E\eta} \, | \phi_{E\eta} \ra \mez , \mez
         c_{E\eta} \; = \; \la \phi_{E\eta} | \phi_{\rm in} \ra \mez .
      \ee
      Converting thus (\ref{LS-take-1-in}) into
      \begin{eqnarray} \label{LS-take-2-in-prelim}
         | \psi \ra & = & \int_{0}^{+\infty} \m {\rm d}E \, \sum_\eta \; c_{E\eta} \, | \phi_{E\eta} \ra \; - \; \frac{i}{\hbar} \, \int_{-\infty}^0 \m {\rm d}t' \;\,
         \int_{0}^{+\infty} \m {\rm d}E \, \sum_\eta \; c_{E\eta} \;\, \hat{\sf U}^\dagger\n(t'\n) \, \hat{\sf V} \, \hat{\sf U}_0(t') \, | \phi_{E\eta} \ra \; = \nonumber\\
         & = & \int_{0}^{+\infty} \m {\rm d}E \, \sum_\eta \; c_{E\eta} \, | \phi_{E\eta} \ra \; - \; \frac{i}{\hbar} \, \int_{-\infty}^0 \m {\rm d}t' \;\,
         \int_{0}^{+\infty} \m {\rm d}E \, \sum_\eta \; c_{E\eta} \;\, \hat{\sf U}^\dagger\n(t'\n) \, \hat{\sf V} \, e^{-(i/\hbar)Et'} \, | \phi_{E\eta} \ra \mz . \mz
      \end{eqnarray}
      An additional refinement can be implemented here,\\
      based upon including an extra factor $e^{+\varepsilon t'\m/\hbar}$ with $\varepsilon \to +0$. Such that
      \begin{eqnarray} \label{LS-take-2-in}
         \hspace*{-1.50cm}
         | \psi \ra & = & \int_{0}^{+\infty} \m {\rm d}E \, \sum_\eta \; c_{E\eta} \, | \phi_{E\eta} \ra \; - \; \frac{i}{\hbar} \, \lim_{\varepsilon \to +0} \, \int_{-\infty}^0 \m {\rm d}t' \;\,
         \int_{0}^{+\infty} \m {\rm d}E \, \sum_\eta \; c_{E\eta} \;\, \hat{\sf U}^\dagger\n(t'\n) \, \hat{\sf V} \, e^{-(i/\hbar)(E+i\varepsilon)t'} \, | \phi_{E\eta} \ra \mz . \mz
      \end{eqnarray}
      An importance of inserting $e^{+\varepsilon t'\m/\hbar}$ becomes evident later on\\
      (see equation (\ref{LS-take-4-in-take-2}) below and the accompanying discussion).\\
      For now it is enough to recognize that the presence of $e^{+\varepsilon t'\m/\hbar}$\\
      does not affect the value of the associated integral in (\ref{LS-take-2-in}) when the limit of $\varepsilon \to +0$ is taken.
      
\newpage
      
{\color{blue}
\dotfill\\
\begin{centering}
{\sl Lecture \#1}\\
\vspace*{-0.20cm}
\end{centering}
\dotfill}

\vspace*{+1.50cm}

      Similarly we shall write
      \be \label{phi-out-expansion}
         | \phi_{\rm out} \ra \; = \; \int_{0}^{+\infty} \m {\rm d}E \, \sum_\eta \; C_{E\eta} \, | \phi_{E\eta} \ra \mez , \mez
         C_{E\eta} \; = \; \la \phi_{E\eta} | \phi_{\rm out} \ra \mez ;
      \ee
      and
      \begin{eqnarray} \label{LS-take-2-out}
         \hspace*{-1.50cm}
         | \psi \ra & = & \int_{0}^{+\infty} \m {\rm d}E \, \sum_\eta \; C_{E\eta} \, | \phi_{E\eta} \ra \; + \; \frac{i}{\hbar} \, \lim_{\varepsilon \to +0} \, \int_0^{+\infty}\m\m {\rm d}t' \;\,
         \int_{0}^{+\infty} \m {\rm d}E \, \sum_\eta \; C_{E\eta} \;\, \hat{\sf U}^\dagger\n(t'\n) \, \hat{\sf V} \, e^{-(i/\hbar)(E-i\varepsilon)t'} \, | \phi_{E\eta} \ra \mz . \mez
      \end{eqnarray}
      Let us look again now at (\ref{LS-take-2-in}) and (\ref{LS-take-2-out}).\\
      Assuming that the integrations $\int {\rm d}t'$, $\int_{0}^{+\infty} \m {\rm d}E \, \sum_\eta$ are interchangeable, one may write
      \be \label{psi-expansion-plus}
         | \psi \ra \; = \; \int_{0}^{+\infty} \m {\rm d}E \, \sum_\eta \; c_{E\eta} \, | \psi_{E\eta}^{+} \ra \mez ;
      \ee
      where by definition
      \be \label{LS-take-4-in}
         | \psi_{E\eta}^{+} \ra \; = \; | \phi_{E\eta} \ra \; - \; \frac{i}{\hbar} \, \lim_{\varepsilon \to +0} \, \int_{-\infty}^0 \m
         \hat{\sf U}^\dagger\n(t'\n) \, e^{-(i/\hbar)(E+i\varepsilon)t'} \, \hat{\sf V} \, | \phi_{E\eta} \ra \; {\rm d}t' \mez .
      \ee
      Similarly one has
      \be \label{psi-expansion-minus}
         | \psi \ra \; = \; \int_{0}^{+\infty} \m {\rm d}E \, \sum_\eta \; C_{E\eta} \, | \psi_{E\eta}^{-} \ra \mez ;
      \ee
      where by definition
      \be \label{LS-take-4-out}
         | \psi_{E\eta}^{-} \ra \; = \; | \phi_{E\eta} \ra \; + \; \frac{i}{\hbar} \, \lim_{\varepsilon \to +0} \, \int_0^{+\infty} \m\m
         \hat{\sf U}^\dagger\n(t'\n) \, e^{-(i/\hbar)(E-i\varepsilon)t'} \, \hat{\sf V} \, | \phi_{E\eta} \ra \; {\rm d}t' \mez .
      \ee
      In summary, elaborations of this paragraph have enabled us\\
      to replace the formulas (\ref{LS-take-1-in}) and (\ref{LS-take-1-out}) by equivalent relations
      (\ref{psi-expansion-plus})-(\ref{LS-take-4-in}) and (\ref{psi-expansion-minus})-(\ref{LS-take-4-out}).\\
      $[\,$Recall also (\ref{phi-in-expansion}), (\ref{phi-out-expansion}) in this context.$\,]$\\
      This is a substantial progress as we shall see shortly.
\end{itemize}
      
\newpage
      
{\color{blue}
\dotfill\\
\begin{centering}
{\sl Lecture \#2}\\
\vspace*{-0.20cm}
\end{centering}
\dotfill}
\begin{itemize}
\item \underline{\bf B.3 The explicit Lippmann-Schwinger equation}\\
      Take (\ref{LS-take-4-in}) and plug in (\ref{hat-sf-U-explicit}). An integration over $t'$ can then be performed explicitly.
      Indeed, one has
      \begin{eqnarray} \label{LS-take-4-in-take-2}
         | \psi_{E\eta}^{+} \ra & = & | \phi_{E\eta} \ra \; - \; \frac{i}{\hbar} \, \lim_{\varepsilon \to +0} \, \int_{-\infty}^0 \m
         e^{-(i/\hbar)(E-\hat{\sf H}+i\varepsilon)t'} \, \hat{\sf V} \, | \phi_{E\eta} \ra \; {\rm d}t' \; = \nonumber\\
         & = & | \phi_{E\eta} \ra \; + \; \lim_{\varepsilon \to +0} \, \frac{1}{E-\hat{\sf H}+i\varepsilon} \,
         \hat{\sf V} \, | \phi_{E\eta} \ra \mez .
      \end{eqnarray}
      Showing that $| \psi_{E\eta}^{+} \ra$ is obtainable from $| \phi_{E\eta} \ra$ without any reference to the time variable.\\
      Moreover, it is evident that the presence of $i\varepsilon$ in the denominator of (\ref{LS-take-4-in-take-2}) is essential,\\
      since it eliminates the singularity of an ill-defined entity $(E-\hat{\sf H})^{-1}$.\\
      $[\,$Recall that $\hat{\sf H}$ has a positive definite continuous spectrum,\\ \phantom{$[\,$}hence $(E-\hat{\sf H})^{-1}$ includes division by zero.$\,]$\\
      The quantity
      \be \label{G-+-well-def}
         \lim_{\varepsilon \to +0} \, \frac{1}{E-\hat{\sf H}+i\varepsilon} \; = \; \frac{1}{E-\hat{\sf H}+i\,0_+}
      \ee
      represents a well defined operator.\\
      $[\,$Well definedness of the formula (\ref{G-+-well-def}) cannot be established just by an immediate inspection.\\
      \phantom{$[\,$}A detailed self contained analysis of an entity (\ref{G-+-well-def}) is supplemented in {\sl Appendix IV}.$\,]$\\
      It is the so called {\color{red} {\sl retarded Green operator}} corresponding to $\hat{\sf H}$.\\
      Completely analogous manipulations can be performed also in the case of (\ref{LS-take-4-out}),\\
      we leave them to the reader. One would encounter there the so called\\
      {\color{red} {\sl advanced Green operator}} $(E-\hat{\sf H}-i\,0_+)^{-1}$.\\
      Stated shortly, it turns out that relations (\ref{LS-take-4-in}) and (\ref{LS-take-4-out}) can be redisplayed\\
      in the final time independent form
      \be \label{LSE-explicit}
         {\color{red}| \psi_{E\eta}^{\pm} \ra \; = \; | \phi_{E\eta} \ra
         \; + \; \frac{1}{E - \hat{\sf H} \pm i\,0_+} \; \hat{\sf V} \, | \phi_{E\eta} \ra \mez .}
      \ee
      This is the so called {\color{red} {\sl Lippmann-Schwinger equation} (LSE, explicit version)}. Most importantly,\\
      {\color{red} LSE represents the fundamental starting point of the time independent scattering theory}.
\item \underline{\bf B.4 The Lippmann-Schwinger equation: Elaborations}\\
      Assume a finite $\varepsilon>0$ and define for convenience
      \be \label{LS-take-5-in-out-varepsilon}
         | \psi_{E\eta}^{\pm,\varepsilon} \ra \; = \; | \phi_{E\eta} \ra
         \; + \; \frac{1}{E - \hat{\sf H} \pm i\,\varepsilon} \; \hat{\sf V} \, | \phi_{E\eta} \ra \mez ;
      \ee
      such that $| \psi_{E\eta}^{\pm} \ra = \lim_{\varepsilon \to +0} | \psi_{E\eta}^{\pm,\varepsilon} \ra$.
      Multiply (\ref{LS-take-5-in-out-varepsilon}) by $(E - \hat{\sf H} \pm i\,\varepsilon)$ to get
      \be
         (E - \hat{\sf H} \pm i\,\varepsilon) \, | \psi_{E\eta}^{\pm,\varepsilon} \ra \; = \;
         (E - \hat{\sf H} \pm i\,\varepsilon) \, | \phi_{E\eta} \ra \; + \; \hat{\sf V} \, | \phi_{E\eta} \ra \mez .
      \ee
      Recalling (\ref{hat-sf-H-H0-V}) one may simplify the r.h.s.~and write down an interesting formula
      \be \label{LSE-interesting}
         (E - \hat{\sf H} \pm i\,\varepsilon) \, | \psi_{E\eta}^{\pm,\varepsilon} \ra \; = \;
         (E - \hat{\sf H}_0 \pm i\,\varepsilon) \, | \phi_{E\eta} \ra \mez .
      \ee
      Equation (\ref{LSE-interesting}) leads towards two important consequences:

\newpage
      
{\color{blue}
\dotfill\\
\begin{centering}
{\sl Lecture \#2}\\
\vspace*{-0.20cm}
\end{centering}
\dotfill}

      \begin{itemize}
      \item[{\it 1)}]
      Firstly, one may take (\ref{LSE-interesting}) and rearrange it into
      \be
         (E - \hat{\sf H}_0 \pm i\,\varepsilon) \, | \psi_{E\eta}^{\pm,\varepsilon} \ra \; = \;
         (E - \hat{\sf H}_0 \pm i\,\varepsilon) \, | \phi_{E\eta} \ra \; + \; \hat{\sf V} \, | \psi_{E\eta}^{\pm,\varepsilon} \ra \mez ;
      \ee
      and further into
      \be
         | \psi_{E\eta}^{\pm,\varepsilon} \ra \; = \; | \phi_{E\eta} \ra \; + \;
         \frac{1}{E - \hat{\sf H}_0 \pm i\,\varepsilon} \, \hat{\sf V} \, | \psi_{E\eta}^{\pm,\varepsilon} \ra \mez .
      \ee
      For $\varepsilon \to +0$ one obtains immediately the so called {\color{red} {\sl implicit Lippmann-Schwinger equation}},
      \be \label{LSE-implicit}
         {\color{red} | \psi_{E\eta}^{\pm} \ra \; = \; | \phi_{E\eta} \ra \; + \;
         \frac{1}{E - \hat{\sf H}_0 \pm i\,0_+} \, \hat{\sf V} \, | \psi_{E\eta}^{\pm} \ra \mez .}
      \ee
      Note formal similarity between the explicit LSE (\ref{LSE-explicit}) and the implicit LSE (\ref{LSE-implicit}).\\
      The implicit LSE (\ref{LSE-implicit}) contains the free Green operator corresponding to $\hat{\sf H}_0$.\\
      $[\,$See {\sl Appendix IV} for more details regarding this free Green operator.$\,]$\\
      Also, the r.h.s.~of (\ref{LSE-implicit}) contains $| \psi_{E\eta}^{\pm} \ra$ rather than $| \phi_{E\eta} \ra$.\\
      While (\ref{LSE-explicit}) can be regarded as an explicit definition of $| \psi_{E\eta}^{\pm} \ra$,\\
      equation (\ref{LSE-implicit}) does determine $| \psi_{E\eta}^{\pm} \ra$ only implicitly.\\
      Yet the implicit LSE (\ref{LSE-implicit}) possesses great theoretical importance, as we shall see later on.\\
      For now let us point out the following:\\
      \phantom{cra} The explicit LSE (\ref{LSE-explicit}) contains the full Green operator $(E - \hat{\sf H} \pm i\,0_+)^{-1}$\m.\\
      \phantom{cra} It can be thus hard to handle (\ref{LSE-explicit}) numerically when one wishes to calculate $| \psi_{E\eta}^{\pm} \ra$.\\
      \phantom{cra} On the other hand, the implicit LSE (\ref{LSE-implicit})\\
      \phantom{cra} contains just the free Green operator $(E - \hat{\sf H}_0 \pm i\,0_+)^{-1}$\m,\\
      \phantom{cra} which is much easier to handle as documented in {\sl Appendix IV}\m.\\
      \phantom{cra} Importantly, the solution of (\ref{LSE-implicit}) can be performed in an iterative fashion.\\
      \phantom{cra} Such that one starts from (\ref{LSE-implicit}) and writes
      \be \label{LSE-Born-prelim-1}
         | \psi_{E\eta}^{\pm} \ra \; = \; | \phi_{E\eta} \ra \; + \;
         \frac{1}{E - \hat{\sf H}_0 \pm i\,0_+} \, \hat{\sf V} \left( | \phi_{E\eta} \ra \; + \;
         \frac{1}{E - \hat{\sf H}_0 \pm i\,0_+} \, \hat{\sf V} \, | \psi_{E\eta}^{\pm} \ra \right) \mez ;
      \ee
      \phantom{cra} giving
      \begin{eqnarray} \label{LSE-Born-prelim-2}
         \mez | \psi_{E\eta}^{\pm} \ra & = & | \phi_{E\eta} \ra
         \; + \; \frac{1}{E - \hat{\sf H}_0 \pm i\,0_+} \, \hat{\sf V} \, | \phi_{E\eta} \ra \; + \;
         \frac{1}{E - \hat{\sf H}_0 \pm i\,0_+} \, \hat{\sf V} \, \frac{1}{E - \hat{\sf H}_0 \pm i\,0_+} \, \hat{\sf V} \,
         | \psi_{E\eta}^{\pm} \ra \; . \mz\mez
      \end{eqnarray}
      \phantom{cra} By proceeding forward in the same spirit one gets
      \begin{eqnarray} \label{LSE-Born}
         | \psi_{E\eta}^{\pm} \ra & = & | \phi_{E\eta} \ra
         \; + \; \frac{1}{E - \hat{\sf H}_0 \pm i\,0_+} \, \hat{\sf V} \, | \phi_{E\eta} \ra \; + \;
         \frac{1}{E - \hat{\sf H}_0 \pm i\,0_+} \, \hat{\sf V} \, \frac{1}{E - \hat{\sf H}_0 \pm i\,0_+} \, \hat{\sf V} \,
         | \psi_{E\eta}^{\pm} \ra \; = \nonumber\\
         & = & | \phi_{E\eta} \ra
         \; + \; \frac{1}{E - \hat{\sf H}_0 \pm i\,0_+} \, \hat{\sf V} \, | \phi_{E\eta} \ra \; + \;
         \frac{1}{E - \hat{\sf H}_0 \pm i\,0_+} \, \hat{\sf V} \, \frac{1}{E - \hat{\sf H}_0 \pm i\,0_+} \, \hat{\sf V} \,
         | \phi_{E\eta} \ra \nonumber\\
         & + & \frac{1}{E - \hat{\sf H}_0 \pm i\,0_+} \, \hat{\sf V} \, \frac{1}{E - \hat{\sf H}_0 \pm i\,0_+} \, \hat{\sf V} \,
         \frac{1}{E - \hat{\sf H}_0 \pm i\,0_+} \, \hat{\sf V} \, | \psi_{E\eta}^{\pm} \ra \mez ;
      \end{eqnarray}
      \phantom{cra} and so on. This is the so called {\sl Born series expansion} for the LSE.\\
      \phantom{cra} Note that successive terms of (\ref{LSE-Born}) contain successively higher powers of $\hat{\sf V}$\m.\\
      \phantom{cra} Born series corresponds thus to the perturbation theory in $\hat{\sf V}$\m.\\
      \phantom{cra} The Born series approach is therefore most convenient\\
      \phantom{cra} when one encounters scattering by weak potentials.

\newpage
      
{\color{blue}
\dotfill\\
\begin{centering}
{\sl Lecture \#2}\\
\vspace*{-0.20cm}
\end{centering}
\dotfill}

      \item[{\it 2)}] Secondly, one may simply perform the limit of $\varepsilon \to +0$ in (\ref{LSE-interesting}),\\
      and take advantage of the property (\ref{hat-sf-H-0-eigenproblem}). This reveals an important insight of having
      \be \label{hat-sf-H-eigenproblem}
         {\color{red} \hat{\sf H} \, | \psi_{E\eta}^{\pm} \ra \; = \; E \, | \psi_{E\eta}^{\pm} \ra \mez .}
      \ee
      Hence {\color{red} $| \psi_{E\eta}^{\pm} \ra$ is an eigenvector of the full Hamiltonian $\hat{\sf H}$,\\
      with the same eigenvalue $E$ as appropriate for the eigenvector $| \phi_{E\eta} \ra$ of $\hat{\sf H}_0$}.\\
      Vectors $| \psi_{E\eta}^{\pm} \ra$, $| \phi_{E\eta} \ra$ and the eigenvalue $E$ are mutually interconnected\\
      through the LSEs (\ref{LSE-explicit}) and (\ref{LSE-implicit}).
      \end{itemize}
\item \underline{\bf B.5 Stationary scattering states $\psi_{E\eta}^\pm\n(x)$}\\
      Let us define $\psi_{E\eta}^\pm\n(x) = \la x | \psi_{E\eta}^{\pm} \ra$, and call:\\
      $\psi_{E\eta}^+\n(x)$ = the retarded \hspace*{+0.0cm} stationary scattering wavefunctions;\\
      $\psi_{E\eta}^-\n(x)$ = the advanced stationary scattering wavefunctions.\\
      Formula (\ref{hat-sf-H-eigenproblem}) shows that the entities $\psi_{E\eta}^\pm\n(x)$ solve the time independent Schr\"{o}dinger equation
      \be \label{TISCHE-psi-E-eta-pm}
         {\color{red}
         \left\{ \, -\,\frac{\hbar^2}{2\,m}\,\partial_{xx} \, + \, V\m(x) \, \right\} \psi_{E\eta}^\pm\n(x) \; = \; E \, \psi_{E\eta}^\pm\n(x) \mez . }
      \ee
      As we know from elementary quantum mechanics,\\ each continuum energy level $E>0$ is twice degenerate.\\
      Thus $\psi_{E\eta}^\pm\n(x)$ is not determined uniquely just by the Schr\"{o}dinger equation (\ref{TISCHE-psi-E-eta-pm}).\\
      An unambiguous specification of $\psi_{E\eta}^\pm\n(x)$ requires two additional pieces of information:\\
      Appropriate boundary conditions and an adequate normalization factor.\\
      These two additional characteristics of $\psi_{E\eta}^\pm\n(x)$ are found by returning to the implicit LSE (\ref{LSE-implicit}).\\
      Namely, the position representation of equation (\ref{LSE-implicit}) takes the form
      \be \label{LSE-implicit-x-posrep}
         \psi_{E\eta}^\pm\n(x) \; = \; \phi_{E\eta}\n(x) \; + \; \int_{a}^{b} \m {\rm d}y \;\,
         \la x | \, \frac{1}{E - \hat{\sf H}_0 \pm i\,0_+} \, | y \ra \; V\m(y) \; \psi_{E\eta}^\pm\n(y) \mez ;
      \ee
      where $\phi_{E\eta}\n(x)=(\ref{phi-E-eta-def})=\sqrt{\frac{m}{2\,\pi\,\hbar^2 K}} \;\, e^{+i \eta Kx}$ and $\hbar K=\sqrt{2\,m\,E}$.\\
      The free Green operator $(E - \hat{\sf H}_0 \pm i\,0_+)^{-1}$ entering into (\ref{LSE-implicit-x-posrep})\\
      has been analyzed in much detail within {\sl Appendix IV}\m,\\
      and the associated equations (\ref{G-0-E-pm-def}) and (\ref{Green-take-4}) supply an explicit result
      \be \label{free-Green-function-x-y}
         \la x | \, \frac{1}{E - \hat{\sf H}_0 \pm i\,0_+} \, | y \ra \; = \; (\mp i)\,\frac{m}{\hbar^2 K} \; e^{\pm iK|x-y|} \mez .
      \ee
      If so, then the implicit LSE (\ref{LSE-implicit-x-posrep}) can be redisplayed as
      \be \label{LSE-implicit-x-take-3}
         \psi_{E\eta}^{\pm}\n(x) \; = \; \sqrt{\frac{m}{2\,\pi\,\hbar^2 K}} \;\, e^{+i \eta K x} \; \mp \; \frac{i\,m}{\hbar^2 K} \, \int_{a}^{b} \m {\rm d}y \;\,
         e^{\pm i K|x-y|} \; V\m(y) \; \psi_{E\eta}^{\pm}\n(y) \mez .
      \ee

\newpage
      
{\color{blue}
\dotfill\\
\begin{centering}
{\sl Lecture \#2}\\
\vspace*{-0.20cm}
\end{centering}
\dotfill}
      
      To simplify our subsequent notations, let us set conveniently
      \be \label{psi-tilde-psi}
         {\color{red} \psi_{E\eta}^{\pm}\n(x) \; = \; \sqrt{\frac{m}{2\,\pi\,\hbar^2 K}} \;\, \tilde{\psi}_{E\eta}^{\pm}\n(x) \mez ;}
      \ee
      this converts the LSE (\ref{LSE-implicit-x-take-3}) into
      \be \label{LSE-implicit-x-take-3-tilde}
         \tilde{\psi}_{E\eta}^{\pm}\n(x) \; = \; e^{+i \eta K x} \; \mp \; \frac{i\,m}{\hbar^2 K} \, \int_{a}^{b} \m {\rm d}y \;\,
         e^{\pm i K|x-y|} \; V\m(y) \; \tilde{\psi}_{E\eta}^{\pm}\n(y) \mez .
      \ee   
      Most importantly,\\ equation (\ref{LSE-implicit-x-take-3-tilde}) enables us to understand now the qualitative behavior of $\tilde{\psi}_{E\eta}^{\pm}\n(x)$\\
      in the left asymptotic region of $x \leq a$ and in the right asymptotic region of $x \geq b$.\\
      Indeed, a brief inspection of formula (\ref{LSE-implicit-x-take-3-tilde}) reveals that, inevitably,
      \begin{eqnarray}
         \label{tilde-psi-bc-1} {\color{red} \tilde{\psi}_{E(+1)}^{+}\n(x > b)} & {\color{red} = } & {\color{red} \hspace*{+1.77cm} T_{E(+1)}^{+} \; e^{+i K x} \mez ;}\\
         \label{tilde-psi-bc-2} {\color{red} \tilde{\psi}_{E(+1)}^{+}\n(x < a)} & {\color{red} = } & {\color{red} e^{+i K x} \; + \; R_{E(+1)}^{+} \; e^{-i K x} \n\mez ;}\\
         \label{tilde-psi-bc-3} {\color{red} \tilde{\psi}_{E(-1)}^{+}\n(x > b)} & {\color{red} = } & {\color{red} e^{-i K x} \; + \; R_{E(-1)}^{+} \; e^{+i K x} \n\mez ;}\\
         \label{tilde-psi-bc-4} {\color{red} \tilde{\psi}_{E(-1)}^{+}\n(x < a)} & {\color{red} = } & {\color{red} \hspace*{+1.77cm} T_{E(-1)}^{+} \; e^{-i K x} \mez .}
      \end{eqnarray}
      Here the $T_{E(\pm 1)}^{+}$ and $R_{E(\pm 1)}^{+}$ are some as yet unknown coefficients, given formally by prescriptions
      \begin{eqnarray}
         \label{T-E-formal}
         T_{E(+1)}^{+} & = & 1 \; - \; \frac{i\,m}{\hbar^2 K} \, \int_{a}^{b} \m {\rm d}y \;\, e^{-iKy} \; V\m(y) \;
         \tilde{\psi}_{E(+1)}^{+}\m(y) \mez \m\m\m ;\\
         \label{R-E-formal}
         R_{E(+1)}^{+} & = & \hspace*{+0.40cm} - \; \frac{i\,m}{\hbar^2 K} \, \int_{a}^{b} \m {\rm d}y \;\, e^{+iKy} \; V\m(y)
         \; \tilde{\psi}_{E(+1)}^{+}\m(y) \mez ;
      \end{eqnarray}
      and analogously for $T_{E(-1)}^{+}$, $R_{E(-1)}^{+}$.\\
      $[\,$Note that the r.h.s.~of equations (\ref{T-E-formal})-(\ref{R-E-formal}) contains $\tilde{\psi}_{E\eta}^{\pm}\n(y)$,\\
      \phantom{$[\,$}hence we cannot use (\ref{T-E-formal})-(\ref{R-E-formal}) to directly access $T_{E(+1)}^{+}$ and $R_{E(+1)}^{+}$ prior to knowing $\tilde{\psi}_{E\eta}^{\pm}\n(y)$.$\,]$\\
      Similarly one finds that
      \begin{eqnarray}
         \label{tilde-psi-bc-5} \tilde{\psi}_{E(+1)}^{-}\n(x > b) & = & e^{+i K x} \; + \; R_{E(+1)}^{-} \; e^{-i K x} \mez \m;\\
         \label{tilde-psi-bc-6} \tilde{\psi}_{E(+1)}^{-}\n(x < a) & = & \hspace*{+1.77cm} T_{E(+1)}^{-} \; e^{+i K x} \mez ;\\
         \label{tilde-psi-bc-7} \tilde{\psi}_{E(-1)}^{-}\n(x > b) & = & \hspace*{+1.77cm} T_{E(-1)}^{-} \; e^{-i K x} \mez ;\\
         \label{tilde-psi-bc-8} \tilde{\psi}_{E(-1)}^{-}\n(x < a) & = & e^{-i K x} \; + \; R_{E(-1)}^{-} \; e^{+i K x} \mez ;
      \end{eqnarray}
      where again the $T_{E(\pm 1)}^{-}$ and $R_{E(\pm 1)}^{-}$ are some as yet unknown coefficients.\\
      Equations (\ref{tilde-psi-bc-1})-(\ref{tilde-psi-bc-4}) and (\ref{tilde-psi-bc-5})-(\ref{tilde-psi-bc-8}) combined with formula (\ref{psi-tilde-psi}) supply\\
      appropriate boundary conditions and an adequate normalization factor\\
      for the continuum wavefunctions $\psi_{E\eta}^\pm\n(x)$ solving the time independent Schr\"{o}dinger equation (\ref{TISCHE-psi-E-eta-pm}).\\
      The stationary scattering states $\psi_{E\eta}^\pm\n(x)$ are thus uniquely characterized now.
      
\newpage
      
{\color{blue}
\dotfill\\
\begin{centering}
{\sl Lecture \#2}\\
\vspace*{-0.20cm}
\end{centering}
\dotfill}

\vspace*{+2.00cm}

      Additional remarks to be taken into account:\\
      Firstly, in our later elaborations we shall assign direct physical meaning to $T_{E(\pm 1)}^{+}$ and $R_{E(\pm 1)}^{+}$,\\
      see equations (\ref{phi-out-T-def})-(\ref{phi-out-R-def}), (\ref{T-probability-final})-(\ref{R-probability-final})
      and the accompanying discussion.\\
      Secondly, moment of reflection validates insightful properties
      \be
         \psi_{E(-1)}^{-}\n(x) \; = \; \left(\psi_{E(+1)}^{+}\n(x)\right)^{\m\m\n *} \mez , \mez
         \psi_{E(-1)}^{+}\n(x) \; = \; \left(\psi_{E(+1)}^{-}\n(x)\right)^{\m\m\n *} \mez .
      \ee
      In other words, the advanced scattering wavefunctions $\psi_{E\eta}^-\n(x)$\\
      are just complex conjugates of their retarded counterparts $\psi_{E\eta}^+\n(x)$.\\
      Thirdly, the retarded stationary scattering wavefunctions $\psi_{E\eta}^+\n(x)$\\
      are analyzed once again in a self contained fashion within {\sl Appendix II}.\\
      In particular, their specific boundary conditions (\ref{tilde-psi-bc-1})-(\ref{tilde-psi-bc-4}) are rederived $[\,$see equations (\ref{tilde-psi-E-+-bc-b})-(\ref{tilde-psi-E---bc-a})$\,]$,\\
      two fundamental properties of the coefficients $T_{E(\pm 1)}^{+}$ and $R_{E(\pm 1)}^{+}$ are demonstrated to hold\\
      $[\,$see equations (\ref{T-R-1})-(\ref{T-R-0})$\,]$, and the corresponding\\
      orthonormality \& closure relations are established $[\,$see equations (\ref{psi-E-eta-onrel-take-1}), (\ref{H-closure})$\,]$.\\
      Also the associated Wronskian is examined,\\
      this leads to an additional insight $[\,$see equations (\ref{T-E-+-both-the-same})-(\ref{w-E-x-explicit})$\,]$.\\
      Fourthly, in {\sl Appendix V} we describe an extremely simple numerical algorithm\\
      for calculating the stationary scattering wavefunction $\psi_{E(+1)}^{+}\n(x)$.\\
      This algorithm is based on solving the boundary value problem (\ref{TISCHE-psi-E-eta-pm}), (\ref{tilde-psi-bc-1})-(\ref{tilde-psi-bc-2})\\
      on a discrete $x$-grid using an elementary finite difference method.\\
      An analogous approach can of course be employed also in the case of $\psi_{E(-1)}^{+}\n(x)$ and $\psi_{E(\pm 1)}^{-}\n(x)$.
      
\newpage
      
{\color{blue}
\dotfill\\
\begin{centering}
{\sl Lecture \#3}\\
\vspace*{-0.20cm}
\end{centering}
\dotfill}
      
\item \underline{\bf B.6 Determination of the scattering operator $\bm\hat{\bm S}$: Part \#1}\\
      Let us return now to the scattering matrix element (\ref{S-matels}), $\la \chi_{\rm out} | \hat{S} | \phi_{\rm in} \ra$.\\
      The given prescribed in-wavepacket $| \phi_{\rm in} \ra$ can be expanded as done above in (\ref{phi-in-expansion}).\\
      The associated $t=0$ physical state $| \psi \ra = (\ref{psi-from-phi-in})$\\
      takes then the form (\ref{psi-expansion-plus}), with $| \psi_{E\eta}^{+} \ra$ related to
      $| \phi_{E\eta} \ra$ via the LSE (\ref{LSE-explicit}) or (\ref{LSE-implicit}).\\
      An arbitrarily chosen out-wavepacket $| \chi_{\rm out} \ra$ can be expanded similarly as above in (\ref{phi-out-expansion}), that is,
      \be \label{chi-out-expansion}
         | \chi_{\rm out} \ra \; = \; \int_{0}^{+\infty} \m {\rm d}E \, \sum_\eta \; b_{E\eta} \, | \phi_{E\eta} \ra \mez , \mez
         b_{E\eta} \; = \; \la \phi_{E\eta} | \chi_{\rm out} \ra \mez .
      \ee
      The associated $t=0$ physical state
      \be \label{varphi-from-chi-out}
         | \varphi \ra \; = \; \lim_{t \to +\infty} \, \hat{\sf U}^\dagger\n(t) \, \hat{\sf U}_0(t) \, | \chi_{\rm out} \ra
      \ee
      $[\,${\sl cf.}~relation (\ref{psi-from-phi-out})$\,]$ takes then the form analogous to (\ref{psi-expansion-minus}), namely,
      \be \label{varphi-expansion-minus}
         | \varphi \ra \; = \; \int_{0}^{+\infty} \m {\rm d}E \, \sum_\eta \; b_{E\eta} \, | \psi_{E\eta}^{-} \ra \mez ;
      \ee
      with $| \psi_{E\eta}^{-} \ra$ related to $| \phi_{E\eta} \ra$ via the LSE (\ref{LSE-explicit}) or (\ref{LSE-implicit}).\\
      Our sought scattering matrix element can now be evaluated as follows:
      \begin{eqnarray} \label{S-matel-chi-out-phi-in-final}
         \la \chi_{\rm out} | \hat{S} | \phi_{\rm in} \ra & = & \lim_{T \to +\infty} \, 
         \la \chi_{\rm out} | \, \hat{\sf U}_0^\dagger\n(+T) \; \hat{\sf U}(+T) \; \hat{\sf U}^\dagger\n(-T) \; \hat{\sf U}_0(-T) \,
         | \phi_{\rm in} \ra \; = \nonumber\\
         & = & \lim_{T \to +\infty} \, \left(
         \la \chi_{\rm out} | \, \hat{\sf U}_0^\dagger\n(+T) \; \hat{\sf U}(+T) \right) \left( \hat{\sf U}^\dagger\n(-T) \;
         \hat{\sf U}_0(-T) \, | \phi_{\rm in} \ra \right) \; = \; \la \varphi | \psi \ra \; = \nonumber\\
         & = & \int_{0}^{+\infty} \m {\rm d}E \, \sum_\eta \, \int_{0}^{+\infty} \m {\rm d}E' \, \sum_{\eta'}
         \;\, b_{E\eta}^* \; c_{E'\n\eta'\n} \; \la \psi_{E\eta}^{-} | \psi_{E'\n\eta'\n}^{+} \ra \mez .
      \end{eqnarray}
      We have used along the way the definition (\ref{hat-S-def}) of $\hat{S}$ together with relations (\ref{psi-from-phi-in}),
      (\ref{varphi-from-chi-out}).\\ Now it is meaningful to introduce the overlap entities
      \be \label{S-matel-basic}
         {\color{red} S_{(E\eta)(E'\n\eta'\n)} \; = \; \la \psi_{E\eta}^{-} | \psi_{E'\n\eta'\n}^{+} \ra \mez .}
      \ee
      Clearly, knowledge of $S_{(E\eta)(E'\n\eta'\n)}$\\ enables us to evaluate $\la \chi_{\rm out} | \hat{S} | \phi_{\rm in} \ra$ for any
      chosen $| \phi_{\rm in} \ra$, $| \chi_{\rm out} \ra$ via (\ref{S-matel-chi-out-phi-in-final}).\\
      Hence the fundamental scattering operator $\hat{S}$ is completely determined by $S_{(E\eta)(E'\n\eta'\n)}$.\\
      In addition to the overlaps $S_{(E\eta)(E'\n\eta'\n)}=(\ref{S-matel-basic})$\\
      we also conveniently introduce similarly looking entities
      \be \label{S-matel-basic-2}
         {\color{red} Q_{(E\eta)(E'\n\eta'\n)}^\pm \; = \; \la \psi_{E\eta}^{\pm} | \psi_{E'\n\eta'\n}^{\pm} \ra \mez .}
      \ee
      Quantities (\ref{S-matel-basic})-(\ref{S-matel-basic-2}) can be analyzed solely using the time independent scattering formalism,\\
      and especially using the LSE (\ref{LSE-explicit}), (\ref{LSE-implicit}).\\
      Such an analysis is pursued in the next two paragraphs.
\end{itemize}
      
\newpage
      
{\color{blue}
\dotfill\\
\begin{centering}
{\sl Lecture \#3}\\
\vspace*{-0.20cm}
\end{centering}
\dotfill}
\begin{itemize}
\item \underline{\bf B.7 Mathematical interlude \#1}\\
      Let us evaluate explicitly the overlap entity $Q_{(E\eta)(E'\n\eta'\n)}^+ \m = \la \psi_{E\eta}^{+} | \psi_{E'\n\eta'\n}^{+} \ra$.\\
      One begins with substituting for $\la \psi_{E\eta}^{+} |$ from the explicit LSE (\ref{LSE-explicit}), to get
      \begin{eqnarray} \label{Q-matel-take-1}
         Q_{(E\eta)(E'\n\eta'\n)}^+ & = & \la \phi_{E\eta} | \psi_{E'\n\eta'\n}^{+} \ra \; + \; \la \phi_{E\eta} | \,
         \hat{\sf V} \, \frac{1}{E - \hat{\sf H} - i\,\varepsilon} \, | \psi_{E'\n\eta'\n}^{+} \ra \; = \nonumber\\
         & = & \la \phi_{E\eta} | \psi_{E'\n\eta'\n}^{+} \ra \; + \; \la \phi_{E\eta} | \,
         \hat{\sf V} \, | \psi_{E'\n\eta'\n}^{+} \ra \; \frac{1}{E - E' - i\,\varepsilon} \mez .
      \end{eqnarray}
      We have exploited here the property (\ref{hat-sf-H-eigenproblem}), i.e., $\hat{\sf H}\,| \psi_{E'\n\eta'\n}^{+} \ra\,=\,E'\,| \psi_{E'\n\eta'\n}^{+} \ra$.\\
      The limit of $\varepsilon \to +0$ is of course implicitly understood in (\ref{Q-matel-take-1}).\\
      Our next step is to substitute for $| \psi_{E'\n\eta'\n}^{+} \ra$ from the implicit LSE (\ref{LSE-implicit})\\
      while taking the same value of $\varepsilon \to +0$. This yields
      \begin{eqnarray} \label{Q-matel-take-2}
         Q_{(E\eta)(E'\n\eta'\n)}^+ & = & \la \phi_{E\eta} | \phi_{E'\n\eta'\n} \ra 
         \; + \; \la \phi_{E\eta} | \, \frac{1}{E' - \hat{\sf H}_0 + i \varepsilon }
         \, \hat{\sf V} \, | \psi_{E'\n\eta'\n}^{+} \ra \; + \; \la \phi_{E\eta} | \,
         \hat{\sf V} \, | \psi_{E'\n\eta'\n}^{+} \ra \; \frac{1}{E - E' - i\,\varepsilon} \; = \nonumber\\
         & = & \delta(E-E'\n) \; \delta_{\eta\eta'} \; + \; \la \phi_{E\eta} | \, \hat{\sf V} \, | \psi_{E'\n\eta'\n}^{+} \ra \,
         \left\{ \frac{1}{E' - E + i\,\varepsilon} \, + \, \frac{1}{E - E' - i \varepsilon } \right\} \mez .
      \end{eqnarray}
      We have exploited here the property (\ref{hat-sf-H-0-eigenproblem}), i.e., $\hat{\sf H}_0\,| \phi_{E\eta} \ra\,=\,E\,| \phi_{E\eta} \ra$.\\
      Also the orthonormality relations (\ref{hat-sf-H-0-oncl}) have been incorporated.\\
      Importantly, the $\{ \bm\cdots \}$ term of (\ref{Q-matel-take-2}) equals to zero, therefore $Q_{(E\eta)(E'\n\eta'\n)}^+=\delta(E-E'\n)\,\delta_{\eta\eta'}$.\\
      The same outcome would result also for $Q_{(E\eta)(E'\n\eta'\n)}^-$. Thus, altogether, we may conclude that
      \be \label{LSE-orthonormality}
         {\color{red} \la \psi_{E\eta}^{\pm} | \psi_{E'\n\eta'\n}^{\pm} \ra \; = \; \delta(E-E'\n) \; \delta_{\eta\eta'} \mez .}
      \ee
      {\color{red} The just derived formula (\ref{LSE-orthonormality}) reveals that the solutions $| \psi_{E\eta}^{\pm} \ra$ of the LSE
      (\ref{LSE-explicit}) or (\ref{LSE-implicit})\\ satisfy orthonormality relations analogous to those possessed by $| \phi_{E\eta} \ra$
      and listed in (\ref{hat-sf-H-0-oncl}).\\ The corresponding eigenvalue equations (\ref{hat-sf-H-0-eigenproblem}) and (\ref{hat-sf-H-eigenproblem})
      are worth recalling in this context.}\\
      In passing we note that retarded version of the orthonormality relations (\ref{LSE-orthonormality})\\
      has been derived by elementary means also in {\sl Appendix II}, see equation (\ref{psi-E-eta-onrel-take-1}).
\end{itemize}
      
\newpage
      
{\color{blue}
\dotfill\\
\begin{centering}
{\sl Lecture \#3}\\
\vspace*{-0.20cm}
\end{centering}
\dotfill}
\begin{itemize}
\item \underline{\bf B.8 Mathematical interlude \#2}\\
      Next, let us evaluate explicitly the fundamental entity $S_{(E\eta)(E'\n\eta'\n)}=(\ref{S-matel-basic})$ relevant for scattering.\\
      We shall proceed analogously as in the previous paragraph. Namely,\\
      one begins with substituting for $\la \psi_{E\eta}^{-} |$ from the explicit LSE (\ref{LSE-explicit}), to get
      \begin{eqnarray} \label{S-matel-take-1}
         S_{(E\eta)(E'\n\eta'\n)} & = & \la \phi_{E\eta} | \psi_{E'\n\eta'\n}^{+} \ra \; + \; \la \phi_{E\eta} | \,
         \hat{\sf V} \, \frac{1}{E - \hat{\sf H} + i\,\varepsilon} \, | \psi_{E'\n\eta'\n}^{+} \ra \; = \nonumber\\
         & = & \la \phi_{E\eta} | \psi_{E'\n\eta'\n}^{+} \ra \; + \; \la \phi_{E\eta} | \,
         \hat{\sf V} \, | \psi_{E'\n\eta'\n}^{+} \ra \; \frac{1}{E - E' + i\,\varepsilon} \mez .
      \end{eqnarray}
      We have exploited here the property (\ref{hat-sf-H-eigenproblem}), i.e., $\hat{\sf H}\,| \psi_{E'\n\eta'\n}^{-} \ra\,=\,E'\,| \psi_{E'\n\eta'\n}^{-} \ra$.\\
      The limit of $\varepsilon \to +0$ is of course implicitly understood in (\ref{S-matel-take-1}).\\
      Our next step is to substitute for $| \psi_{E'\n\eta'\n}^{+} \ra$ from the implicit LSE (\ref{LSE-implicit})\\
      while taking the same value of $\varepsilon \to +0$. This yields
      \begin{eqnarray} \label{S-matel-take-2}
         S_{(E\eta)(E'\n\eta'\n)} & = & \la \phi_{E\eta} | \phi_{E'\n\eta'\n} \ra 
         \; + \; \la \phi_{E\eta} | \, \frac{1}{E' - \hat{\sf H}_0 + i \varepsilon }
         \, \hat{\sf V} \, | \psi_{E'\n\eta'\n}^{+} \ra \; + \; \la \phi_{E\eta} | \,
         \hat{\sf V} \, | \psi_{E'\n\eta'\n}^{+} \ra \; \frac{1}{E - E' + i\,\varepsilon} \; = \nonumber\\
         & = & \delta(E-E'\n) \; \delta_{\eta\eta'} \; + \; \la \phi_{E\eta} | \, \hat{\sf V} \, | \psi_{E'\n\eta'\n}^{+} \ra \,
         \left\{ \frac{1}{E' - E + i\,\varepsilon} \, + \, \frac{1}{E - E' + i \varepsilon } \right\} \mez .
      \end{eqnarray}
      We have exploited here the property (\ref{hat-sf-H-0-eigenproblem}), i.e., $\hat{\sf H}_0\,| \phi_{E\eta} \ra\,=\,E\,| \phi_{E\eta} \ra$.\\
      Also the orthonormality relations (\ref{hat-sf-H-0-oncl}) have been incorporated.\\
      Importantly, the $\{ \bm\cdots \}$ term of (\ref{S-matel-take-2}) does not vanish, but equals to
      \be
         \{ \bm\cdots \} \; = \; -\,2\,i\,\pi \;\, \frac{1}{\pi} \, \frac{\varepsilon}{(E-E'\n)^2+\varepsilon^2} \; = \;
         -\,2\,i\,\pi \;\, \delta_\varepsilon\m(E-E'\n) \mez ;
      \ee
      with the Lorentzian profile function
      \be
         \delta_\varepsilon\m(E-E'\n) \; = \; \frac{1}{\pi} \, \frac{\varepsilon}{(E-E'\n)^2+\varepsilon^2}
      \ee
      reducing to $\delta(E-E'\n)$ in the limit of $\varepsilon \to +0$. If so, then
      \begin{eqnarray} \label{S-matel-take-3}
         {\color{red} S_{(E\eta)(E'\n\eta'\n)}} & {\color{red} = } & {\color{red} \delta(E-E'\n) \; \delta_{\eta\eta'} \; - \; 2\,i\,\pi \;\, \delta(E-E'\n) \;
         T_{(E\eta)(E\eta')} \mez ;}
      \end{eqnarray}
      where by definition
      \begin{eqnarray}
         \label{T-matels-1} {\color{red} T_{(E\eta)(E\eta')}} & {\color{red} = } &
         {\color{red} \la \phi_{E\eta} | \, \hat{\sf V} \, | \psi_{E\eta'}^{+} \ra \; = } \\
         \label{T-matels-2} & {\color{red} = } & {\color{red}
         \la \phi_{E\eta} | \, \hat{\sf V} \, + \, \hat{\sf V} \, \frac{1}{E - \hat{\sf H} + i\,0_+ } \, \hat{\sf V}
         \, | \phi_{E\eta'} \ra \mez . }
      \end{eqnarray}
      In (\ref{T-matels-2}) we have again substituted for $| \psi_{E\eta'}^{+} \ra$ from the explicit LSE (\ref{LSE-explicit}).
\end{itemize}
      
\newpage
      
{\color{blue}
\dotfill\\
\begin{centering}
{\sl Lecture \#3}\\
\vspace*{-0.20cm}
\end{centering}
\dotfill}
\begin{itemize}
\item \underline{\bf B.9 Determination of the scattering operator $\bm\hat{\bm S}$: Part \#2}\\
      We have already shown that the fundamental scattering operator $\hat{S}$\\
      is completely determined by the matrix elements $S_{(E\eta)(E'\n\eta'\n)}=(\ref{S-matel-basic})=(\ref{S-matel-take-3})$.\\
      The r.h.s.~of equation (\ref{S-matel-take-3}) is proportional to $\delta(E-E'\n)$.\\
      This so called on-shell energy factor\\ indicates that the energy $E$ is conserved in scattering processes.\\
      Quantity (\ref{T-matels-2}) represents the so called on-shell $T$-matrix element.\\
      It can be conveniently viewed as
      \be \label{T-matels-using-hat-T}
         {\color{red} T_{(E\eta)(E\eta')} \; = \; \la \phi_{E\eta} | \, \hat{T}(E) \, | \phi_{E\eta'} \ra \mez ;}
      \ee
      where by definition
      \be \label{hat-T-E-def}
         {\color{red} \hat{T}(E) \; = \; \hat{\sf V} \; + \; \hat{\sf V} \, \frac{1}{E - \hat{\sf H} + i\,0_+ } \, \hat{\sf V} \mez .}
      \ee
      Prescription (\ref{hat-T-E-def}) introduces the so called {\color{red} Transition Operator or $\hat{T}$-operator}\\
      which depends parameterically upon the impact energy $E$.\\
      Practical evaluation of $T_{(E\eta)(E\eta')}$ for a concrete physical problem\\
      can be based upon two main strategies:
      \begin{itemize}
      \item[{\it i)}]
         Numerical construction of the retarded stationary scattering states $| \psi_{E\eta}^{+} \ra$,\\
         followed by an application of the formula (\ref{T-matels-1}).\\
         Recall that $\psi_{E(+1)}^{+}\n(x) = \la x | \psi_{E(+1)}^{+} \ra$ represents\\ an unique solution of
         the boundary value problem (\ref{TISCHE-psi-E-eta-pm}), (\ref{tilde-psi-bc-1})-(\ref{tilde-psi-bc-2}).\\
         Numerical treatment of the just mentioned boundary value problem is straightforward,\\
         the corresponding algorithm is described in a self contained fashion in {\sl Appendix V}.\\
         Treatment of $\psi_{E(-1)}^{+}\n(x) = \la x | \psi_{E(-1)}^{+} \ra$ is completely analogous.\\
         The just presented route {\it i)} will be pursued forward in our forthcoming elaborations.
      \item[{\it ii)}]
         An analytic (perturbative) calculation of $| \psi_{E\eta}^{\pm} \ra$ based upon the Born series expansion (\ref{LSE-Born}),\\
         again followed by an application of the formula (\ref{T-matels-1}). As noted already,\\
         the Born series approach is most convenient/justified\\ when one encounters scattering by weak potentials.\\
         In fact, there exists even a quicker route leading towards the Born expansion of $\hat{T}(E)=(\ref{hat-T-E-def})$,\\
         which is based upon the Born series expansion of the retarded Green operator $(E-\hat{\sf H} + i\,0_+)^{-1}$\m.\\
         Such an expansion is derived in {\sl Appendix IV}, see equation (\ref{hat-G-E-LSE-Born}).\\
         Combination of (\ref{hat-G-E-LSE-Born}) and (\ref{hat-T-E-def}) yields the desired perturbative Born series expansion
         for $\hat{T}(E)$:
         \begin{eqnarray}
            \hspace*{-0.50cm}
            \hat{T}(E) & = & \hat{\sf V} \; + \; \hat{\sf V} \, \frac{1}{E - \hat{\sf H}_0 + i\,0_+ } \, \hat{\sf V}
            \; + \; \hat{\sf V} \, \frac{1}{E-\hat{\sf H}_0 + i\,0_+} \, \hat{\sf V} \, \frac{1}{E-\hat{\sf H}_0 + i\,0_+} \,
            \hat{\sf V} \; + \; \bm\ldots \mez .
         \end{eqnarray}
         The just presented route {\it ii)} possesses its important merits,\\
         but we shall not pursue forward this direction of thought in our forthcoming elaborations.
      \end{itemize}
\end{itemize}
      
\newpage
      
{\color{blue}
\dotfill\\
\begin{centering}
{\sl Lecture \#3}\\
\vspace*{-0.20cm}
\end{centering}
\dotfill}
\begin{itemize}
\item \underline{\bf B.10 Determination of the scattering operator $\bm\hat{\bm S}$: Part \#3}\\
      Let us return to our discussion of $\hat{S}$ which began in subsection B.6.\\
      Suppose we aim at determining $| \phi_{\rm out} \ra$ for a given prescribed $| \phi_{\rm in} \ra$.\\
      One may take $| \phi_{\rm in} \ra$ and proceed via (\ref{phi-in-expansion}), (\ref{LSE-explicit}) or (\ref{LSE-implicit}),
      (\ref{psi-expansion-plus}) to get the associated $| \psi \ra$.\\ One finds that
      \be \label{psi-from-phi-in-B10}
         | \psi \ra \; = \; \int_{0}^{+\infty} \m {\rm d}E \, \sum_\eta \; | \psi_{E\eta}^{+} \ra \, \la \phi_{E\eta} | \phi_{\rm in} \ra \mez .
      \ee
      Subsequently we wish to apply (\ref{psi-expansion-minus}), (\ref{LSE-explicit}) or (\ref{LSE-implicit}), (\ref{phi-out-expansion}) and
      calculate the sought $| \phi_{\rm out} \ra$.\\ Equations relevant for this purpose take the following appearance:
      \be \label{psi-from-phi-out-B10}
         | \psi \ra \; = \; \int_{0}^{+\infty} \m {\rm d}E \, \sum_\eta \; | \psi_{E\eta}^{-} \ra \, \la \phi_{E\eta} | \phi_{\rm out} \ra \mez ;
      \ee
      \be \label{phi-out-from-phi-E-eta-B10}
         | \phi_{\rm out} \ra \; = \; \int_{0}^{+\infty} \m {\rm d}E \, \sum_\eta \; | \phi_{E\eta} \ra \, \la \phi_{E\eta} | \phi_{\rm out} \ra \mez .
      \ee
      Formula (\ref{phi-out-from-phi-E-eta-B10}) requires knowledge of an overlap $\la \phi_{E\eta} | \phi_{\rm out} \ra$.\\
      Yet the same overlap appears also in (\ref{psi-from-phi-out-B10}), and can be thus obtained from $| \psi \ra$\\
      by exploiting the orthonormality property (\ref{LSE-orthonormality}) in combination with (\ref{psi-from-phi-in-B10}).\\
      Indeed, one gets
      \be
         \la \phi_{E\eta} | \phi_{\rm out} \ra \; = \; \la \psi_{E\eta}^{-} | \psi \ra \; = \;
         \int_{0}^{+\infty} \m {\rm d}E'\n \, \sum_{\eta'}\n \;\, \la \psi_{E\eta}^{-} | \psi_{E'\n\eta'\n}^{+} \ra \, \la \phi_{E'\n\eta'\n} | \phi_{\rm in} \ra \mez .
      \ee
      Returning back to (\ref{phi-out-from-phi-E-eta-B10}), we have explicitly
      \be \label{phi-out-from-phi-in-B10}
         | \phi_{\rm out} \ra \; = \; \int_{0}^{+\infty} \m {\rm d}E \, \sum_\eta \, \int_{0}^{+\infty} \m {\rm d}E'\n \, \sum_{\eta'}\n \;\, | \phi_{E\eta} \ra \,
         \la \psi_{E\eta}^{-} | \psi_{E'\n\eta'\n}^{+} \ra \, \la \phi_{E'\n\eta'\n} | \phi_{\rm in} \ra \mez .
      \ee
      The just derived equation (\ref{phi-out-from-phi-in-B10}) enables us\\
      to determine the sought outgoing asymptotic state $| \phi_{\rm out} \ra$\\
      for any given incoming asymptotic state $| \phi_{\rm in} \ra$,\\
      provided only that the overlap entity
      $\la \psi_{E\eta}^{-} | \psi_{E'\n\eta'\n}^{+} \ra=(\ref{S-matel-basic})=(\ref{S-matel-take-3})$ has been evaluated.\\
      {\color{red} Thus the map $| \phi_{\rm in} \ra \mapsto | \phi_{\rm out} \ra$ is explicitly constructed now,\\ solely within the time independent scattering formalism based upon LSE.}\\
      Moreover, a comparison between our explicit result (\ref{phi-out-from-phi-in-B10}) and the formal definition (\ref{hat-S-formal}) of $\hat{S}$\\
      implies inevitably
      \be \label{hat-S-final-explicit}
         {\color{red} \hat{S} \; = \; \int_{0}^{+\infty} \m {\rm d}E \, \sum_\eta \, \int_{0}^{+\infty} \m {\rm d}E'\n \, \sum_{\eta'}\n \;\, | \phi_{E\eta} \ra \,
         \la \psi_{E\eta}^{-} | \psi_{E'\n\eta'\n}^{+} \ra \, \la \phi_{E'\n\eta'\n} | \mez ;}
      \ee
      or, in other words,
      \be \label{hat-S-final-explicit-matels}
         {\color{red} \la \phi_{E\eta} | \, \hat{S} \, | \phi_{E'\n\eta'\n} \ra \; = \;
         \la \psi_{E\eta}^{-} | \psi_{E'\n\eta'\n}^{+} \ra \; = \; S_{(E\eta)(E'\n\eta'\n)} \; = \;
         (\ref{S-matel-take-3}) \mez . }
      \ee
      Meaning that {\color{red} the fundamental scattering operator $\hat{S}$ is explicitly and simply determined}\\
      {\color{red} by solutions $| \psi_{E\eta}^{\pm} \ra$ of the LSE (\ref{LSE-explicit}) or (\ref{LSE-implicit})}.

\newpage
      
{\color{blue}
\dotfill\\
\begin{centering}
{\sl Lecture \#3}\\
\vspace*{-0.20cm}
\end{centering}
\dotfill}

\vspace*{+2.00cm}

      Let us look again at the scattering matrix elements (\ref{S-matel-chi-out-phi-in-final}).\\
      Equations (\ref{hat-S-final-explicit-matels}) and (\ref{S-matel-take-3}) enable us to simplify
      (\ref{S-matel-chi-out-phi-in-final}) into the final form
      \begin{eqnarray} \label{S-matel-chi-out-phi-in-finalized-prelim}
         \la \chi_{\rm out} | \hat{S} | \phi_{\rm in} \ra & = &
         \int_{0}^{+\infty} \m {\rm d}E \, \sum_\eta \; b_{E\eta}^* \; c_{E\eta} \; - \; 2\,i\,\pi
         \int_{0}^{+\infty} \m {\rm d}E \, \sum_\eta \, \sum_{\eta'} \;\, b_{E\eta}^* \; c_{E\eta'} \; T_{(E\eta)(E\eta')}
         \; = \nonumber\\
         & = & \la \chi_{\rm out} | \phi_{\rm in} \ra \; - \; 2\,i\,\pi
         \int_{0}^{+\infty} \m {\rm d}E \, \sum_\eta \, \sum_{\eta'} \;\, b_{E\eta}^* \; c_{E\eta'} \; T_{(E\eta)(E\eta')} \mez .
      \end{eqnarray}
      Here we recall that $c_{E\eta}=\la \phi_{E\eta} | \phi_{\rm in} \ra$ and $b_{E\eta}=\la \phi_{E\eta} | \chi_{\rm out} \ra$
      as stated in (\ref{phi-in-expansion}) and (\ref{chi-out-expansion}).
\end{itemize}

\newpage
      
{\color{blue}
\dotfill\\
\begin{centering}
{\sl Lecture \#4}\\
\vspace*{-0.20cm}
\end{centering}
\dotfill}
\begin{itemize}
\item \underline{\bf B.11 Transmission and reflection phenomena: Take \#1}\\
      Our formal treatment of the scattering operator $\hat{S}$ has been finalized.\\
      Let us explore now in more detail what does $\hat{S}$ actually tell us about scattering.\\
      Specifically, we shall focus in the two subsequent paragraphs\\
      on {\color{red} deriving simple expressions for\\
      the probability of transmission/reflection of our quantum particle\\
      through/from the interaction region $(a,b)$ of the nonvanishing potential $V\m(x)$. }\\
      We start by putting together formulas (\ref{S-matel-take-3}), (\ref{phi-out-from-phi-in-B10}),
      (\ref{hat-S-final-explicit-matels}). This yields
      \begin{eqnarray} \label{S-matel-chi-out-phi-in-finalized}
         | \phi_{\rm out} \ra & = & | \phi_{\rm in} \ra \; - \; 2\,i\,\pi
         \int_{0}^{\infty} \m {\rm d}E \; \sum_{\eta} \sum_{\eta'} \; | \phi_{E\eta} \ra \;\, T_{(E\eta)(E\eta')}
         \; \la \phi_{E\eta'\n} | \phi_{\rm in} \ra \mez .
      \end{eqnarray}
      The associated on-shell $T$-matrix element $T_{(E\eta)(E\eta')}$ is expressible as a spatial integral\\
      using equations (\ref{T-matels-1}) and (\ref{phi-E-eta-def}). One has
      \be \label{T-matel-gen}
         T_{(E\eta)(E\eta'\n)} \; = \; \sqrt{\frac{m}{2\,\pi\,\hbar^2K}} \, \int_{-\infty}^{+\infty} \m {\rm d}x \;\,
         e^{+i\eta K x} \; V\m(x) \; \psi_{E\eta'\n}^+\n(x) \mez .
      \ee
      Hereafter we shall assume\\
      that the chosen incoming wavepacket $| \phi_{\rm in} \ra$ moves from the left towards the right,\\
      and is thus populated only by positive momenta. Hence
      \be \label{phi-in-positive-momenta}
         \la \phi_{E(-1)} | \phi_{\rm in} \ra \; = \; 0 \mez ;
      \ee   
      valid for all $E$. Formula (\ref{S-matel-chi-out-phi-in-finalized}) is then reduced into
      \begin{eqnarray} \label{S-matel-chi-out-phi-in-take-1}
         | \phi_{\rm out} \ra & = & | \phi_{\rm in} \ra \; - \; 2\,i\,\pi
         \int_{0}^{\infty} \m {\rm d}E \; \sum_{\eta} \;\, | \phi_{E\eta} \ra \;\, T_{(E\eta)(E(+1))}
         \; \la \phi_{E(+1)} | \phi_{\rm in} \ra \; = \nonumber\\
         & = & | \phi_{\rm in} \ra \; - \; 2\,i\,\pi
         \int_{0}^{\infty} \m {\rm d}E \;\, | \phi_{E(+1)} \ra \;\, T_{(E(+1))(E(+1))}
         \; \la \phi_{E(+1)} | \phi_{\rm in} \ra \\
         &  & \phantom{| \phi_{\rm in} \ra} \; - \; 2\,i\,\pi
         \int_{0}^{\infty} \m {\rm d}E \;\, | \phi_{E(-1)} \ra \;\, T_{(E(-1))(E(+1))}
         \; \la \phi_{E(+1)} | \phi_{\rm in} \ra \mez . \nonumber
      \end{eqnarray}
      Most importantly, the two $T$-matrix elements entering into (\ref{S-matel-chi-out-phi-in-take-1})\\
      are expressible solely in terms of the coefficients $T_{E(+1)}^{+}$ and $R_{E(+1)}^{+}$\\
      which we have encountered before when dealing with the boundary conditions for
      $\psi_{E(+1)}^+\n(x)$,\\ see equations
      (\ref{tilde-psi-bc-1}), (\ref{tilde-psi-bc-2}) above. Namely,\\
      after combining formula (\ref{T-matel-gen}) with
      (\ref{psi-tilde-psi}), (\ref{T-E-formal})-(\ref{R-E-formal}) we find that
      \begin{eqnarray}
         \label{T-matel-1D-take-3}
         T_{(E(+1))(E(+1))} & = & \frac{i}{2\,\pi} \; \Bigl(\,T_{E(+1)}^{+}-1\,\Bigr) \mez ;\\
         \label{T-matel-1D-take-2}
         T_{(E(-1))(E(+1))} & = & \frac{i}{2\,\pi} \; R_{E(+1)}^{+} \hspace*{+1.78cm} .
      \end{eqnarray}
      
\newpage

{\color{blue}
\dotfill\\
\begin{centering}
{\sl Lecture \#4}\\
\vspace*{-0.20cm}
\end{centering}
\dotfill}
     
      An insertion of (\ref{T-matel-1D-take-3})-(\ref{T-matel-1D-take-2}) inside (\ref{S-matel-chi-out-phi-in-take-1})
      provides in turn a compact outcome
      \begin{eqnarray} \label{S-matel-chi-out-phi-in-take-2}
         | \phi_{\rm out} \ra
         & = & | \phi_{\rm in} \ra \; - \; \int_{0}^{\infty} \m {\rm d}E \;\, | \phi_{E(+1)} \ra
         \la \phi_{E(+1)} | \phi_{\rm in} \ra \nonumber\\
         & + & \int_{0}^{\infty} \m {\rm d}E \;\, | \phi_{E(+1)} \ra \;\, T_{E(+1)}^{+} \; \la \phi_{E(+1)} | \phi_{\rm in} \ra \\
         & + & \int_{0}^{\infty} \m {\rm d}E \;\, | \phi_{E(-1)} \ra \;\, R_{E(+1)}^{+} \; \la \phi_{E(+1)} | \phi_{\rm in} \ra
         \mez . \nonumber
     \end{eqnarray}
     The first line on the r.h.s.~of (\ref{S-matel-chi-out-phi-in-take-2}) actually vanishes,\\
     as one may verify immediately by recalling (\ref{phi-in-positive-momenta}) and the closure property (\ref{hat-sf-H-0-oncl}).\\
     Our outgoing asymptotic state $| \phi_{\rm out} \ra$ possesses thus a simply looking final appearance
     \begin{eqnarray} \label{S-matel-chi-out-phi-in-take-3}
        | \phi_{\rm out} \ra
        & = & \int_{0}^{\infty} \m {\rm d}E \;\, | \phi_{E(+1)} \ra \;\, T_{E(+1)}^{+} \; \la \phi_{E(+1)} | \phi_{\rm in} \ra \\
        & + & \int_{0}^{\infty} \m {\rm d}E \;\, | \phi_{E(-1)} \ra \;\, R_{E(+1)}^{+} \; \la \phi_{E(+1)} | \phi_{\rm in} \ra \mez .
        \nonumber
     \end{eqnarray}
      {\color{red} Instead of (\ref{S-matel-chi-out-phi-in-take-3}) one can write even more insightfully}
      \be \label{S-matel-chi-out-phi-in-take-4}
         {\color{red} | \phi_{\rm out} \ra \; = \; | \phi_{\rm out}^{\rm T} \ra \; + \; | \phi_{\rm out}^{\rm R} \ra \mez ;}
      \ee
      where by definition
      \begin{eqnarray}
         \label{phi-out-T-def} {\color{red} | \phi_{\rm out}^{\rm T} \ra } & {\color{red} = } &
         {\color{red} \int_{0}^{\infty} \m {\rm d}E \;\, | \phi_{E(+1)} \ra \;\, T_{E(+1)}^{+} \; \la \phi_{E(+1)} | \phi_{\rm in} \ra \mez ; }\\
         \label{phi-out-R-def} {\color{red} | \phi_{\rm out}^{\rm R} \ra } & {\color{red} = } &
         {\color{red} \int_{0}^{\infty} \m {\rm d}E \;\, | \phi_{E(-1)} \ra \;\, R_{E(+1)}^{+} \; \la \phi_{E(+1)} | \phi_{\rm in} \ra \mez \m\m . }
      \end{eqnarray}
      Clearly, $| \phi_{\rm out}^{\rm T} \ra$ populates only positive momenta (much as $| \phi_{\rm in} \ra$ does),\\
      {\color{red} we recognize thus $| \phi_{\rm out}^{\rm T} \ra$ as the outgoing transmitted wavepacket}.\\
      On the other hand, $| \phi_{\rm out}^{\rm R} \ra$ populates only negative momenta (as opposed to $| \phi_{\rm in} \ra$),\\
      {\color{red} we recognize thus $| \phi_{\rm out}^{\rm R} \ra$ as the outgoing reflected wavepacket}.\\
      Our just presented findings justify to name\\
      {\color{red} an entity $T_{E(+1)}^{+}$ as the transmission coefficient}, and\\
      {\color{red} an entity $R_{E(+1)}^{+}$ as the reflection coefficient}.\\
      Recall that numerical calculation of $T_{E(+1)}^{+}$ and $R_{E(+1)}^{+}$ for a given potential $V\m(x)$\\
      is very straightforward. Indeed, for any prescribed energy $E>0$\\
      one solves the boundary value problem (\ref{TISCHE-psi-E-eta-pm}), (\ref{tilde-psi-bc-1})-(\ref{tilde-psi-bc-2})
      for $\psi_{E(+1)}^{+}\n(x)$,\\ using an algorithm which is described in a self contained fashion in {\sl Appendix V}.\\
      Before proceeding further,\\ let us point out that results equivalent to (\ref{S-matel-chi-out-phi-in-take-4}),
      (\ref{phi-out-T-def})-(\ref{phi-out-R-def})\\ have been derived in {\sl Appendix III} using an alternative and perhaps more
      elementary approach\\ (based upon the stationary phase method). $[\,$See especially equations (\ref{Psi-t-x-right-take-3})
      and (\ref{Psi-t-x-left-take-3-ref}).$\,]$
     
\newpage

{\color{blue}
\dotfill\\
\begin{centering}
{\sl Lecture \#4}\\
\vspace*{-0.20cm}
\end{centering}
\dotfill}
      
\item \underline{\bf B.12 Transmission and reflection phenomena: Take \#2}\\
      Quantum mechanical norm conservation\\
      $[\,$which is encoded into unitarity of $\hat{S}$ in the scattering sector of our quantum state space$\,]$\\
      guarantees that $\la \phi_{\rm out} | \phi_{\rm out} \ra = \la \phi_{\rm in} | \phi_{\rm in} \ra = 1$. Yet formulas
      (\ref{S-matel-chi-out-phi-in-take-4}), (\ref{phi-out-T-def})-(\ref{phi-out-R-def}) imply
      \be \label{T-R-probabilities}
         1 \; = \; \la \phi_{\rm out} | \phi_{\rm out} \ra \; = \; \la \phi_{\rm out}^{\rm T} | \phi_{\rm out}^{\rm T} \ra
         \; + \; \la \phi_{\rm out}^{\rm R} | \phi_{\rm out}^{\rm R} \ra \mez ;
      \ee
      with
      \begin{eqnarray}
         \label{T-probability} \la \phi_{\rm out}^{\rm T} | \phi_{\rm out}^{\rm T} \ra & = &
         \int_{0}^{\infty} \m {\rm d}E \;\, \Bigl| \, T_{E(+1)}^{+} \, \Bigr|^2 \, \Bigl| \, \la \phi_{E(+1)} | \phi_{\rm in} \ra \Bigr|^2
         \mez ;\\
         \label{R-probability} \la \phi_{\rm out}^{\rm R} | \phi_{\rm out}^{\rm R} \ra & = &
         \int_{0}^{\infty} \m {\rm d}E \;\, \Bigl| \, R_{E(+1)}^{+} \, \Bigr|^2 \, \Bigl| \, \la \phi_{E(+1)} | \phi_{\rm in} \ra \Bigr|^2
         \mez \m\m .
      \end{eqnarray}
      Equation (\ref{T-R-probabilities}) assigns an important physical interpretation\\ to the self overlaps
      $\la \phi_{\rm out}^{\rm T} | \phi_{\rm out}^{\rm T} \ra$ and $\la \phi_{\rm out}^{\rm R} | \phi_{\rm out}^{\rm R} \ra$:\\
      {\color{red} Quantity $\la \phi_{\rm out}^{\rm T} | \phi_{\rm out}^{\rm T} \ra = (\ref{T-probability})$ possesses the meaning of the
      overall transmission probability, whereas\\ quantity $\;\la \phi_{\rm out}^{\rm R} | \phi_{\rm out}^{\rm R} \ra =
      (\ref{R-probability})$ possesses the meaning of the overall reflection probability.}\\
      An additional simplification becomes possible when the incoming wavepacket $| \phi_{\rm in} \ra$\\
      populates only a narrow interval of energies, say $E_0 - \Delta_E/2 \leq E \leq E_0 + \Delta_E/2$,
      with $\Delta_E \to 0$.\\ Both quantities $T_{E(+1)}^{+}$ and $R_{E(+1)}^{+}$ are continuous functions of $E$,\\
      and remain thus almost constant inside the mentioned energy interval for small enough $\Delta_E$.\\
      The transmission/reflection probabilities (\ref{T-probability})-(\ref{R-probability}) then boil down into
      \begin{eqnarray}
         \label{T-probability-final} \la \phi_{\rm out}^{\rm T} | \phi_{\rm out}^{\rm T} \ra & = &
         \Bigl| \, T_{E_0(+1)}^{+} \, \Bigr|^2 \, \int_{0}^{\infty} \m {\rm d}E \;\, \Bigl| \, \la \phi_{E(+1)} | \phi_{\rm in} \ra \Bigr|^2
         \; = \; \Bigl| \, T_{E_0(+1)}^{+} \, \Bigr|^2 \mez ;\\
         \label{R-probability-final} \la \phi_{\rm out}^{\rm R} | \phi_{\rm out}^{\rm R} \ra & = &
         \Bigl| \, R_{E_0(+1)}^{+} \, \Bigr|^2 \, \int_{0}^{\infty} \m {\rm d}E \;\, \Bigl| \, \la \phi_{E(+1)} | \phi_{\rm in} \ra \Bigr|^2
         \, = \; \Bigl| \, R_{E_0(+1)}^{+} \, \Bigr|^2 \mez \m\m\m\m .
      \end{eqnarray}
      {\color{red} We have just proven that entities $\Bigl|\,T_{E(+1)}^{+}\,\Bigr|^2$ and $\Bigl|\,R_{E(+1)}^{+}\,\Bigr|^2$\\
      are interpreted physically as the transmission/reflection probabilities\\
      corresponding to the given impact energy $E$.\\
      Probability conservation is granted via equation (\ref{T-R-probabilities}), which ensures having}
      \be \label{probability-conservation}
         {\color{red} \Bigl|\,T_{E(+1)}^{+}\,\Bigr|^2 \; + \; \Bigl|\,R_{E(+1)}^{+}\,\Bigr|^2 \; = \; 1 \mez ;}
      \ee
      {\color{red} valid for all $E$.}\\
      Note that property (\ref{probability-conservation}) has been proven also in {\sl Appendix II}\\
      by alternative and more elementary means, see equation (\ref{T-R-1}).\\
      Note also that {\sl Appendix III} contains results equivalent to (\ref{T-probability})-(\ref{R-probability})
      and (\ref{T-probability-final})-(\ref{R-probability-final})\\
      which are derived using an alternative and perhaps more elementary approach\\
      (stationary phase method). $[\,$See especially equations (\ref{Psi-rightarrow-t-x-sqnrm}) and
      (\ref{Psi-leftarrow-t-x-ref-sqnrm}).$\,]$\\
      Not only amplitudes but also phases of $T_{E(+1)}^{+}$, $R_{E(+1)}^{+}$ are physically significant.\\
      Indeed, it can be shown that the values of ${\rm arg}\,T_{E(+1)}^{+}$ and ${\rm arg}\,R_{E(+1)}^{+}$ are related\\
      to the time which the transmitted/reflected particle spends inside the interaction region $(a,b)$.\\
      $[\,$We shall not delve deeper into these matters within the present course.$\,]$\\
      All what we said above about $T_{E(+1)}^{+}$ and $R_{E(+1)}^{+}$ applies of course also to entities
      $T_{E(-1)}^{+}$, $R_{E(-1)}^{+}$\\ $[\,$These obviously describe a situation when $| \phi_{\rm in}\ra$
      populates only negative momenta.$\,]$
      
\newpage

{\color{blue}
\dotfill\\
\begin{centering}
{\sl Lecture \#4}\\
\vspace*{-0.20cm}
\end{centering}
\dotfill}
      
\item \underline{\bf B.13 Transmission and reflection phenomena: An illustrative numerical example}\\
      All our theoretical elaborations pursued so far can be now conveniently illustrated\\
      by a straightforward numerical calculation, performed for a simple model problem.\\
      Namely, we shall take $m=1$, $\hbar=1$, and consider a potential function of the form
      \be \label{V-Jolanta}
         V\m(x) \; = \; \left( 0.5\,x^2 - 0.8 \right) \, e^{-0.1\,x^2} \mez ;
      \ee
      see Ref.~\cite{NM-book}.
      The profile of $V\m(x)$ is plotted below:
      \vspace*{-0.30cm}
      \begin{figure}[h!]
      \hspace*{+4.50cm} \includegraphics[angle=270,scale=0.5]{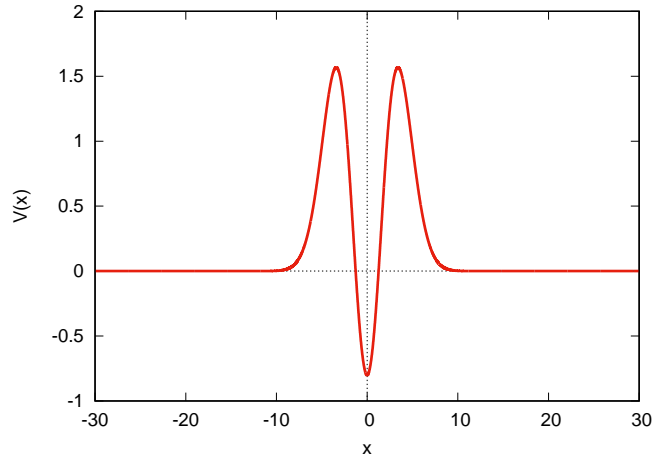}\\
      \vspace*{-0.30cm}
      \caption{ Our model potential $V\m(x)=(\ref{V-Jolanta})$. }
      \end{figure}
      
      One can see that $V\m(x)$ is a symmetric potential consisting of a single well and two barriers.\\
      A simple numerical algorithm of {\sl Appendix V} can now be employed\\
      to calculate the corresponding energy dependent transmission probability $\Bigl|\,T_{E(+1)}^{+}\,\Bigr|^2$\m.\\
      The resulting outcome is depicted below:
      \vspace*{-0.30cm}
      \begin{figure}[h!]
      \hspace*{+3.00cm} \includegraphics[angle=270,scale=0.6]{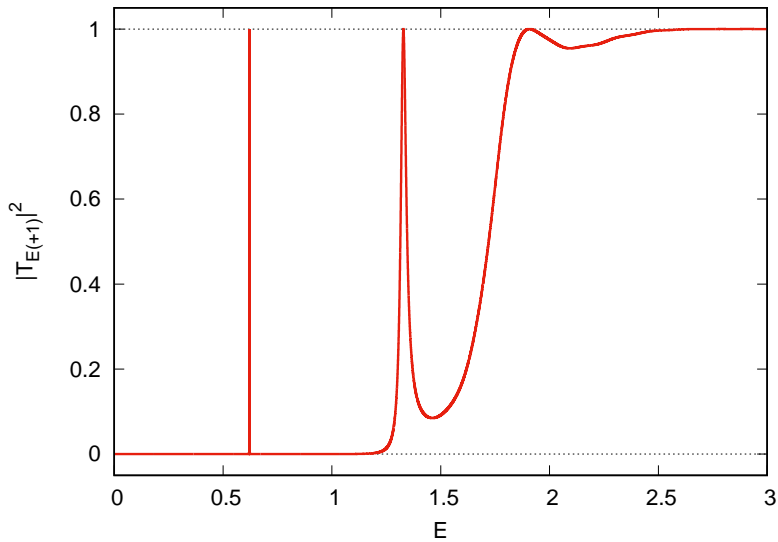}\\
      \vspace*{-0.30cm}
      \caption{ The calculated transmission probability $\Bigl|\,T_{E(+1)}^{+}\,\Bigr|^2$\m. }
      \end{figure}
      
      Note in passing that $T_{E(+1)}^{+} = T_{E(-1)}^{+} = T_{E}^{+}$ as stated by equation (\ref{T-E-+-both-the-same}) of {\sl Appendix II}.
       
\newpage

{\color{blue}
\dotfill\\
\begin{centering}
{\sl Lecture \#4}\\
\vspace*{-0.20cm}
\end{centering}
\dotfill}

      \vspace*{+2.00cm}
      
      Fig.2 shows that the obtained profile of $\Bigl|\,T_{E(+1)}^{+}\,\Bigr|^2$ exhibits the following behavior:
      \begin{itemize}
      \item[{\sf i)}] For $E \to 0$ one encounters $\Bigl|\,T_{E(+1)}^{+}\,\Bigr|^2 \to 0$.
                      This agrees with physics intuition.\\
                      $[\,$Low energy particles approaching our potential are always reflected back.$\,]$
      \item[{\sf ii)}] For $E \to +\infty$ one encounters $\Bigl|\,T_{E(+1)}^{+}\,\Bigr|^2 \to 1$.
                      This agrees with physics intuition, too.\\
                      $[\,$High energy particles approaching our potential are always transmitted.$\,]$
      \item[{\sf iii)}] Most importantly, the profile of $\Bigl|\,T_{E(+1)}^{+}\,\Bigr|^2$ is not smooth,
                        but contains sharp peaks.\\
                        This phenomenon seems counter-intuitive and hardly explainable\\
                        using the theory presented up to now in the present course.\\
                        Indeed, one may wonder why there is an almost unit transmission for $E \doteq 0.621$,\\
                        but almost zero transmission at $E = 0.620$ or $E = 0.622$.\\
                        Similarly for the peak observed at $E \doteq 1.327$ and eventually for the higher peaks.
      \end{itemize}
      The behavior of $\Bigl|\,T_{E(+1)}^{+}\,\Bigr|^2$
      described in {\sf iii)} is an example of a {\color{red} resonance phenomenon}.\\
      An adequate theoretical explanation of the just touched resonance phenomena\\
      is {\color{red} outside the scope of the hermitian scattering theory}.\\
      Desire of physics theoreticians to deeper understand the resonance phenomena\\
      has motivated an intense research activity, leading to the development of the so called\\
      {\color{red} nonhermitian scattering theory} or {\color{red} nonhermitian quantum mechanics}.\\
      This nonhermitian theory\\
      still remains an active field of research which continues to flourish even nowadays.\\
      One elegant version of the nonhermitian scattering theory\\
      is developed in the subsequent Part C of our course.
      
\newpage

{\color{blue}
\dotfill\\
\begin{centering}
{\sl Lecture \#4}\\
\vspace*{-0.20cm}
\end{centering}
\dotfill}
      
\item \underline{\bf B.14 Final summary of the conventional scattering theory in 1D}\\
      Our introduction to the standard or orthodox scattering theory in 1D has been brought to an end.\\
      In Part A we have looked at scattering from the time dependent perspective.\\
      The AC was formulated, leading to entities $| \psi \ra$, $| \phi_{\rm in} \ra$, $| \phi_{\rm out} \ra$ and
      ultimately to $\hat{S}$.\\ We discussed physical meaning of the scattering matrix elements.\\
      Part B was motivated by our desire to switch from the time domain into the energy domain.\\
      In order to reach such a fundamental goal, we needed to proceed in multiple steps.\\
      Stated briefly, a more explicit interconnection between $| \psi \ra$ and $| \phi_{\rm in} \ra$ or $| \phi_{\rm out} \ra$
      has been established,\\ through abstractly looking formulas (\ref{LS-take-1-in}) and (\ref{LS-take-1-out}).
      Subsequently, formulas (\ref{LS-take-1-in}) and (\ref{LS-take-1-out})\\
      were replaced by equivalent yet time independent relations $[\,$(\ref{phi-in-expansion}), (\ref{phi-out-expansion}),
      (\ref{psi-expansion-plus}), (\ref{psi-expansion-minus}), (\ref{LSE-explicit}), (\ref{LSE-implicit})$\,]$.\\
      The LSE (\ref{LSE-explicit}) and (\ref{LSE-implicit}) have led to the concept of stationary scattering states
      $| \psi_{E(\pm 1)}^{\pm} \ra$,\\ which were uniquely characterized by basic equations
      $[\,$(\ref{hat-sf-H-eigenproblem}), (\ref{TISCHE-psi-E-eta-pm}), (\ref{psi-tilde-psi}),
      (\ref{tilde-psi-bc-1})-(\ref{tilde-psi-bc-8})$\,]$,\\
      and which were shown to possess the orthonormality property (\ref{LSE-orthonormality}).\\
      Proceeding further on, by taking an advantage of the LSE (\ref{LSE-explicit}) and (\ref{LSE-implicit}),\\
      we were able to explicitly view the scattering operator $\hat{S}$ (and the closely related operator $\hat{T}$)\\
      solely in terms of the time independent entities $| \psi_{E(\pm 1)}^{\pm} \ra$.\\ See the most important relations
      $[\,$(\ref{S-matel-take-3}), (\ref{T-matels-1}), (\ref{T-matels-2}), (\ref{T-matels-using-hat-T}), (\ref{hat-T-E-def}),
      (\ref{hat-S-final-explicit}), (\ref{hat-S-final-explicit-matels})$\,]$.\\
      We recall also formula (\ref{S-matel-chi-out-phi-in-finalized-prelim}) enabling us to calculate the scattering matrix elements
      (\ref{S-matels}), \\ and the associated (experimentally measurable) transition probabilities (\ref{wp-out-in-def}).\\
      In the next step, we have illuminated the physical contents of $\hat{S}$ and $\hat{T}$\\
      by studying transmission/reflection of our quantum particle\\
      through/from the interaction region of the potential.\\
      Formulas (\ref{phi-out-T-def})-(\ref{phi-out-R-def}) lead us towards the concept of the transmission/reflection
      coefficient,\\ while in equations (\ref{T-probability-final})-(\ref{R-probability-final}) we have determined\\
      the corresponding transmission/reflection probabilities.\\
      Finally, an illustrative numerical example was supplied,\\
      which has also highlighted an existence of scattering resonances.\\
      Thus, all in all, the general time independent scattering theory in 1D has been built up,\\
      and Part B can be closed. We hope that Part B has convinced the readers that\\
      the time independent approach to scattering is not only advantageous computationally,\\
      but leads also to deep physics insights.
\end{itemize}
  
\newpage
      
{\color{blue}
\dotfill\\
\begin{centering}
{\sl Lecture \#5}\\
\vspace*{-0.20cm}
\end{centering}
\dotfill}

\vspace*{+0.30cm}

{\bf C. Nonhermitian Scattering Theory}\\

{\sl Based upon Ref.~\cite{Tolstikhin}.}

\begin{itemize}
\item \underline{\bf C.1 Motivational considerations}\\
      In Part A we have introduced the basic theoretical setup of quantum scattering theory,\\
      via looking at scattering phenomena from the time dependent perspective.\\
      The AC was formulated, leading to entities $| \psi \ra$, $| \phi_{\rm in} \ra$, $| \phi_{\rm out} \ra$ and
      ultimately to $\hat{S}$.\\ Operator $\hat{S}$ was shown to contain all the information about an overall
      scattering outcome.\\ We discussed physical meaning of the matrix elements of $\hat{S}$.\\
      In Part B we have developed the conventional or orthodox scattering theory,\\
      based upon switching from the time domain (wavepackets)\\
      into the energy domain (continuum eigenstates of the total Hamiltonian $\hat{\sf H}$).\\
      Central ingredient of this approach was the LSE (\ref{LSE-explicit}) and (\ref{LSE-implicit}).\\
      The just mentioned conventional or orthodox formulation is undoubtedly valid, exact, general,\\
      and used by vast majority of researchers working in scattering theory.\\
      But, remember, {\color{red} the whole energy domain formalism of Part B\\ (time independent scattering theory, LSE)\\
      represents merely a {\bf tool} for constructing $\hat{S}$ and for understanding its properties}.\\
      In physics we are ultimately interested in the mapping $| \psi_{\rm out} \ra \, = \, \hat{S} \, | \phi_{\rm in} \ra$
      of equation (\ref{hat-S-formal}),\\ not in the stationary eigenstates of $\hat{\sf H}$\m.\\
      By the way, {\color{red} the continuum eigenstates of $\hat{\sf H}$ are not square integrable,\\
      so they do not represent legitimate physical state vectors of quantum mechanics}.\\
      Taking into account these remarks \ldots \,\m one may reasonably ask now:\\
      Is there just a single possible way of studying the scattering phenomena ?\\
      $[\,$$\equiv$ the one outlined above in Part B ?$\,]$\\
      Or, are there other possible ways of getting $\hat{S}$\\
      and the associated physical information about scattering ?\\
      If yes, do these alternative ways possess any merits/offer any benefits\\
      compared to the above outlined conventional approach of Part B ?\\
      Stated concisely, an answer to these questions looks as follows.\\
      Nowadays there exist two alternative routes leading towards $\hat{S}$\\
      and to the associated scattering observables, namely:
      
      \vspace*{+0.50cm}
      
      \begin{itemize}
      \item[{\it 1)}] {\color{blue} {\sl Wavepacket propagation in the time domain}}\\
      Meaning that one solves the time dependent Schr\"{o}dinger equation (\ref{TDSCHE}) numerically for a given incoming state $| \phi_{\rm in} \ra$ and gets thus the corresponding
      outgoing state $| \phi_{\rm out} \ra$.
      No need to deal with the energy domain, or continuum eigenstates of $\hat{\sf H}$, or LSE.
      Such a strategy looks somewhat like a "brute force" approach based upon heavy computing,
      but some pragmatically oriented people prefer this style.
      
\newpage
      
{\color{blue}
\dotfill\\
\begin{centering}
{\sl Lecture \#5}\\
\vspace*{-0.20cm}
\end{centering}
\dotfill}
      
      \vspace*{+2.00cm}
      
      \item[{\it 2)}] {\color{blue} {\sl Nonhermitian scattering theory}} {\color{red} (and this we should study from now on)}\\
      Motivated by our desire to better understand resonance phenomena and related matters.\\
      $[\,$Resonances $\sim$ rather unexpected and often sharp peaks in the transmission profile $\Bigl|\,T_{E(+1)}^{+}\,\Bigr|^2$\m.$\,]$\\
      Based upon solving an eigenvalue problem
      \be \label{TISCHE}
         \left( \, -\,\frac{\hbar^2}{2\,m}\,\partial_{xx} \, + \, V\m(x) \, \right) \varphi_j(x) \; = \; E_j \, \varphi_j(x) \mez ;
      \ee
      where the eigenvalue $E_j$ is now allowed to be complex valued,\\
      and where the pertinent eigenfunction $\varphi_j(x)$ satisfies the so called Siegert boundary conditions\\
      $[\,$see our more detailed discussion below, especially Eqs.~(\ref{Siegert-bc}) and (\ref{Siegert-bc-2})$\,]$.\\      
      Complex valued $E_j$'s indicate where the word {\sl "nonhermitian"} comes from.\\
      $[\,$Hermitian operators of the standard quantum mechanics have only real eigenvalues.$\,]$\\
      As we shall see later on, knowledge of $\{ E_j, \varphi_j(x) \}_j$ enables us to build $\hat{S}$\\
      and describe thus adequately all the scattering phenomena.\\
      Moreover, when the nonhermitian formulation of the problem is adopted,\\ the concept of a resonance becomes sharply defined,\\
      and the role of resonances in scattering processes is then far better understood.\\
      In addition, the nonhermitian formalism itself is valuable by its own right,\\
      since it allows us to look at the scattering problems from an unconventional perspective.\\
      This in turn often leads towards a compelling physical interpretation\\
      and facilitates predictions of novel unexpected phenomena\\
      which are difficult to anticipate when working within the conventional formalism of Part B.
      \end{itemize}
      The three above discussed approaches to scattering theory\\
      $[\,${\it 0)} conventional or orthodox, {\it 1)} wavepacket based, {\it 2)} nonhermitian$\,$]\\
      are complementary, each of them possesses its own advantages and disadvantages.\\
      The real expert should ideally be conversant both with {\it 0)} and {\it 1)} and {\it 2)},\\
      and be able to combine these approaches neatly\\
      while taking into account specific features of a concrete quantum mechanical problem under study.
      
\newpage
      
{\color{blue}
\dotfill\\
\begin{centering}
{\sl Lecture \#5}\\
\vspace*{-0.20cm}
\end{centering}
\dotfill}

\item \underline{\bf C.2 Interlude: Bound states in the standard (hermitian) quantum mechanics}\\
      Before elaborating more on the nonhermitian scattering theory, let us prepare the ground a bit,\\
      by recalling the solution of equation (\ref{TISCHE}) for the case of bound states.\\
      $[\,$This is a standard stuff which can be found in any conventional textbook on quantum mechanics.$\,]$\\
      Let the range $(a,b)$ of our potential $V\m(x)$\\
      be conveniently changed into $(-a,+a)$ from now onwards $[\,$here $a>0$ is a prescribed parameter$\,]$.\\
      Suppose that $E_j$ and $\varphi_j(x)$ is a bound state solution of equation (\ref{TISCHE}). Then $E_j<0$ as we know.\\
      In the right asymptotic region of $x > (+a)$, equation (\ref{TISCHE}) reduces to
      \be \label{TISCHE-asympt}
         -\,\frac{\hbar^2}{2\,m}\,\partial_{xx}\,\varphi_j(x) \; = \; E_j \, \varphi_j(x) \mez .
      \ee
      General solution of (\ref{TISCHE-asympt}) takes of course an explicit form
      \be
         \varphi_j(x) \; = \; {\cal A}_+ \, e^{+\kappa_j x} \; + \; {\cal A}_- \, e^{-\kappa_j x} \mez ;
      \ee
      where
      \be
         \kappa_j \; = \; \sqrt{(-2)\,m\,E_j}/\hbar \mez ;
      \ee
      while ${\cal A}_\pm$ are as yet arbitrary complex coefficients. Note that $\kappa_j>0$.\\
      Since we are dealing with a bound state,\\
      the coefficient ${\cal A}_+$ pertaining to the exploding exponential $e^{+\kappa_j x}$ must vanish, thereby
      \be \label{varphi-j-bound-right}
         \varphi_j(x \to +\infty) \; \sim \; e^{-\kappa_j x} \mez .
      \ee
      Similar sequence of arguments applies also for the left asymptotic region of $x \leq (-a)$.\\
      One encounters again (\ref{TISCHE-asympt}), with the general solution
      $\varphi_j(x) = {\cal B}_+ \, e^{+\kappa_j x} + \, {\cal B}_- \, e^{-\kappa_j x}$ $({\rm here}\;\,{\cal B}_\pm \in {\mathbb C})$.\\
      Since we are dealing with a bound state,\\
      the coefficient ${\cal B}_-$ pertaining to the exploding exponential $e^{-\kappa_j x}$ must vanish, thereby
      \be \label{varphi-j-bound-left}
         \varphi_j(x \to -\infty) \; \sim \; e^{+\kappa_j x} \mez .
      \ee
      {\color{red}
      Properties (\ref{varphi-j-bound-right}) and (\ref{varphi-j-bound-left}) indicate that $\varphi_j(x)$ behaves in the asymptotic region
      of $x \to \pm\infty$\\ as a {\bf single exponential} $e^{\mp \kappa_j x}$\m, not as a superposition of two exponentials.}\\
      This observation will be recalled in the next paragraph,\\
      where it will motivate us to introduce the Siegert boundary conditions.\\
      Before proceeding further, let us make one additional remark regarding the bound state quantization.\\
      Namely, quantization of the bound state energy levels $E_j$ arises due to imposing\\
      {\color{red} {\bf two} boundary conditions} (\ref{varphi-j-bound-right}) and (\ref{varphi-j-bound-left}) on the solutions of the second order
      ODE (\ref{TISCHE}). Indeed,\\ while the ODE (\ref{TISCHE}) possesses always two linearly independent particular solutions\\
      for any trial $E_j$ (even for $E_j$ complex),\\ the two boundary conditions (\ref{varphi-j-bound-right}) and (\ref{varphi-j-bound-left}) can simultaneously be met\\
      only for a discrete set of specific eigenenergies.\\
      $[\,$See again any standard quantum mechanics textbook for details.\\
      \phantom{$[\,$}For bound states we require additionally $E_j$ to be real and negative.$\,]$\\
      
\newpage
      
{\color{blue}
\dotfill\\
\begin{centering}
{\sl Lecture \#5}\\
\vspace*{-0.20cm}
\end{centering}
\dotfill}
      
\item \underline{\bf C.3 Nonhermitian Quantum Mechanics: Introducing the Siegert states}\\
      In the previous paragraph we have dealt with the bound state solutions of equation (\ref{TISCHE}).\\
      Let us conveniently introduce the bound state problem once again using a slightly different language.\\
      Namely, when looking for the bound states of a given short ranged potential $V\m(x)$,\\
      we are interested in solutions $E_j$ and $\varphi_j(x)$ of an eigenvalue problem
      \be \label{TISCHE-take-2}
         {\color{red} \left( \, -\,\frac{\hbar^2}{2\,m}\,\partial_{xx} \, + \, V\m(x) \, \right) \varphi_j(x) \; = \; E_j \, \varphi_j(x) \mez ;}
      \ee
      subjected to the boundary conditions
      \be \label{Siegert-bc}
         {\color{red}
         \varphi_j(x \to +\infty) \; \sim \; e^{-ik_j x} \mez , \mez
         \varphi_j(x \to -\infty) \; \sim \; e^{+ik_j x} \mez ;}
      \ee
      where by definition
      \be \label{k-j-from-E-j}
         k_j \; = \; \sqrt{2\,m\,E_j}/\hbar \mez .
      \ee
      The associated energy eigenvalue $E_j$ is required to be real and negative for the case of bound states.\\
      Instead of (\ref{Siegert-bc}) one may write equivalently      
      \be \label{Siegert-bc-2}
         {\color{red}
         \Bigl(\partial_x \, + \, i\,k_j \Bigr) \, \varphi_j(x) \Bigr|_{x=+a} = \; 0 \mez , \mez
         \Bigl(\partial_x \, - \, i\,k_j \Bigr) \, \varphi_j(x) \Bigr|_{x=-a} = \; 0 \mez . }
      \ee
      Here, as we recall, symbol $a>0$ stands for the potential range $[\,$such that $V\m(x)=0$ for $|x| \ge a$$\,]$.\\
      It is a simple matter to verify that (\ref{Siegert-bc}) implies (\ref{Siegert-bc-2}) and vice versa.\\
      {\color{red} Most importantly, the nonhermitian quantum mechanics\\
      generalizes the notion of a bound state into the so called {\bf Siegert state}.\\
      Namely, when looking for the Siegert states of a given short ranged potential $V\m(x)$,\\
      we are interested in solutions $E_j$ and $\varphi_j(x)$ of an eigenvalue problem (\ref{TISCHE-take-2})\\
      subjected to the {\bf Siegert} boundary conditions (\ref{Siegert-bc}) or (\ref{Siegert-bc-2}). Definition (\ref{k-j-from-E-j}) remains valid,\\
      yet {\bf the eigenvalue ${\bm E}_{\bm j}$ is allowed to be generally complex}.\\
      Hence $k_j=(\ref{k-j-from-E-j})$ is also generally complex (not just negative imaginary as for the bound states),\\
      and the exponentials $e^{\mp ik_j x}$ appearing in (\ref{Siegert-bc}) might be {\bf exploding}! }\\
      Bound states are apparently special cases of the Siegert states.\\
      If a given Siegert state solution $\varphi_j(x)$ of (\ref{TISCHE-take-2})-(\ref{Siegert-bc})/(\ref{Siegert-bc-2}) explodes at $x \to \pm \infty$,\\
      then it is not square integrable and not normalizable in the usual sense.\\
      $[\,$The normalization issue is left unresolved for now, we will come to it later on in the course.$\,]$\\
      Not only the bound states,\\
      but all the Siegert state solutions $E_j$ and $\varphi_j(x)$ defined by equations (\ref{TISCHE-take-2})-(\ref{Siegert-bc})/(\ref{Siegert-bc-2})\\
      possess quantized (discrete) energy levels (some of these being complex).\\
      Indeed, quantization of $E_j$ arises in this context again due to imposing\\
      {\color{red} {\bf two} boundary conditions} (\ref{Siegert-bc})/(\ref{Siegert-bc-2}) on the solutions of the second order ODE (\ref{TISCHE-take-2}).\\
      While the ODE (\ref{TISCHE-take-2}) possesses\\
      always two linearly independent particular solutions for any trial complex $E_j$,\\
      the two boundary conditions (\ref{Siegert-bc})/(\ref{Siegert-bc-2}) can simultaneously be met\\
      only for a discrete set of specific eigenenergies.
      
\newpage
      
{\color{blue}
\dotfill\\
\begin{centering}
{\sl Lecture \#5}\\
\vspace*{-0.20cm}
\end{centering}
\dotfill}

\item \underline{\bf C.4 Looking at the Siegert states once again}\\
      So far we have just defined the Siegert states $\varphi_j(x)$\\
      as solutions of the mathematical eigenvalue problem (\ref{TISCHE-take-2})-(\ref{Siegert-bc})/(\ref{Siegert-bc-2}).\\
      Now it is appropriate to ask the following important questions:
      \begin{itemize}
      \item[{\it i)}] How can one calculate all these Siegert states in practice (numerically)\\ for a given potential $V\m(x)$ ?
      \item[{\it ii)}] How about normalization ? How about the orthonormality and closure relations ?
      \item[{\it iii)}] What are the Siegert states $\varphi_j(x)$ actually good for ?\\
                        How are they related to the scattering theory,
                        and in particular to quantities like $\hat{S}$ or $T_{E(+1)}^{+}$\\
                        (where $E>0$ is of course real valued and continuous) ?
      \item[{\it iv)}] Can the language of Siegert states provide deeper insights into such phenomena\\
                       like e.g.~resonance scattering ?
      \end{itemize}
      All the just posed questions are addressed very explicitly in the paragraphs below,\\
      which describe an elegant and clear cut formulation developed\\ by Oleg Tolstikhin in 1990's and onwards \cite{Tolstikhin}.\\
      Note that the concept of a Siegert state was introduced by Gamow and Siegert back in 1930's.\\
      However, it was only Tolstikhin in 1990's who succeeded\\
      to bridge the gap between the world of Siegert states
      and the standard quantum scattering theory.\\
      It was Tolstikhin who sharpened fundamentally the Siegert state formalism\\ and brought it to practical use.
\item \underline{\bf C.5 Developing the Siegert state formalism: Take \#1}\\
      Let us recall equations (\ref{TISCHE-take-2}), (\ref{Siegert-bc-2})\\
      and define the Siegert states as solutions of the following eigenvalue problem:
      \be \label{ss-ode}
         {\color{red} \left( \, -\,\frac{\hbar^2}{2\,m} \, \partial_{xx} \, + \, V\m(x) \, \right) \, \varphi(x) \; = \; E \, \, \varphi(x) \mez ;}
      \ee
      \be \label{ss-og-bc-1}
         {\color{red} \Bigl( \, \partial_x \, - \, ik \, \Bigr) \, \varphi(x) \, \Bigr|_{x=+a} \; = \; 0 \mez ;}
      \ee
      \be \label{ss-og-bc-2}
         {\color{red} \Bigl( \, \partial_x \, + \, ik \, \Bigr) \, \varphi(x) \, \Bigr|_{x=-a} \; = \; 0 \mez .}
      \ee
      Here $k$ stands for an as yet unknown generally complex parameter\\
      which gives rise to the complex energy $E = \frac{\hbar^2 k^2}{2\,m}$.\\
      Let $\{ {\it b}_\nu(x) \}_\nu$ be an arbitrary complete orthonormal basis set on an interval $x \in [-a,+a]$,\\
      such that
      \be \label{b-j-onrel}
         \int_{-a}^{+a} b_\nu^*\n(x) \, b_{\nu'}\n(x) \; {\rm d}x \; = \; \delta_{\nu\nu'} \mez .
      \ee
      Note that the basis functions ${\it b}_\nu(x)$ are {\sl not} constrained to vanish at the edges $x=\pm a$.\\
      $[\,$Completeness actually requires at least some of the $b_\nu$'s to be nonvanishing at $x=\pm a$.$\,]$\\
      For example one may take
      \be \label{b-j-def}
         b_\nu\n(x) \; = \; \sqrt{\frac{2\,\nu+1}{2\,a}} \; P_{\nu}\m\m\left(\frac{x}{a}\right) \mez , \mez \nu \, = \, 0,1,2,3\ldots \mez ;
      \ee
      with $P_{\nu}\m(u)$ being the conventional Legendre polynomials.\\
      
\newpage
      
{\color{blue}
\dotfill\\
\begin{centering}
{\sl Lecture \#5}\\
\vspace*{-0.20cm}
\end{centering}
\dotfill}

      Alternatively one may employ the Fourier basis set
      \be
         b_\nu\n(x) \; = \; \frac{e^{+i\,\nu\,\pi\,\frac{x}{a}}}{\sqrt{2\,a}} \mez , \mez \nu \, \in \, \{ \ldots,-2,-1,0,+1,+2,\ldots \} \mez .
      \ee
      The as yet unknown Siegert state wavefunction can be for $x \in [-a,+a]$ expanded as
      \be \label{phi-expansion}
         {\color{red} \varphi(x) \; = \; \sum_{\nu} \, c_\nu \, {\it b}_\nu(x) \mez ;}
      \ee
      where the coefficients $c_\nu$ remain to be determined.\\
      $[\,$Boundary condition (\ref{ss-og-bc-1}) implies that\\
      \phantom{$[\,$}$\varphi(x \ge (+a)) \, = \, {\cal C}_+ \, e^{+ikx}$\m, similarly $\varphi(x \le (-a)) \, = \, {\cal C}_- \, e^{-ikx}$ due to (\ref{ss-og-bc-2}).\\
      \phantom{$[\,$}Factors ${\cal C}_\pm$ are fixed by demanding continuity of $\varphi(x)$ at $x = \pm a$.$\,]$\\
      In order to find the as yet unknown expansion coefficients $c_\nu$\\
      we multiply (\ref{ss-ode}) by ${\it b}_\nu^*\n(x)$ and integrate over $x \in [-a,+a]$. One gets
      \be
         - \, \frac{\hbar^2}{2\,m} \, \int_{-a}^{+a} \m {\it b}_\nu^*\n(x) \, \partial_{xx} \, \varphi(x) \; {\rm d}x \;\, + \;\,
         \int_{-a}^{+a} \m {\it b}_\nu^*\n(x) \, \Bigl(V\m(x)-E\Bigr) \, \varphi(x) \; {\rm d}x \; = \; 0 \mez .
      \ee
      Subsequently we integrate the first term by parts,\\ and incorporate the Siegert boundary conditions (\ref{ss-og-bc-1})-(\ref{ss-og-bc-2}).\\
      We also insert the basis set expansion (\ref{phi-expansion}) afterwards.\\
      This yields an intermediate outcome
      \begin{eqnarray} \label{Siegert-genevp-long}
         \hspace*{-1.50cm} & & - \, \frac{i\hbar^2 k}{2\,m} \, \sum_{\nu'} \, c_{\nu'} \, \Bigl( \, {\it b}_\nu^*\n(+a) \, {\it b}_{\nu'}\m(+a)
         \, + \, {\it b}_\nu^*\n(-a) \, {\it b}_{\nu'}\m(-a) \, \Bigr) \; + \;
         \frac{\hbar^2}{2\,m} \; \sum_{\nu'} \, c_{\nu'} \m\m \int_{-a}^{+a} \m \Bigl(\partial_x {\it b}_\nu^*\n(x)\Bigr) \Bigl(\partial_x {\it b}_{\nu'}\m(x)\Bigr) \, {\rm d}x \nonumber\\
         \hspace*{-1.50cm} & & + \; \sum_{\nu'} \, c_{\nu'} \m\m \int_{-a}^{+a} \m {\it b}_\nu^*\n(x) \, V\m(x) \, {\it b}_{\nu'}\m(x) \; {\rm d}x
         \; - \; \frac{\hbar^2 k^2}{2\,m} \; c_\nu \; = \; 0 \mez .
      \end{eqnarray}
      Next, let us conveniently translate formula (\ref{Siegert-genevp-long}) into the language of linear algebra.\\
      For this purpose we shall introduce the following matrix elements:
      \begin{eqnarray}
         \label{Siegert-H-matels}
         {\color{red} H_{\nu\nu'}} & {\color{red} =} & {\color{red} \frac{\hbar^2}{2\,m} \, \int_{-a}^{+a} \m \Bigl(\partial_x {\it b}_\nu^*\n(x)\Bigr) \Bigl(\partial_x {\it b}_{\nu'}\m(x)\Bigr) \, {\rm d}x 
         \; + \; \int_{-a}^{+a} \m {\it b}_\nu^*\n(x) \, V\m(x) \, {\it b}_{\nu'}\m(x) \; {\rm d}x \mez ;}\\
         \label{Siegert-L-matels}
         {\color{red} L_{\nu\nu'}} & {\color{red} =} & {\color{red} \frac{\hbar^2}{2\,m} \, \Bigl( \, {\it b}_\nu^*\n(+a) \, {\it b}_{\nu'}\m(+a) \, + \, {\it b}_\nu^*\n(-a) \, {\it b}_{\nu'}\m(-a) \, \Bigr) \mez .}
      \end{eqnarray}
      Instead of (\ref{Siegert-genevp-long}) one may then write equivalently and concisely
      \be \label{Siegert-genevp-matrix}
         {\color{red}
         \left( {\bm H} \, - \, i\,k \, {\bm L} \, - \, \frac{\hbar^2 k^2}{2\,m} \, {\bm I} \right) {\bm c} \; = \; {\bm 0} \mez ;}
      \ee
      with obvious notations.
      
\newpage
      
{\color{blue}
\dotfill\\
\begin{centering}
{\sl Lecture \#5}\\
\vspace*{-0.20cm}
\end{centering}
\dotfill}

\vspace*{+2.00cm}

      Suppose that the basis set expansion (\ref{phi-expansion}) is truncated to $N$ terms\\
      (where $N$ is arbitrarily large in principle, it is of course restricted in practical calculations).\\
      The matrix equation (\ref{Siegert-genevp-matrix}) is then correspondingly truncated as well.\\
      $[\,$It contains then $N$-by-$N$ matrices ${\bm H}$, ${\bm L}$, ${\bm I}$ and and $N$-dimensional column vector ${\bm c}$.$\,]$\\
      Equation (\ref{Siegert-genevp-matrix})$|_{N\,{\rm finite}}$ can now be classified as a quadratic eigenvalue problem of linear algebra,\\
      to be solved for unknowns $k$ and ${\bm c}$. An adequate solution of (\ref{Siegert-genevp-matrix})$|_{N\,{\rm finite}}$ is performed\\
      by taking advantage of a linearization technique which is described in the next paragraph.\\ For the sake of clarity, we point out here that
      equation (\ref{Siegert-genevp-matrix})$|_{N\,{\rm finite}}$ or (\ref{Siegert-genevp-matrix})$|_{N \to \infty}$ is implied\\ by our original Siegert eigenvalue problem (\ref{ss-ode}), (\ref{ss-og-bc-1})-(\ref{ss-og-bc-2}). On the other hand,\\ it is not obvious whether every solution $\{ k,{\bm c} \}$ of equation
      (\ref{Siegert-genevp-matrix})$|_{N\,{\rm finite}}$\\
      corresponds to some approximate Siegert eigenstate (\ref{phi-expansion})\\
      which would satisfy (\ref{ss-ode}), (\ref{ss-og-bc-1})-(\ref{ss-og-bc-2})
      in the limit of $N \to \infty$.\\ For this reason we prefer to call all the solutions of (\ref{Siegert-genevp-matrix})$|_{N\,{\rm finite}}$ as {\color{red} Siegert pseudostates}.
      
\newpage
      
{\color{blue}
\dotfill\\
\begin{centering}
{\sl Lecture \#6}\\
\vspace*{-0.20cm}
\end{centering}
\dotfill}

\item \underline{\bf C.6 Developing the Siegert state formalism: Take \#2}\\
      The quadratic eigenvalue problem (\ref{Siegert-genevp-matrix})$|_{N\,{\rm finite}}$ can be redisplayed in a more standard appearance
      \be \label{qep}
         {\color{red} \left( {\bm A} \, + \, \lambda \, {\bm B} \, + \, \lambda^2 \, {\bm I} \right) {\bm c} \; = \; {\bm 0}
         \mez ;}
      \ee
      where
      \be \label{bm-A-bm-B-defs}
         {\color{red} {\bm A} \; = \; +\,\frac{2\,m}{\hbar^2} \; {\bm H} \mez , \mez
                      {\bm B} \; = \; -\,\frac{2\,m}{\hbar^2} \; {\bm L} \mez ;}
      \ee
      and
      \be \label{lambda-from-k}
         {\color{red} \lambda \; = \; i\,k \mez .}
      \ee
      To facilitate further progress and linearize the l.h.s.~of (\ref{qep}) with respect to $\lambda$,\\
      we conveniently introduce a $2N$-component column vector
      \be \label{vector-2N-from-N}
         {\color{red} \left( \matrix{ {\bm c} \cr \bm\tilde{\bm c} } \right) \; = \; \left( \matrix{ {\bm c} \cr \lambda \, {\bm c} } \right) \mez .}
      \ee
      Equation (\ref{qep}) is then equivalent to
      \be \label{lep}
         \left( \matrix{ {\bm O} & {\bm I} \cr -{\bm A} & -{\bm B} } \right)
         \left( \matrix{ {\bm c} \cr \bm\tilde{\bm c} } \right) \; = \;
         \lambda \, \left( \matrix{ {\bm c} \cr \bm\tilde{\bm c} } \right) \mez .
      \ee
      Indeed, one may easily check that (\ref{qep}) implies (\ref{lep}) and vice versa.\\
      Importantly, the transformed problem (\ref{lep}) acquires doubled matrix dimension $2N$\m,\\
      yet it is linear in $\lambda$ as opposed to the $N$-dimensional matrix problem (\ref{qep}).\\
      $[\,$One may observe that the replacement (\ref{qep}) $\mapsto$ (\ref{lep}) resembles\\
      \phantom{$[\,$}the well known passage from a single second order ODE to two coupled first order ODEs.$\,]$\\
      Proceeding further, we multiply (\ref{lep}) by matrix
      \be \label{Siegert-lep-matrix-regular}
         \left( \matrix{ {\bm B} & {\bm I} \cr {\bm I} & {\bm O} } \right) \mez ;
      \ee
      while noting that
      \be \label{Siegert-lep-matrix-regular-inv}
         \left( \matrix{ {\bm B} & {\bm I} \cr {\bm I} & {\bm O} } \right)^{\m\m\m -1} \; = \; \left( \matrix{ {\bm 0} & {\bm I} \cr {\bm I} & -{\bm B} } \right) \mez .
      \ee
      $[\,$Due to (\ref{Siegert-lep-matrix-regular-inv}) the matrix (\ref{Siegert-lep-matrix-regular}) is certainly regular.$\,]$
      This converts equation (\ref{lep}) into an equivalent symmetric form
      \be \label{lep-sym}
         {\color{red} 
         \left( \matrix{ -{\bm A} & {\bm O} \cr {\bm O} & {\bm I} } \right)
         \left( \matrix{ {\bm c} \cr \bm\tilde{\bm c} } \right) \; = \;
         \lambda \, \left( \matrix{ {\bm B} & {\bm I} \cr {\bm I} & {\bm O} } \right)
         \left( \matrix{ {\bm c} \cr \bm\tilde{\bm c} } \right) \mez .}
      \ee
      Symmetric linear eigenvalue problems (\ref{lep-sym}) are treatable (even numerically)\\
      by standard well established methods of linear algebra.\\
      Hence also equations (\ref{qep}) and (\ref{Siegert-genevp-matrix})$|_{N\,{\rm finite}}$
      can be discussed and solved in this way.

\newpage
      
{\color{blue}
\dotfill\\
\begin{centering}
{\sl Lecture \#6}\\
\vspace*{-0.20cm}
\end{centering}
\dotfill}

\item \underline{\bf C.7 Developing the Siegert state formalism: Take \#3}\\
      Let us look again at problem (\ref{qep}), and define the corresponding quadratic matrix polynomial
      \be \label{bm-M-lambda-def}
         {\color{red} {\bm M}\m(\lambda) \; = \; {\bm A} \, + \, \lambda \, {\bm B} \, + \, \lambda^2 \,\m {\bm I} \mez .}
      \ee
      Instead of (\ref{qep}) one can write then
      \be \label{qep-M}
         {\bm M}\m(\lambda) \; {\bm c}  \; = \; {\bm 0} \mez .
      \ee
      The matrix equation (\ref{qep-M}) possesses a nontrivial solution ${\bm c}$ iff ${\rm det}[{\bm M}\m(\lambda)]=0$.\\
      But ${\rm det}[{\bm M}\m(\lambda)]$ is a $\lambda$-polynomial of degree $2N$\m.\\
      Polynomials of degree $2N$ generically have $2N$ distinct roots.\\
      Hereafter we shall assume, for the sake of maximum simplicity,\\
      that this is the case also for our polynomial equation ${\rm det}[{\bm M}\m(\lambda)]=0$.\\
      We shall also assume, again for the sake of maximum simplicity,\\ that
      all the $2N$ roots of ${\rm det}[{\bm M}\m(\lambda)]=0$ are nonzero.\\
      $[\,$Correctness of the just made assumptions can always be checked by numerical calculation\\
      \phantom{$[\,$}performed a posteriori for each concrete physical system under study and for concrete $N$\m.$\,]$\\
      If so, equations (\ref{qep-M}), (\ref{lep-sym}), (\ref{qep}) possess exactly $2N$ distinct eigensolutions
      \be \label{lep-eigensolutions}
         {\color{red} \lambda^{(n)} \neq 0 \mez , \mez \left( \matrix{ {\bm c}^{(n)} \cr \bm\tilde{\bm c}^{(n)} } \right) \mez , \mez
         1 \leq n \leq 2N \mez .}
      \ee
      Here the eigenvectors are linearly independent due to (\ref{lep}) and (\ref{lep-sym}).\\
      $[\,$Hence the $2N$ eigenvectors listed in (\ref{lep-eigensolutions}) form a basis
      in the pertinent linear space ${\mathbb C}^{2N}$\m\m.$\,]$\\
      Correspondingly, in the finite basis set representation (\ref{Siegert-genevp-matrix})$|_{N\,{\rm finite}}$\\
      of our original eigenvalue problem (\ref{ss-ode}), (\ref{ss-og-bc-1})-(\ref{ss-og-bc-2})\\
      we have obtained exactly $2N$ distinct linearly independent Siegert pseudostates (\ref{phi-expansion}), that is,
      \be \label{Siegert-phi-n-explicit}
         {\color{red} \varphi^{(n)}\m(x) \; = \; \sum_{\nu} \, c_\nu^{(n)} \, {\it b}_\nu(x) \mez .}
      \ee
      The associated Siegert eigenvalue follows immediately from equation (\ref{lambda-from-k}). One has
      \be \label{Siegert-phi-n-eigenvalues}
         {\color{red} k^{(n)} \; = \; -i\,\lambda^{(n)} \mez , \mez
         E^{(n)} \; = \; \frac{\hbar^2 \m \left(k^{(n)}\right)^{\m 2}}{2\,m} \; = \; -\,\frac{\hbar^2 \m \left(\lambda^{(n)}\right)^{\m 2}}{2\,m} \mez .}
      \ee
      We have just finished answering question {\it i)} raised in the paragraph C.4.\\
      From now on we shall conveniently adopt the notation $\varphi^{(n)}\m(x)$\\
      instead of writing $\varphi_j\n(x)$ as we did tentatively in C.3-C.4.\\
      The same applies also for the pertinent eigenvalues $E^{(n)}$ and the related $k^{(n)}$\m.
      
\newpage
      
{\color{blue}
\dotfill\\
\begin{centering}
{\sl Lecture \#6}\\
\vspace*{-0.20cm}
\end{centering}
\dotfill}
      
\item \underline{\bf C.8 Developing the Siegert state formalism: Take \#4}\\
      Let us establish now appropriate orthonormality relations\\
      for the eigensolutions (\ref{lep-eigensolutions}) of the problem (\ref{qep-M}), (\ref{lep-sym}), (\ref{qep}).\\
      Take any two distinct eigensolutions (\ref{lep-eigensolutions}), such that
      \be \label{lep-sol-n}
         \left( \matrix{ -{\bm A} & {\bm O} \cr {\bm O} & {\bm I} } \right)
         \left( \matrix{ {\bm c}^{(n)} \cr \bm\tilde{\bm c}^{(n)} } \right) \; = \;
         \lambda^{(n)} \, \left( \matrix{ {\bm B} & {\bm I} \cr {\bm I} & {\bm O} } \right)
         \left( \matrix{ {\bm c}^{(n)} \cr \bm\tilde{\bm c}^{(n)} } \right) \mez ;
      \ee
      \be \label{lep-sol-n'}
         \left( \matrix{ -{\bm A} & {\bm O} \cr {\bm O} & {\bm I} } \right)
         \left( \matrix{ {\bm c}^{(n'\m)} \cr \bm\tilde{\bm c}^{(n'\m)} } \right) \; = \;
         \lambda^{(n'\m)} \, \left( \matrix{ {\bm B} & {\bm I} \cr {\bm I} & {\bm O} } \right)
         \left( \matrix{ {\bm c}^{(n'\m)} \cr \bm\tilde{\bm c}^{(n'\m)} } \right) \mez .
      \ee
      Multiply (\ref{lep-sol-n}) by $({\bm c}^{(n'\m)T},\bm\tilde{\bm c}^{(n'\m)T})$ and 
      (\ref{lep-sol-n'}) by $({\bm c}^{(n)T},\bm\tilde{\bm c}^{(n)T})$. Subtract the two to get
      \be \label{lep-scalprod-take-1}
         (\lambda^{(n)}-\lambda^{(n'\m)}) \, \left( \matrix{ {\bm c}^{(n)T} & \bm\tilde{\bm c}^{(n)T} } \right)
         \left( \matrix{ {\bm B} & {\bm I} \cr {\bm I} & {\bm O} } \right)
         \left( \matrix{ {\bm c}^{(n'\m)} \cr \bm\tilde{\bm c}^{(n'\m)} } \right) \; = \; 0 \mez .
      \ee
      Formula (\ref{lep-scalprod-take-1}) reveals that for $\lambda^{(n)} \neq \lambda^{(n'\m)}$ (and thus for $n \neq n'\n$)\\
      the eigensolutions (\ref{lep-eigensolutions}) are orthogonal with respect to the scalar product
      \be \label{lep-scalprod-take-2}
         \left( \matrix{ {\bm c}^{(n)T} & \bm\tilde{\bm c}^{(n)T} } \right)
         \left( \matrix{ {\bm B} & {\bm I} \cr {\bm I} & {\bm O} } \right)
         \left( \matrix{ {\bm c}^{(n'\m)} \cr \bm\tilde{\bm c}^{(n'\m)} } \right) \mez .
      \ee
      Note also that for all values of $n$ $(1 \leq n \leq 2N)$ we necessarily have
      \be \label{lep-scalprod-take-3}
         \left( \matrix{ {\bm c}^{(n)T} & \bm\tilde{\bm c}^{(n)T} } \right)
         \left( \matrix{ {\bm B} & {\bm I} \cr {\bm I} & {\bm O} } \right)
         \left( \matrix{ {\bm c}^{(n)} \cr \bm\tilde{\bm c}^{(n)} } \right) \; \neq \; 0 \mez .
      \ee
      Since if the l.h.s.~of (\ref{lep-scalprod-take-3}) was zero for some $n$, then the vector
      $\left( \matrix{ {\bm B} & {\bm I} \cr {\bm I} & {\bm O} } \right)
      \left( \matrix{ {\bm c}^{(n)} \cr \bm\tilde{\bm c}^{(n)} } \right)$\\
      would be orthogonal to all the $2N$-dimensional vectors of our linear space ${\mathbb C}^{2N}$\m\m,\\
      hence $\left( \matrix{ {\bm B} & {\bm I} \cr {\bm I} & {\bm O} } \right)
      \left( \matrix{ {\bm c}^{(n)} \cr \bm\tilde{\bm c}^{(n)} } \right)$ would vanish, and thus also
      $\left( \matrix{ {\bm c}^{(n)} \cr \bm\tilde{\bm c}^{(n)} } \right)$ would vanish
      $[\,${\sl cf}.~property (\ref{Siegert-lep-matrix-regular-inv})$\,]$.\\
      Our above made observations lead us towards imposing the normalization convention
      \be \label{norm-conv-1}
         {\color{red}
         \left( \matrix{ {\bm c}^{(n)T} & \bm\tilde{\bm c}^{(n)T} } \right)
         \left( \matrix{ {\bm B} & {\bm I} \cr {\bm I} & {\bm O} } \right)
         \left( \matrix{ {\bm c}^{(n'\m)} \cr \bm\tilde{\bm c}^{(n'\m)} } \right) \; = \;
         2 \, \lambda^{(n)} \; \delta_{nn'} \mez .}
      \ee
      The extra factor $2 \, \lambda^{(n)}$ appearing on the r.h.s.~of (\ref{norm-conv-1}) has been inserted\\
      for the sake mathematical convenience. Recall in this context that $\lambda^{(n)} \neq 0$ for all $n$.\\
      An equivalent version of (\ref{norm-conv-1}) is obtained by exploiting (\ref{vector-2N-from-N}) and reads as
      \be \label{norm-conv-2}
         {\color{red} \Bigl( \, \lambda^{(n)} + \lambda^{(n'\m)} \Bigr) \, {\bm c}^{(n)T} {\bm c}^{(n'\m)} \; + \;
         {\bm c}^{(n)T} {\bm B}\,{\bm c}^{(n'\m)}
         \; = \; 2 \, \lambda^{(n)} \; \delta_{nn'} \mez .}
      \ee
      
\newpage
      
{\color{blue}
\dotfill\\
\begin{centering}
{\sl Lecture \#6}\\
\vspace*{-0.20cm}
\end{centering}
\dotfill}
      
\item \underline{\bf C.9 Developing the Siegert state formalism: Take \#5}\\
      Let us establish now appropriate closure relations\\
      for the eigensolutions (\ref{lep-eigensolutions}) of the problem (\ref{qep-M}), (\ref{lep-sym}), (\ref{qep}).
      As mentioned already,\\ the $2N$ eigenvectors listed in (\ref{lep-eigensolutions}) form a basis
      in the pertinent linear space ${\mathbb C}^{2N}$\m\m.\\
      Meaning that every $2N$-dimensional column vector $\left( \matrix{ {\bm u} \cr {\bm v} } \right)$ can be expressed in the form
      \be
         \left( \matrix{ {\bm u} \cr {\bm v} } \right) \; = \; \sum_{n=1}^{2N} \; q_n \left( \matrix{ {\bm c}^{(n)} \cr \bm\tilde{\bm c}^{(n)} } \right) \mez ;
      \ee
      with the coefficients
      \be
         q_n \; = \; \frac{1}{2\,\lambda^{(n)}} \, \left( \matrix{ {\bm c}^{(n)T} & \bm\tilde{\bm c}^{(n)T} } \right)
         \left( \matrix{ {\bm B} & {\bm I} \cr {\bm I} & {\bm O} } \right) \, \left( \matrix{ {\bm u} \cr {\bm v} } \right) \mez ;
      \ee
      as being implied by (\ref{norm-conv-1}). Consider an $2N$-by-$2N$ matrix
      \be \label{bm-Q-def}
         {\bm Q} \; = \; \sum_{n=1}^{2N} \, \frac{1}{2\,\lambda^{(n)}} \, \left( \matrix{ {\bm c}^{(n)} \cr \bm\tilde{\bm c}^{(n)} } \right)
         \left( \matrix{ {\bm c}^{(n)T} & \bm\tilde{\bm c}^{(n)T} } \right) \mez .
      \ee
      Clearly,
      \be
         {\bm Q} \, \left( \matrix{ {\bm B} & {\bm I} \cr {\bm I} & {\bm O} } \right) \left( \matrix{ {\bm u} \cr {\bm v} } \right)
         \; = \; \left( \matrix{ {\bm u} \cr {\bm v} } \right) \mez ;
      \ee
      valid for all $\left( \matrix{ {\bm u} \cr {\bm v} } \right)$. Meaning that
      \be \label{bm-Q-explicit}
         {\bm Q} \; = \; \left( \matrix{ {\bm B} & {\bm I} \cr {\bm I} & {\bm O} } \right)^{\m\m-1} \; = \;
         \left( \matrix{ {\bm O} & {\bm I} \cr {\bm I} & -{\bm B} } \right) \mez ;
      \ee
      {\sl cf}.~equation (\ref{Siegert-lep-matrix-regular-inv}).
      From (\ref{bm-Q-def}) and (\ref{bm-Q-explicit}) one obtains using (\ref{vector-2N-from-N})\\
      the following important closure properties:
      \be \label{closure-1}
         {\color{red} \sum_{n=1}^{2N} \, \frac{1}{\lambda^{(n)}} \; {\bm c}^{(n)} \, {\bm c}^{(n)T} \; = \; {\bm O} \mez ;}
      \ee
      \be \label{closure-2}
         {\color{red} \sum_{n=1}^{2N} \, {\bm c}^{(n)} \, {\bm c}^{(n)T} \; = \; 2\,{\bm I} \mez ;}
      \ee
      \be \label{closure-3}
         {\color{red} \sum_{n=1}^{2N} \, \lambda^{(n)} \; {\bm c}^{(n)} \, {\bm c}^{(n)T} \; = \; -2\,{\bm B} \mez .}
      \ee
      We have just finished answering question {\it ii)} raised in the paragraph C.4.

\newpage
      
{\color{blue}
\dotfill\\
\begin{centering}
{\sl Lecture \#6}\\
\vspace*{-0.20cm}
\end{centering}
\dotfill}

\item \underline{\bf C.10 Developing the Siegert state formalism: Take \#6}\\
      As the last step of our formal mathematical elaborations\\
      we shall derive spectral representation of the inverse of the quadratic matrix polynomial (\ref{bm-M-lambda-def}).\\
      We claim that
      \be \label{bm-M-lambda-inverse}
         {\color{red} {\bm M}^{-1}\m(\lambda) \; = \;
         \sum_{n=1}^{2N} \, \frac{{\bm c}^{(n)}\,{\bm c}^{(n)T}}{2\,\lambda^{(n)}(\lambda-\lambda^{(n)})} \mez ;}
      \ee
      valid for all complex $\lambda$ except of course of the roots $\lambda^{(n)}$ of equation ${\rm det}[{\bm M}\m(\lambda)]=0$.\\
      In order to prove (\ref{bm-M-lambda-inverse}), we shall multiply both sides by ${\bm M}\m(\lambda)$ from the left,\\
      and see what comes out. One has
      \be
         {\bm M}\m(\lambda) \; {\bm M}^{-1}\m(\lambda) \; = \;
         \sum_{n=1}^{2N} \, \frac{{\bm M}\m(\lambda) \, {\bm c}^{(n)}\,{\bm c}^{(n)T}}{2\,\lambda^{(n)}(\lambda-\lambda^{(n)})} \mez .
      \ee
      But ${\bm M}\m(\lambda)\,{\bm c}^{(n)}\,=\,{\bm M}(\lambda_n)\,{\bm c}^{(n)}\,+\,(\lambda-\lambda^{(n)})\,\m\Bigl({\bm B}+(\lambda+\lambda^{(n)})
      {\bm I}\Bigr)\,\n{\bm c}^{(n)}$\m. Hence
      \be
         {\bm M}\m(\lambda) \; {\bm M}^{-1}\m(\lambda) \; = \; \sum_{n=1}^{2N} \, \frac{\Bigl({\bm B}+(\lambda+\lambda^{(n)})
         {\bm I}\Bigr)\,{\bm c}^{(n)}\,{\bm c}^{(n)T}}{2\,\lambda^{(n)}} \; = \; {\bm I} \mez ;
      \ee
      as a direct consequence of (\ref{closure-1})-(\ref{closure-2}). The validity of (\ref{bm-M-lambda-inverse}) is thus established.
\item \underline{\bf C.11 Siegert state representation of the retarded Green function: Take \#1}\\
      Having passed through all the necessary mathematical preparations,\\
      we are slowly moving on towards addressing the question {\it iii)} raised in paragraph C.4.\\
      Most importantly, {\color{red} the gap between\\
      the mathematical world of Siegert states and the physical world of quantum scattering theory\\
      is bridged via examining the retarded (outgoing) Green function $G_{\m E}^+\n(x,y)$}\\
      which has been introduced and discussed at length in {\sl Appendix IV}.\\
      According to equations (\ref{G-E-pm-SCHE-ihg}) and (\ref{G-E-+-asympt}) from {\sl Appendix IV},\\
      an entity $G_{\m E}^+\n(x,y)$ is defined as an unique solution of the following boundary value problem:
      \be \label{G-E-pm-SCHE-ihg-Siegert}
         {\color{red} \left\{ \, -\,\frac{\hbar^2}{2\,m}\,\partial_{xx} \, + \, V\m(x) \, - \, E \, \right\} \, G_{\m E}^+\n(x,y) \; = \; -\,\delta(x-y) \mez ;}
      \ee
      \be \label{G-E-+-asympt-Siegert}
         G_{\m E}^+\n(x\,\m_{<\,\min(-a,y)}^{>\,\max(+a,y)}\m,y) \; \simeq \; e^{\pm i k x} \mez .
      \ee
      Here $E>0$ stands for the physical impact energy and $k = \sqrt{2\,m\,E}/\hbar$ as usual.\\
      Let us hereafter assume $y \in [-a,+a]$. Requirement (\ref{G-E-+-asympt-Siegert})\\
      can then be replaced by the Siegert type boundary conditions analogous to (\ref{ss-og-bc-1})-(\ref{ss-og-bc-2}), that is,
      \be \label{Siegert-G-bc-1}
         {\color{red} \Bigl( \, \partial_x \, - \, ik \, \Bigr) \, G_{\m E}^+\n(x,y) \, \Bigr|_{x=+a} \; = \; 0 \mez ;}
      \ee
      \be \label{Siegert-G-bc-2}
         {\color{red} \Bigl( \, \partial_x \, + \, ik \, \Bigr) \, G_{\m E}^+\n(x,y) \, \Bigr|_{x=-a} \; = \; 0 \mez .}
      \ee
      If so, then our boundary value problem (\ref{G-E-pm-SCHE-ihg-Siegert}), (\ref{Siegert-G-bc-1})-(\ref{Siegert-G-bc-2}) may be for $|y| \leq a$ analyzed and solved\\
      using the same method as we applied above for the Siegert eigenproblem (\ref{ss-ode}), (\ref{ss-og-bc-1})-(\ref{ss-og-bc-2}).

\newpage
      
{\color{blue}
\dotfill\\
\begin{centering}
{\sl Lecture \#6}\\
\vspace*{-0.20cm}
\end{centering}
\dotfill}

      Namely, let us restrict $x$ like we did for $y$, such that $x,y \in [-a,+a]$,\\
      and expand $G_{\m E}^+\n(x,y)$ similarly as done in (\ref{phi-expansion}) for the case of $\varphi(x)$. This amounts to set
      \be \label{G-expansion-varphi}
         G_{\m E}^+\n(x,y) \; = \; \sum_{\nu \nu'} \, G_{\m E,\nu \nu'}^{+} \, {\it b}_\nu(x) \, {\it b}_{\nu'}\m\n(y) \mez ;
      \ee
      where the as yet unknown expansion coefficients $G_{\m E,\nu \nu'}^{+}$ remain to be determined.\\
      Much as before in (\ref{phi-expansion}),\\
      the summation over $\nu$ is truncated in (\ref{G-expansion-varphi}) just to $N$ terms,
      and similarly for $\nu'$\m\m.\\
      In order to find the $N^2$ matrix elements $G_{\m E,\nu \nu'}^{+}$, we multiply (\ref{G-E-pm-SCHE-ihg-Siegert}) by ${\it b}_\nu^*(x) \, {\it b}_{\nu'}^*\m\n(y)$,\\
      and integrate over both spatial coordinates $x,y \in [-a,+a]$.\\ This yields, with the help of (\ref{b-j-onrel}), a preliminary formula
      \begin{eqnarray} \label{Siegert-G-take-1}
         \hspace*{-2.00cm} & & - \, \frac{\hbar^2}{2\,m} \, \int_{-a}^{+a} \m {\rm d}x \int_{-a}^{+a} \m {\rm d}y \;\,
         {\it b}_\nu^*\n(x) \, {\it b}_{\nu'}^*\m(y) \; \partial_{xx} \, G_{\m E}^+\n(x,y) \\
         \hspace*{-2.00cm} & & + \; \int_{-a}^{+a} \m {\rm d}x \int_{-a}^{+a} \m {\rm d}y \;\, {\it b}_\nu^*\n(x) \, {\it b}_{\nu'}^*\m(y) \, \Bigl(V\m(x)-E\Bigr) \, G_{\m E}^+\n(x,y)
         \; = \; - \, \delta_{\nu\nu'} \mez . \nonumber
      \end{eqnarray}
      Subsequently we integrate the first term of (\ref{Siegert-G-take-1}) by parts,\\ and incorporate the Siegert boundary conditions (\ref{Siegert-G-bc-1})-(\ref{Siegert-G-bc-2}).\\
      We also insert the basis set expansion (\ref{G-expansion-varphi}) afterwards,\\
      and take again an advantage of (\ref{b-j-onrel}).\\ All the mentioned manipulations provide altogether an intermediate outcome
      \begin{eqnarray} \label{Siegert-G-take-2}
         \hspace*{-1.50cm} & & - \, \frac{i\hbar^2 k}{2\,m} \, \sum_{\nu''} \, G_{\m E,\nu''\n\nu'}^{+} \,\m \Bigl( \, {\it b}_\nu^*\n(+a) \, {\it b}_{\nu''}\m(+a)
         \, + \, {\it b}_\nu^*\n(-a) \, {\it b}_{\nu''}\m(-a) \, \Bigr) \; + \;
         \frac{\hbar^2}{2\,m} \; \sum_{\nu''} \, G_{\m E,\nu''\n\nu'}^{+} \m \int_{-a}^{+a} \m \Bigl(\partial_x {\it b}_\nu^*\n(x)\Bigr) \Bigl(\partial_x {\it b}_{\nu''}\m(x)\Bigr) \, {\rm d}x \nonumber\\
         \hspace*{-1.50cm} & & + \; \sum_{\nu''} \, G_{\m E,\nu''\n\nu'}^{+} \m \int_{-a}^{+a} \m {\it b}_\nu^*\n(x) \, V\m(x) \, {\it b}_{\nu''}\m(x) \; {\rm d}x
         \; - \; \frac{\hbar^2 k^2}{2\,m} \; G_{\m E,\nu\nu'}^{+} \; = \; - \, \delta_{\nu\nu'} \mez ;
      \end{eqnarray}   
      which means however nothing else than
      \be \label{Siegert-bm-G-eq-1}
         {\color{red} {\bm G}_{\m E}^+ \,\m
         \left( {\bm H} \, - \, i\,k \, {\bm L} \, - \, \frac{\hbar^2 k^2}{2\,m} \, {\bm I} \right) \; = \; -\;{\bm I} \mez .}
      \ee
      Here ${\bm G}_{\m E}^+$ stands of course for an $N$-by-$N$ matrix of elements $G_{\m E,\nu\nu'}^{+}$.\\
      Matrices ${\bm H}$ and ${\bm L}$ are defined above in (\ref{Siegert-H-matels})-(\ref{Siegert-L-matels}).\\
      Instead of (\ref{Siegert-bm-G-eq-1}) one may write equivalently
      \be \label{Siegert-bm-G-eq-2}
         {\color{red} {\bm G}_{\m E}^+ \,\m
         \left( {\bm A} \, + \, \lambda \, {\bm B} \, + \, \lambda^2 \, {\bm I} \right) \; = \; -\,\frac{2\,m}{\hbar^2}\,{\bm I}
         \mez ;}
      \ee
      with the matrices ${\bm A}$ and ${\bm B}$ being given by prescription (\ref{bm-A-bm-B-defs}),
      and $\lambda=ik$ as in (\ref{lambda-from-k}).\\ Even more concisely we can state that
      \be \label{Siegert-bm-G-eq-3}
         {\color{red} {\bm G}_{\m E}^+ \; {\bm M}\m(\lambda) \; = \; -\,\frac{2\,m}{\hbar^2}\,{\bm I} \mez ;}
      \ee
      where the matrix polynomial ${\bm M}\m(\lambda)=(\ref{bm-M-lambda-def})$.
      
\newpage
      
{\color{blue}
\dotfill\\
\begin{centering}
{\sl Lecture \#6}\\
\vspace*{-0.20cm}
\end{centering}
\dotfill}
      
\item \underline{\bf C.12 Siegert state representation of the retarded Green function: Take \#2}\\
      Our introduction to nonhermitian scattering theory approaches an epic final.\\
      In this paragraph and in the next one,\\ all the above pursued technical developments of Part C come to fruition.\\
      Most importantly, equation (\ref{Siegert-bm-G-eq-3}) enables us to conclude that
      \be
         {\bm G}_{\m E}^+ \; = \; -\,\frac{2\,m}{\hbar^2} \; {\bm M}^{-1}\m(\lambda) \mez .
      \ee
      Our previously obtained formula (\ref{bm-M-lambda-inverse}) implies in turn
      \be \label{Siegert-bm-G-finres}
         {\color{red} {\bm G}_{\m E}^+ \; = \; \frac{m}{\hbar^2} \;
         \sum_{n=1}^{2N} \, \frac{{\bm c}^{(n)}\,{\bm c}^{(n)T}}{\lambda^{(n)}(\lambda^{(n)}-\lambda)} \mez ;}
      \ee
      or, in other words,
      \be \label{Siegert-G-matels-finres}
         G_{\m E,\nu\nu'}^+ \; = \; \frac{m}{\hbar^2} \;
         \sum_{n=1}^{2N} \, \frac{c_\nu^{(n)}\,c_{\nu'}^{(n)}}{\lambda^{(n)}(\lambda^{(n)}-\lambda)} \mez .
      \ee
      Subsequently one may combine (\ref{Siegert-G-matels-finres}) with (\ref{G-expansion-varphi}) and
      (\ref{Siegert-phi-n-explicit}).\\ This yields a simple, fully explicit, and compelling result
      \be \label{Siegert-retarded-Green-function-finres}
         {\color{red} G_{\m E}^+\n(x,y) \; = \; \frac{m}{\hbar^2} \;
         \sum_{n=1}^{2N} \, \frac{\varphi^{(n)}\m(x)\;\varphi^{(n)}\m(y)}{\lambda^{(n)}(\lambda^{(n)}-\lambda)} \; = \;
         \frac{m}{\hbar^2} \; \sum_{n=1}^{2N} \, \frac{\varphi^{(n)}\m(x)\;\varphi^{(n)}\m(y)}{k^{(n)}(k-k^{(n)})}
         \mez . \mez [\,-a \leq x,y \leq +a\,]}
      \ee
      Consistency of the whole mathematical formulation leading to (\ref{Siegert-retarded-Green-function-finres})\\
      excludes an existence of singularities in the denominator term $(k-k^{(n)})^{-1}$\m.\\
      $[\,$Another remark: Boundary condition (\ref{Siegert-G-bc-1}) implies that\\
      \phantom{$[\,$}$G_{\m E}^+\n(x\ge(+a),y) = {\cal G}_+ \, e^{+ikx}$\m, similarly $G_{\m E}^+\n(x\le(-a),y) = {\cal G}_- \, e^{-ikx}$
                     due to (\ref{Siegert-G-bc-2}). \hfill $(\#)$\\
      \phantom{$[\,$}This holds of course assuming $|y| \leq a$.\\
      \phantom{$[\,$}Factors ${\cal G}_\pm$ are fixed by demanding continuity of $G_{\m E}^+\n(x,y)$ at $x = \pm a$.$\,]$\\
      {\color{red} Equation (\ref{Siegert-retarded-Green-function-finres}) and our comment $(\#)$ express\\
      the retarded Green function $G_{\m E}^+\n(x,y)$ of the conventional quantum scattering theory\\
      solely in terms of the nonhermitian Siegert pseudostates (\ref{Siegert-phi-n-explicit}) and their pertinent eigenvalues (\ref{Siegert-phi-n-eigenvalues}).}\\
      $[\,$We tacitly assume $(-a) \leq y \leq (+a)$ and $x \in (-\infty,+\infty)$ in the just presented statements.$\,]$\\
      We recall from Part B that knowledge of $G_{\m E}^+\n(x,y)$ for $|y| \leq a$\\
      enables us to straightforwardly determine\\ the $\hat{T}$-operator and thus also the fundamental scattering operator $\hat{S}$\\
      $[\,$see equations (\ref{S-matel-take-3}), (\ref{T-matels-using-hat-T}), (\ref{hat-T-E-def})$\,]$. If so, then\\
      {\color{red} {\bf the entire quantum scattering theory can be reformulated}\\
      {\bf solely in terms of the Siegert pseudostates} $\left\{ \lambda^{(n)}\m,\, \varphi^{(n)}\m(x) \right\}_{n=1}^{2N}$\\
      {\bf which contain complete physical information about all the scattering phenomena}.}\\
      Valid of course as a well defined approximation,\\
      which becomes exact in the complete basis set limit of $N \to \infty$.\\
      As one can see, the fundamental equation (\ref{Siegert-retarded-Green-function-finres}) of Tolstikhin contains merely\\
      a discrete sum over all the $2N$ calculated Siegert pseudostates.\\
      The stationary scattering states $\psi_{E(\pm 1)}^{+}\m(x)$ of Part B are not referenced at all
      in (\ref{Siegert-retarded-Green-function-finres}).
      
\newpage
      
{\color{blue}
\dotfill\\
\begin{centering}
{\sl Lecture \#6}\\
\vspace*{-0.20cm}
\end{centering}
\dotfill}

\item \underline{\bf C.13 Siegert state representation of the transmission coefficient}\\
      Let us supply now\\ a concrete example of an application of the nonhermitian Siegert pseudostate formalism,\\
      namely, a Siegert based calculation of the transmission coefficient $T_{E}^{+}=T_{E(\pm 1)}^{+}$.\\
      $[\,$For the sake of clarity we recall in this context equation (\ref{T-E-+-both-the-same}) of {\sl Appendix II}.$\,]$\\
      We shall take advantage of the formula (\ref{G-E-+-a-b-final}) which was derived in {\sl Appendix IV}.\\
      When translated into our notation of Part C, equation (\ref{G-E-+-a-b-final}) looks as follows:
      \be \label{Siegert-G-E-+-a-b-final}
         G_{\m E}^+\n(-a,+a) \; = \; \frac{(-i)\,m}{\hbar^2 k} \;\, T_{E}^{+} \;\, e^{+2ika} \mez .
      \ee
      Combination of (\ref{Siegert-G-E-+-a-b-final}) and (\ref{Siegert-retarded-Green-function-finres})
      provides immediately\\ an interesting, important, and explicit result
      \be \label{Siegert-T-E-+-final}
         {\color{red} T_{E}^{+} \; = \; i \; k \; e^{-2ika} \, \sum_{n=1}^{2N} \,
         \frac{\varphi^{(n)}\m(-a)\;\varphi^{(n)}\m(+a)}{k^{(n)}(k-k^{(n)})} \mez . }
      \ee
      This demonstrates that the transmission coefficient $T_{E}^{+}$ of the conventional scattering theory\\
      is expressible solely in terms of the Siegert pseudostates.\\
      Formula (\ref{Siegert-T-E-+-final}) lends itself well to a numerical test.\\
      $[\,$As a benchmark one may take here our conventional calculation of $T_{E}^{+}=T_{E(\pm 1)}^{+}$\\
      \phantom{$[\,$}which was performed above in the paragraph B.13.\\
      \phantom{$[\,$}The just sketched program will be implemented in a very explicit fashion during our {\sl Project}.$\,]$\\
      It is worthy to look again at equation (\ref{Siegert-T-E-+-final}) and examine its physical contents.\\
      In (\ref{Siegert-T-E-+-final}), the transmission coefficient $T_{E}^{+}$ is determined\\
      as a discrete sum over separate contributions of all the corresponding $2N$ Siegert pseudostates.\\
      For a particular physical system under study, and for a prescribed impact energy $E = \frac{\hbar^2 k^2}{2m}$,\\
      it may sometimes happen that the physical value of $k$\\ almost coincides with $k^{(n)}$ at some single $n$ (say at $n=1$).\\
      If so, then the $n=1$ term often dominates in the summation (\ref{Siegert-T-E-+-final}),\\ and hence $T_{E}^{+}$ can be approximated just by
      \be \label{Siegert-T-E-+-approx}
         T_{E}^{+} \; \doteq \; i \; k \; e^{-2ika} \; \frac{\varphi^{(1)}\m(-a)\;\varphi^{(1)}\m(+a)}{k^{(1)}(k-k^{(1)})} \mez .
      \ee
      Suppose that $\varphi^{(1)}\m(x)$ solves the Siegert boundary value problem (\ref{ss-ode}), (\ref{ss-og-bc-1})-(\ref{ss-og-bc-2}).\\
      Then it is not hard to see that, necessarily, $\Re\,k^{(1)} > 0$ and $\Im\,k^{(1)} < 0$.\\
      $[\,$Indeed, since $k>0$ is close to $k^{(1)}$\m, then $\Re\,k^{(1)} > 0$ and $\Im\,k^{(1)}$ is small in magnitude.\\
      \phantom{$[\,$}Furthermore, as we know, $\varphi^{(1)}\m(x \,\m ^{>}_{<} \pm a) = {\cal C}_{\pm} \, e^{\pm ik^{(1)}x}$\m.\\
      \phantom{$[\,$}For $\Im\,k^{(1)} > 0$ the exponential $e^{\pm ik^{(1)}x}$ would decay with $x \to \pm\infty$,\\
      \phantom{$[\,$}thus $\varphi^{(1)}\m(x)$ would be a bound state pertinent to real $E^{(1)}$ and thus to real $k^{(1)}$, contradiction.\\
      \phantom{$[\,$}For $\Im\,k^{(1)} = 0$, $\varphi^{(1)}\m(x)$ would be an ordinary continuum eigenstate of $\hat{\sf H}$ having real $E^{(1)}>0$,\\
      \phantom{$[\,$}yet this is inconsistent with the Siegert boundary conditions (\ref{ss-og-bc-1})-(\ref{ss-og-bc-2})\\
      \phantom{$[\,$}due to properties (\ref{probability-conservation}), (\ref{T-R-1}).$\,]$\\
      Correspondingly, the associated complex energy eigenvalue $E^{(1)} = \,\m\hbar^2(k^{(1)})^2\m/(2\,m)$\\
      takes the form $E^{(1)}={\cal E}_{\rm res}^{(1)}-i\,\Gamma_{\rm res}^{(1)}/2$ where both ${\cal E}_{\rm res}^{(1)}>0$ and $\Gamma_{\rm res}^{(1)}>0$.\\
      We call $\varphi^{(1)}\m(x)$ as a {\color{red} resonance state}.
      
\newpage
      
{\color{blue}
\dotfill\\
\begin{centering}
{\sl Lecture \#6}\\
\vspace*{-0.20cm}
\end{centering}
\dotfill}

      Equation (\ref{Siegert-T-E-+-approx}) can be now further transformed, into
      \be \label{Siegert-T-E-+-approx-2}
         \hspace*{-1.00cm}
         T_{E}^{+} \; \doteq \; i \; \frac{\hbar^2}{2\,m} \left( \frac{k(k+k^{(1)})}{k^{(1)}} \right) e^{-2ika} \; \frac{\varphi^{(1)}\m(-a)\;\varphi^{(1)}\m(+a)}{E-E^{(1)}}
         \; \doteq \; i \; \frac{\hbar^2 k^{(1)}}{m} \; e^{-2ika} \; \frac{\varphi^{(1)}\m(-a)\;\varphi^{(1)}\m(+a)}{E-{\cal E}_{\rm res}^{(1)}+i\,\Gamma_{\rm res}^{(1)}/2} \mez .
      \ee
      The just obtained approximation (\ref{Siegert-T-E-+-approx-2}) of the transmission coefficient $T_{E}^{+} = (\ref{Siegert-T-E-+-final})$\\
      is expected to work very accurately in an energy range of $E \approx {\cal E}_{\rm res}^{(1)}$,\\
      provided that $\Gamma_{\rm res}^{(1)}$ is small in magnitude compared to ${\cal E}_{\rm res}^{(1)}$, and provided that\\
      there are no other resonances $E^{(n)}={\cal E}_{\rm res}^{(n)}-i\,\Gamma_{\rm res}^{(n)}/2$ located nearby $E^{(1)}={\cal E}_{\rm res}^{(1)}-i\,\Gamma_{\rm res}^{(1)}/2$.\\
      Assuming that (\ref{Siegert-T-E-+-approx-2}) holds reasonably well,\\
      the resulting transmission probability possesses for $E \approx {\cal E}_{\rm res}^{(1)}$ a Lorentzian form
      \be \label{Siegert-T-E-+-approx-3}
         {\color{red} \Bigl|\,T_{E}^{+}\,\Bigr|^2 \; \doteq \; {\cal Q} \;
         \frac{\left(\frac{\Gamma_{\rm res}^{(1)}}{2}\right)^{\m\m\m\n 2}}{\left(E-{\cal E}_{\rm res}^{(1)}\right)^{\m\m\m\n 2}+\left(\frac{\Gamma_{\rm res}^{(1)}}{2}\right)^{\m\m\m\n 2}}
         \mez ;}
      \ee
      with a prefactor
      \be
         {\cal Q} \; = \; \left| \, \frac{\hbar^2 k^{(1)}}{m} \, \frac{2}{\Gamma_{\rm res}^{(1)}} \, \varphi^{(1)}\m(-a)\;\varphi^{(1)}\m(+a) \, \right|^2 \mez .
      \ee
      Formula (\ref{Siegert-T-E-+-approx-3}) represents the so called\\
      {\color{red} Breit-Wigner profile} of a resonance peak in the transmission probability.\\
      Let us revisit now our illustrative scattering calculation discussed in the paragraph B.13.\\
      We claim that each of the two sharp peaks of $\Bigl|\,T_{E}^{+}\,\Bigr|^2$ encountered in Fig.2 of B.13\\
      appears merely due to the {\color{red} resonance transmission phenomenon},\\
      which is described adequately by the nonhermitian equation (\ref{Siegert-T-E-+-approx-3}).\\
      $[\,$Here apparently ${\cal Q} \doteq 1$, an explanation exists but reaches beyond the scope of our present course.$\,]$\\
      In other words, we anticipate that the two mentioned narrow transmission peaks of Fig.2\\
      arise due to {\color{red} presence of isolated resonances in our model potential} $V\m(x)$\\
      $[\,$defined by equation (\ref{V-Jolanta}) and plotted in Fig.1$\,]$.\\
      The just presented claim will be verified by an explicit numerical calculation\\
      during our computational {\sl Project}.\\
      In the paragraphs C.12-C.13 we have answered, arguably quite convincingly,\\
      the fundamental questions {\it iii)} and {\it iv)} raised before in C.4.\\
      Our short introduction to nonhermitian scattering theory can thus be brought to an end.
\end{itemize}

\newpage
      
{\color{orange}
\dotfill\\
\begin{centering}
{\sl Project}\\
\vspace*{-0.20cm}
\end{centering}
\dotfill}

\vspace*{+2.00cm}

{\bf D. Mini-project: Transmission phenomena studied using the conventional scattering theory\\
                      and using the nonhermitian Siegert based approach}
\vspace*{+0.20cm}
\begin{itemize}
\item The aim of this tutorial mini-project will be to investigate numerically\\
      the problem of scattering of a quantum particle in 1D.\\
      We shall focus on evaluating the transmission probability $\Bigl|\,T_{E}^{+}\,\Bigr|^2$\\
      as a function of the incident particle energy $E$, using two approaches:
      \begin{itemize}
      \item[{\it a)}] The standard hermitian time independent scattering theory of Part B.\\
                      See especially the paragraph B.13, which contains also the definition of our model problem.
      \item[{\it b)}] The nonhermitian scattering theory, based upon the Siegert pseudostate formalism\\
                      as outlined in Part C. See especially the paragraph C.13.
      \end{itemize}
\item Our first major goal will be to demonstrate that both theoretical frameworks {\it a)} and {\it b)}\\
      provide the same transmission profiles. $[\,$Which should be the same as plotted in Fig.2.$\,]$\\
      Our second goal will be to highlight fundamental advantage of the nonhermitian formalism\\
      $[\,$approach {\it b)}$\,]$ for the description of resonance scattering.\\
      This would amount to calculate explicitly the corresponding resonance wavefunctions $\varphi^{(n)}\m(x)$,\\
      and to identify their role in forming Lorentzian peaks (\ref{Siegert-T-E-+-approx-3}) in the obtained transmission profile.
\item Our third (optional) goal might be to proceed further\\
      by considering other suitable 1D potentials, and to try to understand the role\\
      of the so called exceptional points and/or anti-bound states in scattering phenomena.
\end{itemize}

\newpage

{\color{magenta}
\dotfill\\
\begin{centering}
{\sl Appendix I}\\
\vspace*{-0.20cm}
\end{centering}
\dotfill}

\vspace*{+0.50cm}

{\bf Appendix I: Continuity equation and related matters}\\

{\sl Based upon Refs.~\cite{Taylor,Roman}.}

\begin{itemize}
\item Consider the time dependent Schr\"{o}dinger equation (\ref{TDSCHE}) in the position representation,
      \be \label{TDSCHE-posrep}
         i \hbar \; \partial_t \, \psi(t,x) \; = \; -\,\frac{\hbar^2}{2\,m} \, \partial_{xx} \, \psi(t,x) \; + \; V\m(x) \, \psi(t,x) \mez .
      \ee
      Let $\psi(t,x)$ be any particular solution of (\ref{TDSCHE-posrep}),\\
      it can be a wavepacket but also a spatially delocalized function,\\
      depending upon the prescribed initial condition $\psi(t_{\rm init},x)$.\\
      A combination of formulas
      \begin{eqnarray}
         (+i)\hbar \; \partial_t \, \psi^{\phantom{*}}\n(t,x) & = & -\,\frac{\hbar^2}{2\,m} \, \partial_{xx} \, \psi^{\phantom{*}}\n(t,x) \; + \; V\m(x) \, \psi^{\phantom{*}}\n(t,x) \\
         (-i)\hbar \; \partial_t \, \psi^{*}\n(t,x) & = & -\,\frac{\hbar^2}{2\,m} \, \partial_{xx} \, \psi^{*}\n(t,x) \; + \; V\m(x) \, \psi^{*}\n(t,x)
      \end{eqnarray}
      provides the following outcome,
      \begin{eqnarray} \label{continuity-equation-take-1}
         \partial_t \Bigl( \psi^{*}\n(t,x) \, \psi(t,x) \Bigr) & = & \frac{i\,\hbar}{2\,m} \, \Bigl( \, \psi^{*}\n(t,x) \, \partial_{xx} \, \psi(t,x) \, - \,
         \psi(t,x) \, \partial_{xx} \, \psi^{*}\n(t,x) \, \Bigr) \; = \nonumber\\
         & = & \frac{i\,\hbar}{2\,m} \, \partial_x \Bigl( \, \psi^{*}\n(t,x) \, \partial_{x} \, \psi(t,x) \, - \,
         \psi(t,x) \, \partial_{x} \, \psi^{*}\n(t,x) \, \Bigr) \mez .
      \end{eqnarray}
      Subsequently one gets
      \be \label{continuity-equation-take-2}
         \partial_t \Bigl( \psi^{*}\n(t,x) \, \psi(t,x) \Bigr) \; = \; - \, \partial_x \, \frac{1}{2}
         \left\{ \, \psi^{*}\n(t,x) \left( \frac{\hat{\sf p}}{m} \, \psi(t,x) \right) \, - \,
         \psi(t,x) \left( \frac{\hat{\sf p}}{m} \, \psi(t,x) \right)^{\m\m\m\m *} \, \right\} \mez .
      \ee
      After setting
      \be
         \varrho(t,x) \; = \; \psi^{*}\n(t,x) \, \psi(t,x) \mz , \mz J(t,x) \; = \; \frac{1}{2}
         \left\{ \, \psi^{*}\n(t,x) \left( \frac{\hat{\sf p}}{m} \, \psi(t,x) \right) \, - \,
         \psi(t,x) \left( \frac{\hat{\sf p}}{m} \, \psi(t,x) \right)^{\m\m\m\m *} \, \right\}
      \ee
      one converts (\ref{continuity-equation-take-2}) into the desired continuity equation, namely,
      \be \label{continuity-equation-take-3}
         \partial_t \, \varrho(t,x) \; + \; \partial_x \, J(t,x) \; = \; 0 \mez .
      \ee
      Recall that the continuity equation (\ref{continuity-equation-take-3})\\ expresses mathematically the law of probability conservation.\\
      It is valid for an arbitrary solution $\psi(t,x)$ of the time dependent Schr\"{o}dinger equation (\ref{TDSCHE-posrep}).\\
      In particular, $\psi(t,x)$ does not need to be square integrable or spatially localized.
      
\newpage

{\color{magenta}
\dotfill\\
\begin{centering}
{\sl Appendix I}\\
\vspace*{-0.20cm}
\end{centering}
\dotfill}

\item Equation (\ref{continuity-equation-take-1}) admits an additional and rather important generalization.\\
      Namely, let $\psi_1\n(t,x)$ and $\psi_2\n(t,x)$ be two arbitrary (generally distinct) particular solutions of (\ref{TDSCHE-posrep}).\\
      A combination of formulas
      \begin{eqnarray}
         (+i)\hbar \; \partial_t \, \psi_2^{\phantom{*}}\n(t,x) & = & -\,\frac{\hbar^2}{2\,m} \, \partial_{xx} \, \psi_2^{\phantom{*}}\n(t,x) \; + \; V\m(x) \, \psi_2^{\phantom{*}}\n(t,x) \\
         (-i)\hbar \; \partial_t \, \psi_1^{*}\n(t,x) & = & -\,\frac{\hbar^2}{2\,m} \, \partial_{xx} \, \psi_1^{*}\n(t,x) \; + \; V\m(x) \, \psi_1^{*}\n(t,x)
      \end{eqnarray}
      provides the following outcome,
      \begin{eqnarray} \label{continuity-equation-take-1-off-diag}
         \partial_t \Bigl( \psi_1^{*}\n(t,x) \, \psi_2(t,x) \Bigr) & = & \frac{i\,\hbar}{2\,m} \, \Bigl( \, \psi_1^{*}\n(t,x) \, \partial_{xx} \, \psi_2\n(t,x) \, - \,
         \psi_2\n(t,x) \, \partial_{xx} \, \psi_1^{*}\n(t,x) \, \Bigr) \; = \nonumber\\
         & = & \frac{i\,\hbar}{2\,m} \, \partial_x \Bigl( \, \psi_1^{*}\n(t,x) \, \partial_{x} \, \psi_2\n(t,x) \, - \,
         \psi_2\n(t,x) \, \partial_{x} \, \psi_1^{*}\n(t,x) \, \Bigr) \mez .
      \end{eqnarray}
      Valid again even if the solutions $\psi_{1,2}\n(t,x)$ of (\ref{TDSCHE-posrep}) are not square integrable or spatially localized.
\item Let us revisit equation (\ref{continuity-equation-take-1-off-diag}), and assume that both $\psi_1\n(t,x)$ and $\psi_2\n(t,x)$
      are stationary states of $\hat{\sf H}$.\\
      Thus $\psi_{1,2}\n(t,x) \, = \, \psi_E^{(1,2)}\m(x) \; e^{-(i/\hbar)Et}$ with obvious notations.\\
      Equation (\ref{continuity-equation-take-1}) is correspondingly simplified into
      \be
         \partial_x \Bigl( \, \psi_E^{(1)*}\m(x) \, \partial_{x} \, \psi_E^{(2)}\m(x) \, - \, \psi_E^{(2)}\m(x) \, \partial_{x} \, \psi_E^{(1)*}\m(x) \, \Bigr) \; = \; 0 \mez .
      \ee
      Subsequent integration over $x \in (a,b)$ yields then
      \be \label{continuity-equation-take-4}
         \Bigl( \, \psi_E^{(1)*}\m(x) \, \partial_{x} \, \psi_E^{(2)}\m(x) \, - \,
         \psi_E^{(2)}\m(x) \, \partial_{x} \, \psi_E^{(1)*}\m(x) \, \Bigr)\Bigr|_{x=a}^{x=b} \; = \; 0 \mez .
      \ee
      Formula (\ref{continuity-equation-take-4}) will be exploited in {\sl Appendix II}.
\end{itemize}

\newpage

{\color{magenta}
\dotfill\\
\begin{centering}
{\sl Appendix II}\\
\vspace*{-0.20cm}
\end{centering}
\dotfill}

\vspace*{+0.50cm}

{\bf Appendix II: One-dimensional stationary Schr\"{o}dinger equation}\\

{\sl Based upon Refs.~\cite{Taylor,Roman}.}

\begin{itemize}
\item Let us consider an eigenvalue problem of the full Hamiltonian $\hat{\sf H}=(\ref{hat-sf-H-H0-V})$ in the position representation.\\
      One has
      \be \label{hat-sf-H-evp-posrep}
         \left\{ \, -\,\frac{\hbar^2}{2\,m}\,\partial_{xx} \, + \, V\m(x) \, \right\} \psi_E\n(x) \; = \; E \, \psi_E\n(x) \mez .
      \ee
      Here $\psi_E\n(x)$ is a provisional notation for the eigenfunctions.\\
      In what follows, we shall be concerned only with the continuum eigenstates of $\hat{\sf H}$,\\ associated with $E>0$.
      We omit any discussion of eventual bound states.
\item At any given $E>0$,\\ the problem (\ref{hat-sf-H-evp-posrep}) represents a second order ordinary differential equation (ODE) for $\psi_E\n(x)$.\\
      Since our potential $V\m(x)$ vanishes for all $x \ge b$,\\
      it is legitimate to set $\psi_E\n(x \ge b) = e^{+iKx}$ with $\hbar K = \sqrt{2\,m\,E}$.\\
      Subsequent back propagation of the ODE (\ref{hat-sf-H-evp-posrep}) yields uniquely $\psi_E\n(x)$ also for all $x<b$.\\
      Since $V\m(x)$ vanishes for all $x \le a$, the wavefunction $\psi_E\n(x \le a)$\\
      is necessarily some linear combination of two exponentials $e^{\pm iKx}$\m\n.\\
      Thus $\psi_E\n(x \le a) = A(E)\,e^{+iKx} + B(E)\,e^{-iKx}$\m,\\ where $A(E)$ and $B(E)$ are some uniquely determined coefficients.\\
      Let us recall now formula (\ref{continuity-equation-take-4}) from {\sl Appendix I} and set $\psi_E^{(1)}\m(x)=\psi_E^{(2)}\m(x)=\psi_E\n(x)$, such that
      \be \label{continuity-equation-take-5}
         \Bigl( \, \psi_E^*\n(x) \, \partial_{x} \, \psi_E\n(x) \, - \,
         \psi_E\n(x) \, \partial_{x} \, \psi_E^*\n(x) \, \Bigr)\Bigr|_{x=a}^{x=b} \; = \; 0 \mez .
      \ee
      After plugging into (\ref{continuity-equation-take-5}) the above discussed simple explicit forms of $\psi_E\n(x \le a)$ and $\psi_E\n(x \ge b)$,\\
      one finds that $|A(E)|^2 = |B(E)|^2 + 1$. Showing that $A(E)$ is certainly nonzero.\\
      If so, one may redefine $\psi_E\n(x)$ via dividing it by $A(E)$.\\
      Then $\psi_E\n(x \le a) = e^{+iKx} + R(E)\,e^{-iKx}$ and $\psi_E\n(x \ge b) = T(E)\,e^{+iKx}$\m,\\
      where $R(E)=B(E)/A(E)$ and $T(E)=1/A(E)$ are uniquely determined coefficients.\\
      Since $|A(E)|^2 = |B(E)|^2 + 1$, one inevitably has $|T(E)|^2 + |R(E)|^2 = 1$.
\item Let us summarize the findings of the above paragraph\\
      using a slightly modified notation (updated to match our notation in the main text).\\
      We have shown that there exists an unique particular solution $\tilde{\psi}_{E(+1)}^+\n(x)$ of the ODE (\ref{hat-sf-H-evp-posrep})\\
      which exhibits the following kind of behavior in the asymptotic regions of $x \leq a$ and $x \geq b$:
      \begin{eqnarray}
         \label{tilde-psi-E-+-bc-b} \tilde{\psi}_{E(+1)}^{+}\n(x \geq b) & = & \hspace*{+1.776cm} T_{E(+1)}^{+} \; e^{+i K x} \mez ;\\
         \label{tilde-psi-E-+-bc-a} \tilde{\psi}_{E(+1)}^{+}\n(x \leq a) & = & e^{+i K x} \; + \; R_{E(+1)}^{+} \; e^{-i K x} \mez .
      \end{eqnarray}      
      Here $\hbar K = \sqrt{2\,m\,E}$ as before, while $T_{E(+1)}^{+}$ and $R_{E(+1)}^{+}$ are some uniquely determined coefficients\\
      (these carry a lot of physical meaning, as demonstrated in much detail within the main text).\\
      A completely analogous reasoning reveals that there exists also another\\
      unique particular solution $\tilde{\psi}_{E(-1)}^+\n(x)$ of the ODE (\ref{hat-sf-H-evp-posrep}),\\
      which behaves in the asymptotic regions of $x \leq a$ and $x \geq b$ as follows:
      \begin{eqnarray}
         \label{tilde-psi-E---bc-b} \tilde{\psi}_{E(-1)}^{+}\n(x \geq b) & = & e^{-i K x} \; + \; R_{E(-1)}^{+} \; e^{+i K x} \mez ;\\
         \label{tilde-psi-E---bc-a} \tilde{\psi}_{E(-1)}^{+}\n(x \leq a) & = & \hspace*{+1.805cm} T_{E(-1)}^{+} \; e^{-i K x} \mez ;
      \end{eqnarray}

\newpage

{\color{magenta}
\dotfill\\
\begin{centering}
{\sl Appendix II}\\
\vspace*{-0.20cm}
\end{centering}
\dotfill}

      here again $T_{E(-1)}^{+}$ and $R_{E(-1)}^{+}$ are some uniquely determined coefficients.\\
      In the main text we have assigned direct physical meaning to $T_{E(\pm 1)}^{+}$ and $R_{E(\pm 1)}^{+}$,\\
      see equations (\ref{phi-out-T-def})-(\ref{phi-out-R-def}), (\ref{T-probability-final})-(\ref{R-probability-final})
      and the accompanying discussion.
\item Solutions $\tilde{\psi}_{E(+1)}^+\n(x)$ and $\tilde{\psi}_{E(-1)}^+\n(x)$ of the ODE (\ref{hat-sf-H-evp-posrep}) are linearly independent,\\
      and span thus the two-dimensional eigenspace of $\hat{\sf H}$ assigned to the given continuum energy level $E$.\\
      The entities $T_{E(\pm 1)}^{+}$ and $R_{E(\pm 1)}^{+}$ possess two important properties, namely,
      \be \label{T-R-1}
         \Bigl| \, T_{E(\pm 1)}^{+} \, \Bigr|^2 \; + \; \Bigl| \, R_{E(\pm 1)}^{+} \, \Bigr|^2 \; = \; 1 \mez ;
      \ee
      and
      \be \label{T-R-0}
         T_{E(\pm 1)}^{+*}\,R_{E(\mp 1)}^{+} \; + \; R_{E(\pm 1)}^{+*}\,T_{E(\mp 1)}^{+} \; = \; 0 \mez .
      \ee
      The normalization property (\ref{T-R-1}) is a direct consequence of the law of probability conservation.\\
      It can be quicky verified by taking equation (\ref{continuity-equation-take-4}) and choosing $\psi_E^{(1)}\m(x)=\psi_E^{(2)}\m(x)=\tilde{\psi}_{E(\pm 1)}^{+}\n(x)$.\\
      $[\,$Recall that the diagonal version $\psi_E^{(1)}\m(x)=\psi_E^{(2)}\m(x)$ of equation (\ref{continuity-equation-take-4}) is indeed closely related\\
      \phantom{$[\,$}to the law of probability conservation for stationary states, see {\sl Appendix I}.$\,]$\\
      The orthogonality property (\ref{T-R-0}) follows immediately from equation (\ref{continuity-equation-take-4})\\
      after setting $\psi_E^{(1)}\m(x)=\tilde{\psi}_{E(\pm 1)}^{+}\n(x)$ and $\psi_E^{(2)}\m(x)=\tilde{\psi}_{E(\mp 1)}^{+}\n(x)$.
\item Proceeding further,\\
      let us establish appropriate orthonormality relations for the eigenfunctions $\tilde{\psi}_{E(\pm 1)}^{+}\n(x)$.\\
      To accomplish this goal, we shall set $\eta \in \{ -1,+1 \}$, and examine an overlap
      \be \label{I-overlap-def}
         I_{(E\eta)(E'\n\eta'\n)}^{L} \; = \; \int_{-L}^{+L} \m \tilde{\psi}_{E\eta}^{+*}\n(x) \, \tilde{\psi}_{E'\n\eta'}^{+}\m(x) \; {\rm d}x \mez ; \mez [\,L>0\,]
      \ee
      with an intention of taking the box length $L \to +\infty$ later on.\\
      One has
      \begin{eqnarray}
         \label{I-overlap-analysis-1}
         \left\{ \, -\,\frac{\hbar^2}{2\,m}\,\partial_{xx} \, + \, V\m(x) \, \right\} \;\,\m\n \tilde{\psi}_{E\eta}^{+*}\n(x) & = & E \; \tilde{\psi}_{E\eta}^{+*}\n(x) \mez ;\\
         \label{I-overlap-analysis-2}
         \left\{ \, -\,\frac{\hbar^2}{2\,m}\,\partial_{xx} \, + \, V\m(x) \, \right\} \tilde{\psi}_{E'\n\eta'\n}^{+}\m(x) & = & E' \m \, \tilde{\psi}_{E'\n\eta'\n}^{+}\m(x) \mez \m\m .
      \end{eqnarray}
      Multiply (\ref{I-overlap-analysis-1}) by $\tilde{\psi}_{E'\n\eta'\n}^{+}\m(x)$, and (\ref{I-overlap-analysis-2}) by $\tilde{\psi}_{E\eta}^{+*}\n(x)$.\\
      Subtract (\ref{I-overlap-analysis-2}) from (\ref{I-overlap-analysis-1}), and integrate both sides of the resulting equation over $x \in (-L,+L)$.\\
      This yields an intermediate outcome
      \be \label{I-overlap-analysis-3}
         (E-E'\n) \, I_{(E\eta)(E'\n\eta'\n)}^{L} \; = \; \frac{\hbar^2}{2\,m} \, \int_{-L}^{+L}
         \Bigl\{ \tilde{\psi}_{E\eta}^{+*}\n(x) \, \partial_{xx} \, \tilde{\psi}_{E'\n\eta'\n}^{+}\m(x) \, - \,
         \tilde{\psi}_{E'\n\eta'\n}^{+}\m(x) \, \partial_{xx} \, \tilde{\psi}_{E\eta}^{+*}\n(x) \Bigr\} \; {\rm d}x \mez .
      \ee
      After integrating by parts one gets
      \be \label{I-overlap-analysis-4}
         (E-E'\n) \, I_{(E\eta)(E'\n\eta'\n)}^{L} \; = \; \frac{\hbar^2}{2\,m} \, \left.
         \Bigl\{ \tilde{\psi}_{E\eta}^{+*}\n(x) \, \partial_{x} \, \tilde{\psi}_{E'\n\eta'\n}^{+}\m(x) \, - \,
         \tilde{\psi}_{E'\n\eta'\n}^{+}\m(x) \, \partial_{x} \, \tilde{\psi}_{E\eta}^{+*}\n(x) \Bigr\} \right|_{x=-L}^{x=+L} \mez .
      \ee
      
\newpage

{\color{magenta}
\dotfill\\
\begin{centering}
{\sl Appendix II}\\
\vspace*{-0.20cm}
\end{centering}
\dotfill}

      Assume now that the interval $(-L,+L)$ is large enough\\ as to cover the whole interaction region $x \in (a,b)$.\\
      In that case, the r.h.s.~of (\ref{I-overlap-analysis-4}) can be worked out further\\
      using the asymptotic forms (\ref{tilde-psi-E-+-bc-b})-(\ref{tilde-psi-E-+-bc-a}), (\ref{tilde-psi-E---bc-b})-(\ref{tilde-psi-E---bc-a}) of the involved wavefunctions.\\
      Leading to the following expressions:
      \begin{eqnarray}
         \label{I-overlap-analysis-5}
         \hspace*{-2.00cm} & & \left(\frac{\hbar^2 K^2}{2\,m}-\frac{\hbar^2 K'^2}{2\,m}\n\right) \m I_{(E(+1))(E'\n(+1))}^{L} \; = \\
         \hspace*{-2.00cm} & = & \frac{\hbar^2}{2\,m} \, \Biggl\{ \,i\,T_{E(+1)}^{+*}\,T_{E'\n(+1)}^{+}\,(K+K'\n)\,e^{-i(K-K'\n)L}
         \; + \; i\,R_{E(+1)}^{+*}\,R_{E'\n(+1)}^{+}\,(K+K'\n)\,e^{-i(K-K'\n)L} \; - \;
         i\,(K+K'\n)\,e^{+i(K-K'\n)L} \nonumber\\ \hspace*{-2.00cm} & & \phantom{\frac{\hbar^2}{2\,m} \, \Biggl\{ } + \; i\,R_{E(+1)}^{+*}\,(K-K'\n)\,e^{-i(K+K'\n)L}
         - \; i\,R_{E'\n(+1)}^{+}\,(K-K'\n)\,e^{+i(K+K'\n)L} \, \Biggr\} \mez ; \nonumber\\
         \label{I-overlap-analysis-6}
         \hspace*{-2.00cm} & & \left(\frac{\hbar^2 K^2}{2\,m}-\frac{\hbar^2 K'^2}{2\,m}\n\right) \m I_{(E(+1))(E'\n(-1))}^{L} \; = \\
         \hspace*{-2.00cm} & = & \frac{\hbar^2}{2\,m} \, \Biggl\{ \,i\,T_{E(+1)}^{+*}\,R_{E'\n(-1)}^{+}\,(K+K'\n)\,e^{-i(K-K'\n)L}
         \; + \; i\,R_{E(+1)}^{+*}\,T_{E'\n(-1)}^{+}\,(K+K'\n)\,e^{-i(K-K'\n)L} \nonumber\\
         \hspace*{-2.00cm} & & \phantom{\frac{\hbar^2}{2\,m} \, \Biggl\{ } + \; i\,T_{E(+1)}^{+*}\,(K-K'\n)\,e^{-i(K+K'\n)L}
         - \; i\,T_{E'\n(-1)}^{+}\,(K-K'\n)\,e^{+i(K+K'\n)L} \, \Biggr\} \mez . \nonumber
      \end{eqnarray}
      Here of course $K'\n=\sqrt{2\,m\,E'}$.\\
      Analogous expressions result also for $I_{(E(-1))(E'\n(+1))}^{L}$ and $I_{(E(-1))(E'\n(-1))}^{L}$,\\
      we do not display them here to save space.\\
      Consider now $I_{(E\eta)(E'\n\eta'\n)}^{L}=(\ref{I-overlap-def})$ to be a distributional entity, whose\\
      only physical meaning arises due to integrals
      ${\cal I}_{\eta\eta'}^L\m[\chi_1,\chi_2] = \int_{0}^{+\infty} \m {\rm d}E \int_{0}^{+\infty} \m {\rm d}E' \, I_{(E\eta)(E'\n\eta'\n)}^{L}\,\chi_1\n(E)\,\chi_2\n(E'\n)$,\\
      where $\chi_{1,2}\n(E)$ are continuous test functions of the energy $E>0$ or equivalently of $K>0$.\\
      $[\,$We also tacitly assume that $\chi_{1,2}\n(E)$ falls off to zero at $E \to +\infty$ faster than any power of $K$\m,\\
      \phantom{$[\,$}this makes ${\cal I}_{\eta\eta'}^L\m[\chi_1,\chi_2]$ always well defined.$\,]$\\
      In the limit of $L \to +\infty$, the exponential $e^{\pm i(K+K'\n)L}$ appearing in (\ref{I-overlap-analysis-5})-(\ref{I-overlap-analysis-6})\\
      oscillates rapidly with $K$ and $K'$ unless both $K$ and $K'$ are close to zero.\\
      Meaning that all the terms of (\ref{I-overlap-analysis-5})-(\ref{I-overlap-analysis-6}) containing $(K-K'\n)\,e^{\pm i(K+K'\n)L}$\\
      do not contribute to ${\cal I}_{\eta\eta'}^L\m[\chi_1,\chi_2]$, and can thus be discarded for large $L$.\\
      On the other hand, in the limit of $L \to +\infty$, the exponential $e^{\pm i(K-K'\n)L}$ appearing in (\ref{I-overlap-analysis-5})-(\ref{I-overlap-analysis-6})\\
      oscillates rapidly with $K$ and $K'$ unless $(K-K'\n)$ is close to zero.\\
      Hence (\ref{I-overlap-analysis-5})-(\ref{I-overlap-analysis-6}) can be correspondingly simplified as follows:
      \begin{eqnarray}
         \label{I-overlap-analysis-7}
         \hspace*{-1.50cm} +i\,(K-K'\n) \; I_{(E(+1))(E'\n(+1))}^{L} & = & e^{+i(K-K'\n)L}
         \; - \; \Bigl|\,\m T_{E(+1)}^{+}\Bigr|^2\,e^{-i(K-K'\n)L} \; - \; \Bigl|\,\m R_{E(+1)}^{+}\Bigr|^2\,e^{-i(K-K'\n)L} \mez ;\\
         \label{I-overlap-analysis-8}
         \hspace*{-1.50cm} -i\,(K-K'\n) \; I_{(E(+1))(E'\n(-1))}^{L} & = & T_{E(+1)}^{+*}\,R_{E'\n(-1)}^{+}\,e^{-i(K-K'\n)L} \; + \;
         R_{E(+1)}^{+*}\,T_{E'\n(-1)}^{+}\,e^{-i(K-K'\n)L} \hspace*{+1.55cm} .
      \end{eqnarray}
      $[\,$Note that we have kept $E'$ in (\ref{I-overlap-analysis-8}) for our later convenience, {\sl cf}.~equation (\ref{I-overlap-analysis-11}) below.$\,]$\\
      Analogous expressions would result also for $I_{(E(-1))(E'\n(+1))}^{L}$ and $I_{(E(-1))(E'\n(-1))}^{L}$.
      
\newpage

{\color{magenta}
\dotfill\\
\begin{centering}
{\sl Appendix II}\\
\vspace*{-0.20cm}
\end{centering}
\dotfill}

      An additional dramatic simplification is possible due to properties (\ref{T-R-1})-(\ref{T-R-0}).\\
      Namely, after exploiting (\ref{T-R-1}), equation (\ref{I-overlap-analysis-7}) boils down into
      \be \label{I-overlap-analysis-9}
         I_{(E(+1))(E'\n(+1))}^{L} \; = \; \frac{2\,\sin((K-K'\n)L)}{(K-K'\n)} \mez .
      \ee
      For $L \to +\infty$, the familiar textbook formula reveals that
      \be \label{I-overlap-analysis-10}
         I_{(E(+1))(E'\n(+1))}^{\infty} \; = \; 2\,\pi \; \delta(K-K'\n) \mez .
      \ee
      Furthermore, equation (\ref{I-overlap-analysis-8}) leads to
      \begin{eqnarray}
         \label{I-overlap-analysis-11}
         \hspace*{-1.50cm} -i\,(K-K'\n) \; I_{(E(+1))(E'\n(-1))}^{L} & = &
         \left( \, T_{E(+1)}^{+*}\,R_{E(-1)}^{+} \, + \, R_{E(+1)}^{+*}\,T_{E(-1)}^{+} \right) e^{-i(K-K'\n)L} \nonumber\\
         \hspace*{-1.50cm} & + & \left( \, T_{E(+1)}^{+*}\,\partial_E\,R_{E(-1)}^{+} \, + \, R_{E(+1)}^{+*}\,\partial_E\,T_{E(-1)}^{+} \right) (E'\n-E) \; e^{-i(K-K'\n)L} \mez .
      \end{eqnarray}
      The first term on the r.h.s.~of (\ref{I-overlap-analysis-11}) vanishes because of (\ref{T-R-0}), hence
      \begin{eqnarray}
         \label{I-overlap-analysis-12}
         \hspace*{-1.50cm} i \; I_{(E(+1))(E'\n(-1))}^{L} & = &
         \left( \, T_{E(+1)}^{+*}\,\partial_E\,R_{E(-1)}^{+} \, + \, R_{E(+1)}^{+*}\,\partial_E\,T_{E(-1)}^{+} \right) \frac{\hbar^2}{2\,m} \; (K'\n+K) \; e^{-i(K-K'\n)L} \mez .
      \end{eqnarray}
      The r.h.s.~of (\ref{I-overlap-analysis-12}) is a smooth bounded function of $(E,E'\n)$ or equivalently of $(K,K'\n)$.\\
      For $L \to +\infty$, the exponential $e^{-i(K-K'\n)L}$ rapidly oscillates when varying $(K,K'\n)$,\\
      unless $K$ is extremely close to $K'$\m\m. Meaning that ${\cal I}_{(+1)(-1)}^{L \to +\infty}\n[\chi_1,\chi_2]=0$ for all the test functions.\\
      Thus in short
      \be \label{I-overlap-analysis-13}
         I_{(E(+1))(E'\n(-1))}^{\infty} \; = \; 0 \mez .
      \ee
      Analogous conclusions would be found also for $I_{(E(-1))(E'\n(+1))}^{\infty}$ and $I_{(E(-1))(E'\n(-1))}^{\infty}$.\\
      If so, we may state now in summary\\
      that our sought orthonormality relations for the eigenfunctions $\tilde{\psi}_{E\eta}^{+}\n(x)$\\
      take the following simple appearance:
      \be \label{tilde-psi-E-eta-onrel-take-1}
         I_{(E\eta)(E'\n\eta'\n)}^{\infty} \; \equiv \; \int_{-\infty}^{+\infty} \m \tilde{\psi}_{E\eta}^{+*}\n(x) \, \tilde{\psi}_{E'\n\eta'}^{+}\m(x) \; {\rm d}x
         \; = \; 2\,\pi \; \delta(K-K'\n) \; \delta_{\eta\eta'} \mez .
      \ee
      Since $\delta(E-E'\n)=\delta\m\m\m\left(\frac{\hbar^2 K^2}{2\,m}-\frac{\hbar^2 K'^2}{2\,m}\right) = \frac{2\,m}{\hbar^2} \; \delta((K+K'\n)(K-K'\n)) =
      \frac{m}{\hbar^2 K} \; \delta(K-K'\n)$,\\ we may write also alternatively
      \be \label{tilde-psi-E-eta-onrel-take-2}
         \int_{-\infty}^{+\infty} \m \tilde{\psi}_{E\eta}^{+*}\n(x) \, \tilde{\psi}_{E'\n\eta'}^{+}\m(x) \; {\rm d}x
         \; = \; 2\,\pi \; \frac{\hbar^2 K}{m} \; \delta(E-E'\n) \; \delta_{\eta\eta'} \; = \; 2\,\pi \; \frac{\hbar}{m} \; \sqrt{2\,m\,E} \;\, \delta(E-E'\n) \; \delta_{\eta\eta'} \mez .
      \ee
      The above displayed orthonormality relations (\ref{tilde-psi-E-eta-onrel-take-1})-(\ref{tilde-psi-E-eta-onrel-take-2})
      look even simpler\\ when one uses renormalized wavefunctions
      \be \label{psi-E-eta-+-x-def}
         \psi_{E\eta}^{+}\m(x) \; = \; \sqrt{\frac{m}{2\,\pi\,\hbar^2 K}} \; \tilde{\psi}_{E\eta}^{+}\m(x) \mez .
      \ee
      Indeed, we have
      \be \label{psi-E-eta-onrel-take-1}
         \int_{-\infty}^{+\infty} \m \psi_{E\eta}^{+*}\n(x) \, \psi_{E'\n\eta'}^{+}\m(x) \; {\rm d}x
         \; = \; \delta(E-E'\n) \; \delta_{\eta\eta'} \mez .
      \ee
      In an absence of the potential, or for energies $E$ much larger than the magnitude of the potential,\\
      our wavefunction $\psi_{E\eta}^{+}\m(x)$ reduces to $\phi_{E\eta}\m(x)=(\ref{phi-E-eta-def})$.

\newpage

{\color{magenta}
\dotfill\\
\begin{centering}
{\sl Appendix II}\\
\vspace*{-0.20cm}
\end{centering}
\dotfill}

\item Having established appropriate orthonormality relations\\
      for the eigenfunctions $\tilde{\psi}_{E\eta}^{+}\n(x)$ or $\psi_{E\eta}^{+}\n(x)$, let us briefly comment on completeness and closure.\\
      We claim that our quantum state space of all the square integrable functions of $x$\\ is spanned by the following orthonormal basis,
      \be \label{H-basis}
         \Bigl\{ \, \psi_n\n(x) \, \Bigr\}_{\m n=1}^{n_{\rm bound}} \mez \& \mez \Bigl\{ \, \psi_{E\eta}^{+}\n(x) \, \Bigr\}_{E>0}^{\eta=\pm 1} \mez .
      \ee
      Here $\{ \psi_n\n(x) \}_{n=1}^{n_{\rm bound}}$ are all the bound state wavefunctions of the studied problem\\ (unit normalization is assumed).\\
      The corresponding closure relation takes then an explicit form
      \be \label{H-closure}
         \sum_{n=1}^{n_{\rm bound}} \, \psi_n\n(x) \, \psi_n^*\n(x'\n) \; + \;
         \int_{0}^{\infty} \m {\rm d}E \, \sum_\eta \, \psi_{E\eta}^{+}\n(x) \; \psi_{E\eta}^{+*}\n(x'\n) \; = \; \delta(x-x'\n) \mez .
      \ee
      Our above presented closure statement (\ref{H-basis})-(\ref{H-closure}) must of course be validated.\\
      Probably the most quick and elementary way of accomplishing this goal is the following one.\\
      Consider an eigenvalue problem (\ref{hat-sf-H-evp-posrep}) with the coordinate $x$ restricted to a finite box, $-L \leq x \leq +L$.\\
      Such an eigenproblem possesses just a discrete non-degenerate energy spectrum.\\
      $[\,$Belonging to physical bound states and an infinite number of discretized quasi-continuum states,\\
      \phantom{$[\,$}to be denoted collectively as $\{ \psi_\nu^L\n(x) \}_{\nu>0}$.$\,]$\\
      These discrete eigenstates $\{ \psi_\nu^L\n(x) \}_{\nu>0}$ form a complete orthonormal basis on ${\mathbb L}^2(-L,+L)$,\\
      as explained in every elementary quantum mechanics textbook. Meaning also that\\
      every square integrable function of $x \in (-\infty,+\infty)$ can be arbitrarily accurately expanded\\
      in the orthonormal basis set $\{ \psi_\nu^L\n(x) \}_{\nu>0}$, provided only that the box size $L$ is large enough.\\
      In the limit of $L \to +\infty$, the lowest lying eigenstates $\{ \psi_\nu^L\n(x) \}_{\nu=1}^{n_{\rm bound}}$ approach $\{ \psi_n\n(x) \}_{n=1}^{n_{\rm bound}}$\m, whereas\\
      the rest $\{ \psi_\nu^L\n(x) \}_{\nu>n_{\rm bound}}$ resembles closer and closer the continuum eigenstates of our problem.\\
      This shows in turn that the wavefunctions (\ref{H-basis}) indeed do form a complete set\\
      on our quantum state space of all the square integrable functions of $x \in (-\infty,+\infty)$.\\
      If so, the closure property (\ref{H-closure}) results immediately as a consequence of (\ref{psi-E-eta-onrel-take-1}).
\item It is instructive to examine also the Wronskian of the two eigenfunctions $\psi_{E(\pm1)}^{+}\n(x)$,\\
      i.e., the quantity
      \be \label{w-E-x-def}
         w(E,x) \; = \;
         \psi_{E(-1)}^{+}\n(x) \; \partial_x \, \psi_{E(+1)}^{+}\n(x) \; - \;
         \psi_{E(+1)}^{+}\n(x) \; \partial_x \, \psi_{E(-1)}^{+}\n(x) \mez .
      \ee
      An elementary theory of second order ODEs teaches us that $w(E,x)$ is surely nonzero\\
      $[\,$since $\psi_{E(+1)}^{+}\n(x)$ is linearly independent upon $\psi_{E(-1)}^{+}\n(x)$$\,]$,
      and even independent of $x$.\\ Hence $w(E,x)=(\ref{w-E-x-def})$ can be evaluated e.g.~at $x=b$ using
      (\ref{tilde-psi-E-+-bc-b}), (\ref{tilde-psi-E---bc-b}) and (\ref{psi-E-eta-+-x-def}).\\
      One gets $w(E,b) = \frac{i\,m}{\pi\,\hbar^2} \, T_{E(+1)}^{+}$. On the other hand,\\
      we may equivalently evaluate $w(E,x)$ at $x=a$ using (\ref{tilde-psi-E-+-bc-a}), (\ref{tilde-psi-E---bc-a}) and (\ref{psi-E-eta-+-x-def}).\\
      This yields $w(E,a) = \frac{i\,m}{\pi\,\hbar^2} \, T_{E(-1)}^{+}$. Since $w(E,a)=w(E,b)=w(E,x)$,\\ one must inevitably conclude that
      \be \label{T-E-+-both-the-same}
         T_{E(+1)}^{+} \; = \; T_{E(-1)}^{+} \; = \; T_{E}^{+} \mez ;
      \ee
      and
      \be \label{w-E-x-explicit}
         w(E,x) \; = \; \frac{i\,m}{\pi\,\hbar^2} \; T_{E}^{+} \mez .
      \ee
      Thus $T_{E}^{+}$ can never be zero, simply because $w(E,x)$ is nonvanishing.

\newpage

{\color{magenta}
\dotfill\\
\begin{centering}
{\sl Appendix II}\\
\vspace*{-0.20cm}
\end{centering}
\dotfill}

\item We shall conclude {\sl Appendix II}\\ by making another auxiliary notational update, needed mainly for {\sl Appendix III}.\\
      Namely, let us replace the above used subscript index $_{E\eta}$\\
      by mere $k$, with $k = \eta\,K = \eta\,\sqrt{2\,m\,E}$. Clearly $|k|=K$ and ${\rm sgn}\,k=\eta$.\\
      Then $\tilde{\psi}_{E\eta}^{+}\n(x)$ is equivalent to writing just $\tilde{\psi}_{k}^{+}\n(x)$,\\
      similarly instead of $T_{E(\pm 1)}^{+}$ and $R_{E(\pm 1)}^{+}$ one may write just $T(k)$ and $R(k)$.\\
      Formulas (\ref{tilde-psi-E-+-bc-b})-(\ref{tilde-psi-E-+-bc-a}) are then redisplayed in the following alternative appearance,
      \begin{eqnarray}
         \label{tilde-psi-k>0-bc-b} \tilde{\psi}_{k>0}^{+}\n(x \geq b) & = & \hspace*{+1.640cm} T(k) \; e^{+i k x} \mez ;\\
         \label{tilde-psi-k>0-bc-a} \tilde{\psi}_{k>0}^{+}\n(x \leq a) & = & e^{+i k x} \; + \; R(k) \; e^{-i k x} \mez .
      \end{eqnarray}
      Similarly, formulas (\ref{tilde-psi-E---bc-b})-(\ref{tilde-psi-E---bc-a}) can be redisplayed as
      \begin{eqnarray}
         \label{tilde-psi-k<0-bc-b} \tilde{\psi}_{k<0}^{+}\n(x \geq b) & = & e^{+i k x} \; + \; R(k) \; e^{-i k x} \mez ;\\
         \label{tilde-psi-k<0-bc-a} \tilde{\psi}_{k<0}^{+}\n(x \leq a) & = & \hspace*{+1.640cm} T(k) \; e^{+i k x} \mez .
      \end{eqnarray}
      Properties (\ref{T-R-1})-(\ref{T-R-0}) take an alternative form
      \be \label{T-R-1-k}
         \Bigl| \, T(k) \, \Bigr|^2 \; + \; \Bigl| \, R(k) \, \Bigr|^2 \; = \; 1 \mez ;
      \ee
      and
      \be \label{T-R-0-k}
         T^{*}\m\n(+k) \, R(-k) \; + \; R^{*}\m\n(+k) \, T(-k) \; = \; 0 \mez ;
      \ee
      valid for any nonzero $k$. Also $T(+k) = T(-k)$ as stated in (\ref{T-E-+-both-the-same}).\\
      Finally, the orthonormality relations (\ref{tilde-psi-E-eta-onrel-take-1}) are rewritten
      simply as
      \be \label{tilde-psi-E-eta-onrel-take-3}
         \int_{-\infty}^{+\infty} \m \tilde{\psi}_{k}^{+*}\n(x) \, \tilde{\psi}_{k'}^{+}\m(x) \; {\rm d}x
         \; = \; 2\,\pi \; \delta(k-k'\n) \mez ;
      \ee
      and the associated closure property (\ref{H-closure}) becomes
      \be \label{H-closure-k}
         \sum_{n=1}^{n_{\rm bound}} \, \psi_n\n(x) \, \psi_n^*\n(x'\n) \;\, + \;\, \frac{1}{2\,\pi}
         \int_{-\infty}^{+\infty} \m {\rm d}k \;\, \tilde{\psi}_{k}^{+}\m\n(x) \; \tilde{\psi}_{k}^{+*}\m\n(x'\n) \; = \; \delta(x-x'\n) \mez .
      \ee
\end{itemize}

\newpage

{\color{magenta}
\dotfill\\
\begin{centering}
{\sl Appendix III}\\
\vspace*{-0.20cm}
\end{centering}
\dotfill}

\vspace*{+0.50cm}

{\bf Appendix III: The Asymptotic Condition}
\vspace*{+0.10cm}
\begin{itemize}
\item Let $\psi(0,x)$ be a general square integrable wavepacket prepared at the time instant $t=0$.\\
      Assume that $\psi(0,x)$ does not populate any bound states. Then $\psi(0,x)$ is expressible\\
      as a linear combination of the continuum wavefunctions $\tilde{\psi}_{k}^{+}\m(x)$ from {\sl Appendix II}.\\
      Without loss of generality one may set
      \be \label{Psi-x-def}
         \psi(x) \; = \; \int_{-\infty}^{+\infty} \m {\cal F}(k) \; \tilde{\psi}_k^+\n(x) \; {\rm d}k \mez ;
      \ee
      where ${\cal F}(k)$ is an uniquely determined smooth square integrable function of $k$.\\
      Squared norm of $\psi(0,x)=(\ref{Psi-x-def})$ is easily calculated using the orthonormality relations (\ref{tilde-psi-E-eta-onrel-take-3}).\\ One has
      \be \label{Psi-x-sqnrm}
         \int_{-\infty}^{+\infty} \Bigl| \psi(x) \Bigr|^2 {\rm d}x \; = \; 2\,\pi \, \int_{0}^{+\infty} \Bigl| {\cal F}(k) \Bigr|^2 {\rm d}k \mez .
      \ee
      Dynamical time evolution governed by the full Hamiltonian $\hat{\sf H}$ converts $\psi(0,x)$ into
      \be \label{Psi-t-x}
         \psi(t,x) \; = \; \int_{-\infty}^{+\infty} \m {\cal F}(k) \; e^{-\frac{i}{\hbar} \frac{\hbar^2 k^2}{2\,m}\,t} \; \tilde{\psi}_k^+\n(x) \; {\rm d}k \mez .
      \ee
      The norm of $\psi(t,x)$ is conserved in time, thus (\ref{Psi-x-sqnrm}) can be generalized by writing
      \be \label{Psi-t-x-sqnrm}
         \int_{-\infty}^{+\infty} \Bigl| \psi(t,x) \Bigr|^2 {\rm d}x \; = \; 2\,\pi \, \int_{0}^{+\infty} \Bigl| {\cal F}(k) \Bigr|^2 {\rm d}k \mez .
         \mez [\,{\rm for}\;{\rm all}\;t\,]
      \ee
      The task of the present {\sl Appendix III} is to find out\\
      what happens with $\psi(t,x)$ in the limit of $t \to \pm \infty$.
\item Basic observation:\\
      Suppose that $t$ acquires a large magnitude (approaching $\pm \infty$).\\
      Then the dynamical phase factor $e^{-\frac{i}{\hbar} \frac{\hbar^2 k^2}{2\,m}\,t}$ rapidly oscillates with $k$. \hfill $(\star)$
\item For $x \in (a,b)$, the term ${\cal F}(k) \, \tilde{\psi}_k^+\n(x)$ varies smoothly with $k$ and falls off to zero at large $|k|$.\\
      The corresponding $k$-integral of equation (\ref{Psi-t-x}) thus vanishes for $t \to \pm\infty$ due to $(\star)$.\\
      In other words, it turns out that the wavefunction $\psi(t,x) = (\ref{Psi-t-x})$\\
      vanishes in the interaction region for $t \to \pm\infty$.\\
      But this is precisely the statement of the {\sl Asymptotic Condition} (AC) which we wanted to prove.\\
      We also recall in this context that $\psi(t,x) = (\ref{Psi-t-x})$ remains unit normalized at all times.
\item What remains to be done\\
      is to analyze the wavefunction $\psi(t,x) = (\ref{Psi-t-x})$ in the asymptotic regions of $x \not\in (a,b)$.\\
      For the sake of simplicity, and without a significant loss of generality,\\
      we shall hereafter restrict ourselves to the case when ${\cal F}(k)$ populates only positive $k$'s $(\eta=+1)$.\\
      If so, one may replace $\tilde{\psi}_k^+\n(x)$ in equation (\ref{Psi-t-x})\\ by the corresponding asymptotic form 
      (\ref{tilde-psi-k>0-bc-b}) or (\ref{tilde-psi-k>0-bc-a}) while ignoring (\ref{tilde-psi-k<0-bc-b})-(\ref{tilde-psi-k<0-bc-a}).\\
      Further progress is then facilitated by means of the stationary phase approximation (SPA),\\
      as detailed below.

\newpage

{\color{magenta}
\dotfill\\
\begin{centering}
{\sl Appendix III}\\
\vspace*{-0.20cm}
\end{centering}
\dotfill}

\item Assume $x > b$, then
      \be \label{Psi-t-x-right-take-1}
         \psi_>^{\rm tr}\n(t,x) \; = \; \int_{0}^{+\infty} \m {\cal F}(k) \; e^{-\frac{i}{\hbar} \frac{\hbar^2 k^2}{2\,m}\,t} \; T(k) \; e^{+i k x} \; {\rm d}k \; = \;
         \int_{0}^{+\infty} \m {\cal F}(k) \; T(k) \; e^{-i \left( \frac{\hbar k^2}{2\,m}\,t \, - \, k \, x \right)} \; {\rm d}k \mez ;
      \ee
      where we wrote $\psi_>^{\rm tr}\n(t,x)$ instead of $\psi(t,x)$.\\
      Since $t$ (and possibly also $x$) are large in magnitude, the exponential
      \be
         e^{-i \left( \frac{\hbar k^2}{2\,m}\,t \, - \, k \, x \right)}
      \ee
      rapidly oscillates with $k$, except eventually of certain small region(s) of $k$ for which the phase
      \be
         {\cal S}\n(k) \; = \; \frac{\hbar\,k^2}{2\,m}\,t \, - \, k \, x
      \ee
      is stationary. Stationarity condition $(\partial {\cal S}\n(k)/\partial k)\bigl|_{k=k_s}\m=0$ reveals that, for given $(t,x)$,\\
      there is exactly one such region, namely, a region ${\mathbb K}_s$ surrounding a specific momentum
      \be
         k_s \; = \; \frac{m\,x}{\hbar\,t} \mez .
      \ee
      Let us conveniently set
      \be
         \kappa \; = \; k \; - \; k_s \mez ;
      \ee
      then one has
      \be
         {\cal S}\n(k) \; = \; \frac{\hbar (\kappa + k_s)^2}{2\,m}\,t \, - \, (\kappa+k_s)x \; = \;
         -\,\frac{m\,x^2}{2\,\hbar\,t} \; + \; \frac{\hbar\,t}{2\,m} \, \kappa^2 \mez .
      \ee
      Correspondingly, our wavefunction prescription $\psi_>^{\rm tr}\n(t,x)=(\ref{Psi-t-x-right-take-1})$ can be simplified\\
      within the SPA approach as follows:
      \begin{eqnarray} \label{Psi-t-x-right-take-2}
         \psi_>^{\rm tr}\n(t,x) & = & {\cal F}\m\m\m\left(\frac{m\,x}{\hbar\,t}\right) \; T\m\m\m\left(\frac{m\,x}{\hbar\,t}\right)
         \int_{{\mathbb K}_s} \m e^{-i \left( \frac{\hbar k^2}{2\,m}\,t \, - \, k \, x \right)} \; {\rm d}k \; = \nonumber\\
         & = & {\cal F}\m\m\m\left(\frac{m\,x}{\hbar\,t}\right) \; T\m\m\m\left(\frac{m\,x}{\hbar\,t}\right) \;
         e^{+i\,\frac{m\,x^2}{2\,\hbar\,t}} \int_{-\infty}^{+\infty} \m e^{-i\,\frac{\hbar\,t}{2\,m} \, \kappa^2}
         \, {\rm d}\kappa \mez .
      \end{eqnarray}
      Since $x>b>0$, the sign of $\frac{m\,x}{\hbar\,t}$ equals to the sign of $t$.\\
      Hence ${\cal F}\m\m\m\left(\frac{m\,x}{\hbar\,t}\right)=0$ for the case of $t<0$,\\
      due to our previous assumption that ${\cal F}(k)$ populates only negative momenta.\\
      Thus also $\psi_>^{\rm tr}\n(t,x)=0$ for $x>b$ and $t \to -\infty$.\\
      Assume now $t>0$, then ${\cal F}\m\m\m\left(\frac{m\,x}{\hbar\,t}\right)$ is allowed to be nonzero,\\
      and an integration over $\kappa$ in (\ref{Psi-t-x-right-take-2}) yields the final outcome
      \begin{eqnarray} \label{Psi-t-x-right-take-3}
         \psi_>^{\rm tr}\n(t,x) & = & e^{-i\pi/4} \, \sqrt{\frac{2\,\pi\,m}{\hbar\,t}} \; e^{+i\,\frac{m\,x^2}{2\,\hbar\,t}} \,
         {\cal F}\m\m\m\left(\frac{m\,x}{\hbar\,t}\right) \; T\m\m\m\left(\frac{m\,x}{\hbar\,t}\right) \mez . \mez [\,x>b,\,t\to+\infty\,]
      \end{eqnarray}
      
\newpage

{\color{magenta}
\dotfill\\
\begin{centering}
{\sl Appendix III}\\
\vspace*{-0.20cm}
\end{centering}
\dotfill}

      Since ${\cal F}(k)$ is a smooth square integrable function populating positive momenta,\\
      an entity $\psi_>^{\rm tr}\n(t,x)=(\ref{Psi-t-x-right-take-3})$ behaves as a wavepacket\\
      which is spatially localized in some region of large $x$,\\
      and which propagates in time towards the right\\
      (as the wavepacket transmitted through the region of interaction).\\
      Its squared norm can be evaluated easily:
      \be \label{Psi-rightarrow-t-x-sqnrm}
         \hspace*{-1.20cm} \int_{-\infty}^{+\infty} \n \Bigl| \psi_>^{\rm tr}\n(t,x) \Bigr|^2 {\rm d}x \; = \; \int_{0}^{+\infty} \m {\rm d}x \;\,
         \frac{2\,\pi\,m}{\hbar\,t} \; \left|\,{\cal F}\m\m\m\left(\frac{m\,x}{\hbar\,t}\right)\right|^2 \, \left|T\m\m\m\left(\frac{m\,x}{\hbar\,t}\right)\right|^2 \; = \;
         2\,\pi \, \int_{0}^{+\infty} \m {\rm d}k \;\, \Bigl| {\cal F}(k) \Bigr|^2 \, \Bigl| T(k) \Bigr|^2 \mz . \mz
      \ee
      It is worthy to compare (\ref{Psi-rightarrow-t-x-sqnrm}) with the squared norm (\ref{Psi-t-x-sqnrm}) of the whole wavepacket $\psi(t,x)$.\\
      One can see immediately that $\psi_>^{\rm tr}\n(t,x)$ represents just one part (branch) of $\psi(t,x)$.\\
      Note also that any structure of $T(k)$ (e.g.~resonance peaks) is reflected in $\psi_>^{\rm tr}\n(t,x)$ through (\ref{Psi-t-x-right-take-3}).\\
      This applies not only to the modulus of $T(k)$ but also to its phase.\\
      The right asymptotic region of $x>b$ is now resolved.
\item Treatment of the left asymptotic region $x<a$ is analogous. One has
      \be \label{Psi-t-x-left-take-1}
         \psi_<\n(t,x) \; = \; \psi_<^{\rm in}\n(t,x) \; + \; \psi_<^{\rm ref}\n(t,x) \mez ;
      \ee
      where by definition
      \be \label{Psi-t-x-left-take-1-in}
         \psi_<^{\rm in}\n(t,x) \; = \; \int_{0}^{+\infty} \m {\cal F}(k) \; e^{-\frac{i}{\hbar} \frac{\hbar^2 k^2}{2\,m}\,t} \; e^{+i k x} \; {\rm d}k \; = \;
         \int_{0}^{+\infty} \m {\cal F}(k) \; e^{-i \left( \frac{\hbar k^2}{2\,m}\,t \, - \, k \, x \right)} \; {\rm d}k \mez ;
      \ee
      and
      \be \label{Psi-t-x-left-take-1-ref}
         \psi_<^{\rm ref}\n(t,x) \; = \; \int_{0}^{+\infty} \m {\cal F}(k) \; e^{-\frac{i}{\hbar} \frac{\hbar^2 k^2}{2\,m}\,t} \; R(k) \; e^{-i k x} \; {\rm d}k \; = \;
         \int_{0}^{+\infty} \m {\cal F}(k) \; R(k) \; e^{-i \left( \frac{\hbar k^2}{2\,m}\,t \, + \, k \, x \right)} \; {\rm d}k \mez .
      \ee
      The used notations should be self explanatory.\\ Let us consider now separately $\psi_<^{\rm in}\n(t,x)$ and $\psi_<^{\rm ref}\n(t,x)$:
      \vspace*{-0.20cm}
      \begin{itemize}
      \item[$\circ$]
      Contribution $\psi_<^{\rm in}\n(t,x)$ is worked out in the same way as in the previous paragraph,\\ {\sl cf}.~equation (\ref{Psi-t-x-right-take-2}).
      We get
      \be \label{Psi-t-x-left-take-2-in}
         \psi_<^{\rm in}\n(t,x) \; = \; {\cal F}\m\m\m\left(\frac{m\,x}{\hbar\,t}\right) \;
         e^{+i\,\frac{m\,x^2}{2\,\hbar\,t}} \int_{-\infty}^{+\infty} \m e^{-i\,\frac{\hbar\,t}{2\,m} \, \kappa^2} \, {\rm d}\kappa \mez .
      \ee
      Since $x<a<0$, the sign of $\frac{m\,x}{\hbar\,t}$ equals to the sign of $(-t)$.\\
      Hence ${\cal F}\m\m\m\left(\frac{m\,x}{\hbar\,t}\right)=0$ for the case of $t>0$,\\
      due to our previous assumption that ${\cal F}(k)$ populates only negative momenta.\\
      Thus also $\psi_<^{\rm in}\n(t,x)=0$ for $x<a$ and $t \to +\infty$.\\
      Assume now $t<0$, then ${\cal F}\m\m\m\left(\frac{m\,x}{\hbar\,t}\right)$ is allowed to be nonzero,\\
      and an integration over $\kappa$ in (\ref{Psi-t-x-left-take-2-in}) yields the final outcome
      \begin{eqnarray} \label{Psi-t-x-left-take-3-in}
         \psi_<^{\rm in}\n(t,x) & = & e^{+i\pi/4} \, \sqrt{\frac{2\,\pi\,m}{\hbar\,|t|}} \; e^{+i\,\frac{m\,x^2}{2\,\hbar\,t}} \,
         {\cal F}\m\m\m\left(\frac{m\,x}{\hbar\,t}\right) \; \mez . \mez [\,x<a,\,t\to-\infty\,]
      \end{eqnarray}
      
\newpage

{\color{magenta}
\dotfill\\
\begin{centering}
{\sl Appendix III}\\
\vspace*{-0.20cm}
\end{centering}
\dotfill}

      Entity $\psi_<^{\rm in}\n(t,x)=(\ref{Psi-t-x-left-take-3-in})$ behaves as a wavepacket\\
      which is spatially localized in some region of largely negative $x$,\\
      and which propagates with increasing time towards the right\\
      (as the wavepacket approaching the region of interaction).\\
      Its squared norm is easily found to be
      \be \label{Psi-leftarrow-t-x-in-sqnrm}
         \hspace*{-1.00cm} \int_{-\infty}^{+\infty} \n \Bigl| \psi_<^{\rm in}\n(t,x) \Bigr|^2 {\rm d}x \; = \;
         2\,\pi \, \int_{0}^{+\infty} \m {\rm d}k \;\, \Bigl| {\cal F}(k) \Bigr|^2 \; = \; \int_{-\infty}^{+\infty} \n \Bigl| \psi(t,x) \Bigr|^2 {\rm d}x
         \; = \; (\ref{Psi-t-x-sqnrm}) \mez .
      \ee
      This is expected since $\psi_<^{\rm in}\n(t,x)=(\ref{Psi-t-x-left-take-3-in})$ represents the incoming wavepacket $\psi(t\to-\infty,x)$.
      \item[$\circ$]
      Contribution $\psi_<^{\rm ref}\n(t,x)$ is treated again in a similar spirit,\\
      {\sl cf}.~manipulations leading towards equation (\ref{Psi-t-x-right-take-2}). One gets
      \begin{eqnarray} \label{Psi-t-x-left-take-2-ref}
         \psi_<^{\rm ref}\n(t,x) & = & {\cal F}\m\m\m\left(-\,\frac{m\,x}{\hbar\,t}\right) \; R\m\m\m\left(-\,\frac{m\,x}{\hbar\,t}\right) \;
         e^{+i\,\frac{m\,x^2}{2\,\hbar\,t}} \int_{-\infty}^{+\infty} \m e^{-i\,\frac{\hbar\,t}{2\,m} \, \kappa^2} \, {\rm d}\kappa \mez .
      \end{eqnarray}
      Since $x<a<0$, the sign of $-\,\frac{m\,x}{\hbar\,t}$ equals to the sign of $(+t)$.\\
      Hence ${\cal F}\m\m\m\left(-\frac{m\,x}{\hbar\,t}\right)=0$ for the case of $t<0$,\\
      due to our previous assumption that ${\cal F}(k)$ populates only negative momenta.\\
      Thus also $\psi_<^{\rm ref}\n(t,x)=0$ for $x<a$ and $t \to -\infty$.\\
      Assume now $t>0$, then ${\cal F}\m\m\m\left(-\frac{m\,x}{\hbar\,t}\right)$ is allowed to be nonzero,\\
      and an integration over $\kappa$ in (\ref{Psi-t-x-left-take-2-ref}) yields the final outcome
      \begin{eqnarray} \label{Psi-t-x-left-take-3-ref}
         \hspace*{-1.00cm} \psi_<^{\rm ref}\n(t,x) & = & e^{-i\pi/4} \, \sqrt{\frac{2\,\pi\,m}{\hbar\,t}} \; e^{+i\,\frac{m\,x^2}{2\,\hbar\,t}} \,
         {\cal F}\m\m\m\left(-\,\frac{m\,x}{\hbar\,t}\right) \; R\m\m\m\left(-\,\frac{m\,x}{\hbar\,t}\right) \mez . \mez [\,x<a,\,t\to+\infty\,]
      \end{eqnarray}
      Entity $\psi_<^{\rm ref}\n(t,x)=(\ref{Psi-t-x-left-take-3-ref})$ behaves as a wavepacket\\
      which is spatially localized in some region of largely negative $x$,\\
      and which propagates with increasing time towards the left\\
      (as a wavepacket reflected back from the region of interaction by the potential).\\
      Its squared norm is easily found to be
      \be \label{Psi-leftarrow-t-x-ref-sqnrm}
         \hspace*{-1.00cm} \int_{-\infty}^{+\infty} \n \Bigl| \psi_<^{\rm ref}\n(t,x) \Bigr|^2 {\rm d}x \; = \;
         2\,\pi \, \int_{0}^{+\infty} \m {\rm d}k \;\, \Bigl| {\cal F}(k) \Bigr|^2 \, \Bigl| R(k) \Bigr|^2 \mez .
      \ee
      It is worthy to compare the squared norms (\ref{Psi-leftarrow-t-x-ref-sqnrm}) and  (\ref{Psi-rightarrow-t-x-sqnrm})\\
      with the squared norm (\ref{Psi-t-x-sqnrm}) of the whole wavepacket $\psi(t,x)$\\
      $[\,$or, equivalently, with the squared norm (\ref{Psi-leftarrow-t-x-in-sqnrm}) of the incoming wavepacket $\psi_<^{\rm in}\n(t,x)$$\,]$.\\
      Since $|T(k)|^2+|R(k)|^2=1$ as stated above in (\ref{T-R-1-k}),\\
      one recovers the correct and expected result
      \be
         \hspace*{-0.50cm}
         \int_{-\infty}^{+\infty} \n \Bigl| \psi_>^{\rm tr}\n(t,x) \Bigr|^2 {\rm d}x \; + \; \int_{-\infty}^{+\infty} \n \Bigl| \psi_<^{\rm ref}\n(t,x) \Bigr|^2 {\rm d}x
         \; = \; \int_{-\infty}^{+\infty} \n \Bigl| \psi_<^{\rm in}\n(t,x) \Bigr|^2 {\rm d}x \; = \; \int_{-\infty}^{+\infty} \n \Bigl| \psi(t,x) \Bigr|^2 {\rm d}x \mz . \mz
      \ee
      In words, the transmitted population + the reflected population = the incoming population.
      \end{itemize}
      \vspace*{-0.20cm}
      The left asymptotic region of $x<a$ is now resolved.
\newpage

{\color{magenta}
\dotfill\\
\begin{centering}
{\sl Appendix III}\\
\vspace*{-0.20cm}
\end{centering}
\dotfill}

\vspace*{+2.00cm}

\item Summarizing the contents of {\sl Appendix III},\\
      we have given an elementary proof of the {\sl Asymptotic Condition} for the case of 1D scattering,\\
      which was based upon the stationary phase approximation (SPA).\\
      Moreover, we have derived simple analytic closed form prescriptions\\
      for the incoming, transmitted, and reflected wavepacket.
\end{itemize}

\newpage

{\color{magenta}
\dotfill\\
\begin{centering}
{\sl Appendix IV}\\
\vspace*{-0.20cm}
\end{centering}
\dotfill}

\vspace*{+0.50cm}

{\bf Appendix IV: Green functions}\\

{\sl Based upon Appendix 4 of Ref.~\cite{Roman}.}

\begin{itemize}
\item Let us make first some preparatory remarks.\\
      The closure property (\ref{H-closure}) implies that our Hamiltonian $\hat{\sf H}$\\ possesses the following spectral resolution,
      \be \label{H-spectral}
         \hat{\sf H} \; = \; \sum_{n=1}^{n_{\rm bound}} \, | \psi_n \ra \, E_n \, \la \psi_n | \; + \;
         \int_{0}^{\infty} \m {\rm d}E \, \sum_\eta \, | \psi_{E\eta}^{+} \ra \; E \; \la \psi_{E\eta}^{+} | \mez ;
      \ee
      here of course $| \psi_n \ra = \int_{-\infty}^{+\infty} \psi_n\n(x) \, | x \ra \, {\rm d}x$,
      $| \psi_{E\eta}^{+} \ra = \int_{-\infty}^{+\infty} \psi_{E\eta}^{+}\n(x) \, | x \ra \, {\rm d}x$,\\
      and $\{ E_n \}_{n=1}^{n_{\rm bound}}$ are the bound state energies. Let 
      \be \label{psi-wavepacket}
         | \psi \ra \; = \; \sum_{n=1}^{n_{\rm bound}} \, | \psi_n \ra \, \la \psi_n | \psi \ra \; + \;
         \int_{0}^{\infty} \m {\rm d}E \, \sum_\eta \, | \psi_{E\eta}^{+} \ra \, \la \psi_{E\eta}^{+} | \psi \ra
      \ee
      be any square integrable wavepacket, such that $\sum_{n=1}^{n_{\rm bound}} \Bigl| \la \psi_n | \psi \ra \Bigr|^2 \, + \, \int_{0}^{\infty} \m {\rm d}E \, \sum_\eta \,
      \Bigl| \la \psi_{E\eta}^{+} | \psi \ra \Bigr|^2 \m = 1$.\\ An action of $\hat{\sf H}=(\ref{H-spectral})$ on vector $| \psi \ra = (\ref{psi-wavepacket})$ is then
      determined by formula
      \be \label{hat-sf-H-on-psi-wavepacket}
         \hat{\sf H} \, | \psi \ra \; = \; \sum_{n=1}^{n_{\rm bound}} \, | \psi_n \ra \, E_n \, \la \psi_n | \psi \ra \; + \;
         \int_{0}^{\infty} \m {\rm d}E \, \sum_\eta \, | \psi_{E\eta}^{+} \ra \, E \, \la \psi_{E\eta}^{+} | \psi \ra \mez .
      \ee
\item We shall continue our discussion by introducing the Green operators.\\
      The Green operator $(E-\hat{\sf H} \pm i\,0_+)^{-1}$
      $[\,$retarded for the $+i\,0_+$ choice, advanced for the $-i\,0_+$ choice$\,]$\\
      can be defined by its spectral resolution analogous to (\ref{H-spectral}).
      This amounts to set
      \be \label{Green-spectral-def}
         \frac{1}{E-\hat{\sf H} \pm i\,0_+} \; = \; \sum_{n=1}^{n_{\rm bound}} \, | \psi_n \ra \, \frac{1}{E - E_n} \, \la \psi_n | \; + \;
         \int_{0}^{\infty} \m {\rm d}E'\n \, \sum_{\eta'\n} \, | \psi_{E'\n\eta'\n}^{+} \ra \; \frac{1}{E-E'\n \pm i\,0_+} \; \la \psi_{E'\n\eta'\n}^{+} | \mez .
      \ee
      For the sake of clarity we recall that $E>0$ is a given parameter.\\
      $[\,$Since all the bound state energies $E_n$ are negative, no singularity can occur in the $(E - E_n)^{-1}$ term.$\,]$\\
      Well definedness of an operator $(E-\hat{\sf H} \pm i\,0_+)^{-1}=(\ref{Green-spectral-def})$ can be demonstrated\\
      by letting it act on the wavepacket $| \psi \ra=(\ref{psi-wavepacket})$. One gets
      \be \label{Green-on-psi}
         \hspace*{-1.00cm} \frac{1}{E-\hat{\sf H} \pm i\,0_+} \, | \psi \ra \; = \; \sum_{n=1}^{n_{\rm bound}} \, | \psi_n \ra \, \frac{1}{E - E_n} \, \la \psi_n | \psi \ra \; + \;
         \int_{0}^{\infty} \m {\rm d}E'\n \, \sum_{\eta'\n} \, | \psi_{E'\n\eta'\n}^{+} \ra \; \frac{1}{E-E'\n \pm i\,0_+} \; \la \psi_{E'\n\eta'\n}^{+} | \psi \ra \mz\;\; ; \mz
      \ee
      or, more conveniently,
      \be \label{Green-on-psi-x}
         \la x | \, \frac{1}{E-\hat{\sf H} \pm i\,0_+} \, | \psi \ra \; = \; \sum_{n=1}^{n_{\rm bound}} \, \frac{\la \psi_n | \psi \ra}{E - E_n} \, \psi_n\n(x) \; + \;
         \int_{0}^{\infty} \m {\rm d}E'\n \, \sum_{\eta'\n} \, \frac{\la \psi_{E'\n\eta'\n}^{+} | \psi \ra}{E-E'\n \pm i\,0_+} \; \psi_{E'\n\eta'\n}^{+}\n(x) \mez .
      \ee
      
\newpage

{\color{magenta}
\dotfill\\
\begin{centering}
{\sl Appendix IV}\\
\vspace*{-0.20cm}
\end{centering}
\dotfill}

      Furthermore, the familiar textbook formula
      \be \label{PP-textbook}
         \frac{1}{\zeta \pm i\,0_+} \; = \; {\rm P}\,\frac{1}{\zeta} \; \mp \; i\,\pi\,\delta(\zeta)
      \ee
      $[\,$where ${\rm P}$ stands for the Cauchy principal part$\,]$ enables us to write
      \be \label{PP-term}
         \int_{0}^{\infty} \m \frac{\la \psi_{E'\n\eta'\n}^{+} | \psi \ra\,\psi_{E'\n\eta'\n}^{+}\n(x)}{E-E'\n \pm i\,0_+} \; {\rm d}E'\n \; = \;
         {\rm P} \m \int_{0}^{\infty} \m \frac{\la \psi_{E'\n\eta'\n}^{+} | \psi \ra\,\psi_{E'\n\eta'\n}^{+}\n(x)}{E-E'\n} \; {\rm d}E'\n \; \mp \;
         i \, \pi \, \la \psi_{E\eta'\n}^{+} | \psi \ra \, \psi_{E\eta'\n}^{+}\n(x) \mez .
      \ee
      The associated numerator $\la \psi_{E'\n\eta'\n}^{+} | \psi \ra\,\psi_{E'\n\eta'\n}^{+}\n(x)$\\
      is apparently a bounded continuous function of $E'\n$ falling off to zero for $E'\n\to+\infty$.\\
      $[\,$Recall in this context that our wavepacket $| \psi \ra$ is square integrable.$\,]$\\
      Hence the corresponding Cauchy principal integral converges,\\ and the r.h.s.~of equation (\ref{Green-on-psi-x}) is thus well defined.\\
      If so, even the r.h.s.~of the spectral resolution formula (\ref{Green-spectral-def}) acquires a well defined meaning.\\
      An action of $(E-\hat{\sf H} \pm i\,0_+)$ on both sides of equation (\ref{Green-on-psi}) can be worked out using (\ref{hat-sf-H-on-psi-wavepacket}),\\
      one obtains back the original wavepacket $| \psi \ra=(\ref{psi-wavepacket})$.\\ 
      Similarly one verifies an identity $(E-\hat{\sf H} \pm i\,0_+)^{-1} \, (E-\hat{\sf H} \pm i\,0_+) \, | \psi \ra = | \psi \ra$.\\
      Showing that the Green operator $(E-\hat{\sf H} \pm i\,0_+)^{-1}=(\ref{Green-spectral-def})$ is indeed\\
      a legitimate inverse of $(E-\hat{\sf H} \pm i\,0_+)$. Formulas (\ref{Green-on-psi-x}), (\ref{PP-term}) reveal that\\
      $(E-\hat{\sf H} + i\,0_+)^{-1}$ is acting on $| \psi \ra$ differently than $(E-\hat{\sf H} - i\,0_+)^{-1}$ does,\\
      due to the terms $\mp \, i \, \pi \, \la \psi_{E\eta'\n}^{+} | \psi \ra \, \psi_{E\eta'\n}^{+}\n(x)$.\\
      Thus $(E-\hat{\sf H} + i\,0_+)^{-1} \neq (E-\hat{\sf H} - i\,0_+)^{-1}$.
\item The main purpose of the present {\sl Appendix IV} is to analyze the Green functions.\\
      By definition, the Green function $G_{\m E}^\pm\n(x,y)$
      $[\,$retarded for the $^+$ choice, advanced for the $^-$ choice$\,]$\\
      is identified with position representation of the Green operator $(E-\hat{\sf H} \pm i\,0_+)^{-1}$\m\n. Thus
      \begin{eqnarray}
         \label{G-E-pm-def} G_{\m E}^\pm\n(x,y) & = & \la x | \, \frac{1}{E-\hat{\sf H} \pm i\,0_+} \, | y \ra \; = \\
         \label{G-E-pm-spectral} & = & \sum_{n=1}^{n_{\rm bound}} \, \frac{\psi_n\n(x)\,\psi_n^*\n(y)}{E - E_n} \; + \;
         \int_{0}^{\infty} \m {\rm d}E'\n \, \sum_{\eta'\n} \, \frac{\psi_{E'\n\eta'\n}^{+}\n(x)\,\psi_{E'\n\eta'\n}^{+*}\n(y)}{E-E'\n \pm i\,0_+} \mez .
      \end{eqnarray}
      The numerator term $\psi_{E'\n\eta'\n}^{+}\n(x)\,\psi_{E'\n\eta'\n}^{+*}\n(y)$ appearing in (\ref{G-E-pm-spectral})\\
      behaves for large $E'$ as $\sqrt{\frac{m}{2\,\pi\,\hbar^2 K'\n}}\;e^{+i\eta'\n K'\n (x-y)}$ where of course $\hbar K'=\sqrt{2\,m\,E'}$\n.\\
      $[\,$Clarification is provided by our remark following immediately equation (\ref{psi-E-eta-onrel-take-1}) of {\sl Appendix II}.$\,]$\\
      This observation, along with property (\ref{PP-textbook}), makes $G_{\m E}^\pm\n(x,y) = (\ref{G-E-pm-def}) = (\ref{G-E-pm-spectral})$ well defined.\\
      For the same reason, $G_{\m E}^\pm\n(x,y)$ turns out to be a continuos function of variables $x$ and $y$.
      
\newpage

{\color{magenta}
\dotfill\\
\begin{centering}
{\sl Appendix IV}\\
\vspace*{-0.20cm}
\end{centering}
\dotfill}

\item Position representation of an operator identity $(E-\hat{\sf H} \pm i\,0_+) \, (E-\hat{\sf H} \pm i\,0_+)^{-1} \, = \, \hat{\sf 1}$ reveals that\\
      the Green function $G_{\m E}^\pm\n(x,y)$ satisfies an an inhomogeneous Schr\"{o}dinger-like equation
      \be \label{G-E-pm-SCHE-ihg}
         \left\{ \, -\,\frac{\hbar^2}{2\,m}\,\partial_{xx} \, + \, V\m(x) \, - \, E \, \right\} \, G_{\m E}^\pm\n(x,y) \; = \; -\,\delta(x-y) \mez .
      \ee
      Note that $G_{\m E}^+\n(x,y)$ and $G_{\m E}^-\n(x,y)$\\ are two distinct particular solutions of the second order ODE (\ref{G-E-pm-SCHE-ihg}).\\
      Indeed, one has $G_{\m E}^+\n(x,y) \neq G_{\m E}^-\n(x,y)$,\\
      since $(E-\hat{\sf H} + i\,0_+)^{-1} \neq (E-\hat{\sf H} - i\,0_+)^{-1}$ as already pointed out above.\\
      The difference $G_{\m E}^{\rm diff}\m(x,y) \equiv G_{\m E}^+\n(x,y) - G_{\m E}^-\n(x,y)$ solves obviously\\
      the homogeneous counterpart of equation (\ref{G-E-pm-SCHE-ihg}),
      \be \label{G-E-pm-SCHE-hmg}
         \left\{ \, -\,\frac{\hbar^2}{2\,m}\,\partial_{xx} \, + \, V\m(x) \, - \, E \, \right\} \, G_{\m E}^{\rm diff}\m(x,y) \; = \; 0 \mez .
      \ee
      Showing that $G_{\m E}^{\rm diff}\m(x,y)$ must inevitably be a linear combination\\ of the two pertinent eigenfunctions $\psi_{E(\pm 1)}^{+}\n(x)$,\\
      i.e., $G_{\m E}^{\rm diff}\m(x,y) \, = \, Q_{E(+1)}\n(y) \; \psi_{E(+1)}^{+}\n(x) \, + \, Q_{E(-1)}\n(y) \; \psi_{E(-1)}^{+}\n(x)$,\\
      where the $Q$'s are certain as yet unspecified factors.\\ Clearly, $G_{\m E}^{\rm diff}\m(x,y)$ accounts in this way for different boundary conditions\\
      imposed on the particular solutions $G_{\m E}^\pm\n(x,y)$ of the inhomogeneous problem (\ref{G-E-pm-SCHE-ihg}).\\
      $[\,$This issue of boundary conditions is clarified much more explicitly in the paragraphs below.$\,]$
\item In an absence of the potential, the Hamiltonian $\hat{\sf H}$ is replaced by $\hat{\sf H}_0$,\\
      and the entity $G_{\m E}^\pm\n(x,y) = (\ref{G-E-pm-def}) = (\ref{G-E-pm-spectral})$ becomes the free Green function
      \begin{eqnarray}
         \label{G-0-E-pm-def} G_{0,E}^\pm\n(x,y) & = & \la x | \, \frac{1}{E-\hat{\sf H}_0 \pm i\,0_+} \, | y \ra \; = \;
         \int_{0}^{\infty} \m {\rm d}E'\n \, \sum_{\eta'\n} \, \frac{\phi_{E'\n\eta'\n}\n(x)\,\phi_{E'\n\eta'\n}^{*}\n(y)}{E-E'\n \pm i\,0_+} \mez .
      \end{eqnarray}
      Here $\phi_{E'\n\eta'\n}\n(x)$ is given by formula (\ref{phi-E-eta-def}),\\
      see again our remark following immediately equation (\ref{psi-E-eta-onrel-take-1}) of {\sl Appendix II}.\\
      The inhomogeneous Schr\"{o}dinger-like equation (\ref{G-E-pm-SCHE-ihg}) boils down into
      \be \label{G-0-E-pm-SCHE-ihg}
         \left\{ \, -\,\frac{\hbar^2}{2\,m}\,\partial_{xx} \, - \, E \, \right\} \, G_{0,E}^\pm\n(x,y) \; = \; -\,\delta(x-y) \mez .
      \ee
      One has again $G_{0,E}^+\n(x,y) \neq G_{0,E}^-\n(x,y)$ since $(E-\hat{\sf H}_0 + i\,0_+)^{-1} \neq (E-\hat{\sf H}_0 - i\,0_+)^{-1}$\m\n.\\
      Importantly, the simply looking free Green function $G_{0,E}^\pm\n(x,y)=(\ref{G-0-E-pm-def})$\\ lends itself to an explicit analytic evaluation,\\
      such a task is worked out in detail in the next paragraph.
      
\newpage

{\color{magenta}
\dotfill\\
\begin{centering}
{\sl Appendix IV}\\
\vspace*{-0.20cm}
\end{centering}
\dotfill}

\item Let us plug (\ref{phi-E-eta-def}) into (\ref{G-0-E-pm-def}), and set conveniently $K = +\,\sqrt{2\,m\,E}$, $k = \eta' \sqrt{2\,m\,E'}$. One gets
      \begin{eqnarray}
         \label{G-0-E-pm-evaluation-take-1}
         \hspace*{-1.00cm} G_{0,E}^\pm\n(x,y) & = & \int_{0}^{\infty} \m {\rm d}E'\n \, \sum_{\eta'\n} \; \frac{m}{2\,\pi\,\hbar^2|k|} \, \frac{e^{+ik(x-y)}}
         {\left(\frac{\hbar^2 K^2}{2\,m}-\frac{\hbar^2 k^2}{2\,m} \pm i\,0_+\right)} \; = \nonumber\\
         \hspace*{-1.00cm} & = & \frac{1}{2\,\pi} \, \int_{-\infty}^{+\infty} \m \frac{e^{+ik(x-y)}}{\left(\frac{\hbar^2 K^2}{2\,m}-\frac{\hbar^2 k^2}{2\,m} \pm i\,0_+\right)} \; {\rm d}k
         \; = \; \frac{m}{\pi\,\hbar^2} \, \int_{-\infty}^{+\infty} \m \frac{e^{+ik(x-y)}}{\left(K^2 - k^2 \pm i\,\varepsilon \right)} \; {\rm d}k \mez ;
      \end{eqnarray}
      where of course $\varepsilon \to +0$. An obvious symmetry property
      \be \label{G-0-symmetry}
         G_{0,E}^\pm\n(y,x) \; = \; G_{0,E}^{\mp*}\n(x,y)
      \ee
      enables us to restrict our forthcoming analysis of an integral (\ref{G-0-E-pm-evaluation-take-1}) just to cases of $x > y$.\\
      Indeed, knowledge of $G_{0,E}^\pm\n(x>y,y)$ allows us to access also $G_{0,E}^\pm\n(x<y,y)$ via (\ref{G-0-symmetry}),\\
      whereas $G_{0,E}^\pm\n(y,y) = \lim_{x \to 0}\,G_{0,E}^\pm\n(x,y)$.\\
      We shall thus conveniently assume $x > y$ until stated otherwise,\\
      and evaluate the $k$-integral of equation (\ref{G-0-E-pm-evaluation-take-1}) via contour integration ($k \in {\mathbb C}$).\\
      Our closed contour consists of the real $k$-axis\\ closed over the upper half plane of complex $k$ (i.e., $\Im k \ge 0$).\\
      Such a choice of the contour is dictated by nature of the integrand of equation (\ref{G-0-E-pm-evaluation-take-1}).\\
      Namely, for $x>y$, the exponential $e^{+ik(x-y)}$\\ decays to zero in the remote upper half $k$-plane of $\Im k > 0$.\\
      Since we wish to take advantage of the residue theorem,\\
      we need to identify all poles of the term $(K^2 - k^2 \pm i \varepsilon)^{-1}$\m.\\
      Let us set as usual $k = k_R + i\,k_I$ and solve an equation $K^2 - k^2 \pm i \varepsilon = 0$ for unknowns $k_R$, $k_I$.\\
      We find that
      \be \label{k-R-k-I-equations}
         K^2 \, - \, k_R^2 \, + \, k_I^2 \; = \; 0 \mez , \mez 2\,k_R\,k_I \; = \; \pm\,\varepsilon \mez .
      \ee
      One can see now immediately\\ that function $(K^2 - k^2 \pm i \varepsilon)^{-1}$ has just two poles in the complex $k$-plane,\\ their approximate location is
      \be \label{poles-approx}
         k_a \; \doteq \; \left(+K,\frac{\pm\varepsilon}{2\,K}\right) \mez , \mez k_b \; \doteq \; \left(-K,\frac{\mp\varepsilon}{2\,K}\right) \mez .
      \ee
      For the sake of completeness and clarity,\\ we prefer to spell out here explicitly the fully exact expressions\\
      for the mentioned two poles, as obtained by solving equations (\ref{k-R-k-I-equations}). One has
      \be \label{poles-exact-plus}
         k_a \; = \; ( k_{R,a}, k_{I,a} ) \; = \; \left( +\sqrt{\frac{K^2+\sqrt{K^4+\varepsilon^2}}{2}} , \frac{\pm\varepsilon}{2\,k_{R,a}} \right) \mez ;
      \ee
      \be \label{poles-exact-minus}
         k_b \; = \; ( k_{R,b}, k_{I,b} ) \; = \; \left( -\sqrt{\frac{K^2+\sqrt{K^4+\varepsilon^2}}{2}} , \frac{\mp\varepsilon}{2\,k_{R,b}} \right) \mez .
      \ee

\newpage

{\color{magenta}
\dotfill\\
\begin{centering}
{\sl Appendix IV}\\
\vspace*{-0.20cm}
\end{centering}
\dotfill}

      Assume now pickup of the upper sign in the above used $\pm$ and $\mp$ symbols.\\
      In that case only the pole $k_a$ is encircled by our closed contour.\\
      For $k \to k_a$, the integrand of (\ref{G-0-E-pm-evaluation-take-1}) can be recast into
      \be \label{U-+-res}
         \frac{e^{+ik(x-y)}}{(K+k)(K-k) + i \varepsilon } \; = \; \frac{-\,e^{+ik_a(x-y)}}{(K+k_a)(k-k_a+k_a-K) + i \varepsilon } \; \approx \;
         \frac{-\,e^{+iK(x-y)}}{+2\,K\,(k-k_a)} \mez ;
      \ee
      and the residue theorem yields in turn
      \be \label{U-+-finres}
         G_{0,E}^+\n(x,y) \; = \; (-i)\,\frac{m}{\hbar^2 K} \; e^{+iK(x-y)} \mez . \mez [\,x>y\,]
      \ee
      Assume on the other hand pickup of the lower sign in the above used $\pm$ and $\mp$ symbols.\\
      In that case only the pole $k_b$ is encircled by our closed contour.\\
      For $k \to k_b$, the integrand of (\ref{G-0-E-pm-evaluation-take-1}) can be recast into
      \be \label{U---res}
         \frac{e^{+ik(x-y)}}{(K+k)(K-k) + i \varepsilon } \; = \; \frac{+\,e^{+ik_b(x-y)}}{(K+k_b-k_b+k)(K-k_b) + i \varepsilon } \; \approx \;
         \frac{+\,e^{-iK(x-y)}}{+2\,K\,(k-k_b)} \mez ;
      \ee
      and the residue theorem yields in turn
      \be \label{U---finres}
         G_{0,E}^-\n(x,y) \; = \; (+i)\,\frac{m}{\hbar^2 K} \; e^{-iK(x-y)} \mez . \mez [\,x>y\,]
      \ee
      Let us lift from now on the restriction of $x > y$.\\
      Combination of (\ref{G-0-symmetry}), (\ref{U-+-finres}), (\ref{U---finres}) yields a compactly looking final formula
      \be \label{U-final}
         G_{0,E}^\pm\n(x,y) \; = \; (\mp i)\,\frac{m}{\hbar^2 K} \; e^{\pm iK|x-y|} \; = \; G_{0,E}^{\pm}\n(y,x) \; = \; G_{0,E}^{\mp*}\n(x,y) \mez .
      \ee
      Redisplayed once again, we have just found that
      \be \label{Green-take-4}
         G_{0,E}^\pm\n(x,y) \; = \; (\mp i)\,\frac{m}{\hbar^2 K} \; e^{\pm iK|x-y|} \mez , \mez
         K = \sqrt{2\,m\,E}/\hbar \mez .
      \ee
      As an independent check of the whole calculation one can easily verify the property
      \be
         \Bigl( \, \partial_{xx} \, + \, K^2 \, \Bigr) \, G_{0,E}^\pm\n(x,y) \; = \; \frac{2\,m}{\hbar^2} \, \delta(x-y) \mez ;
      \ee
      which is equivalent to (\ref{G-0-E-pm-SCHE-ihg}). Before proceeding further,\\
      let us look again at (\ref{Green-take-4}) and comment on the difference between $G_{0,E}^+\n(x,y)$ and $G_{0,E}^-\n(x,y)$.\\
      One observes that $G_{0,E}^+\n(x\,\m_<^>\,y,y) \simeq e^{\pm i K x}$ while $G_{0,E}^-\n(x\,\m_<^>\,y,y) \simeq e^{\mp i K x}$\m. Thus\\
      $G_{0,E}^+\n(x,y)$ satisfies the outgoing boundary conditions for $x \to \pm\infty$ $[\,y\;{\rm constant}\,]$, whereas\\
      $G_{0,E}^-\n(x,y)$ satisfies the incoming boundary conditions for $x \to \pm\infty$ $[\,y\;{\rm constant}\,]$.
\item Having in hand $G_{0,E}^\pm\n(x,y)=(\ref{Green-take-4})$,\\
      we may turn our atttention back to the full Green function $G_{\m E}^\pm\n(x,y)=$ (\ref{G-E-pm-def})-(\ref{G-E-pm-spectral}).\\
      Instead of analyzing $G_{\m E}^\pm\n(x,y)$ via contour integration we prefer to take a different route.\\
      Namely, let us take a general operator identity
      \be
         \hat{A}^{-1} \; = \; \hat{B}^{-1} \; + \; \hat{B}^{-1} \, (\hat{B}-\hat{A}) \, \hat{A}^{-1} \mez ;
      \ee
      and substitute $\hat{A}=(E-\hat{\sf H} \pm i\,0_+)$, $\hat{B}=(E-\hat{\sf H}_0 \pm i\,0_+)$. This yields an equality
      \be \label{hat-G-E-LSE-implicit}
         \frac{1}{E-\hat{\sf H} \pm i\,0_+} \; = \; \frac{1}{E-\hat{\sf H}_0 \pm i\,0_+} \; + \;
         \frac{1}{E-\hat{\sf H}_0 \pm i\,0_+} \, \hat{\sf V} \, \frac{1}{E-\hat{\sf H} \pm i\,0_+} \mez .
      \ee
      
\newpage

{\color{magenta}
\dotfill\\
\begin{centering}
{\sl Appendix IV}\\
\vspace*{-0.20cm}
\end{centering}
\dotfill}

      Relation (\ref{hat-G-E-LSE-implicit}) can be interpreted as\\
      an {\sl implicit Lippmann-Schwinger equation} (LSE) for the Green operator $(E-\hat{\sf H} \pm i\,0_+)^{-1}$\m.\\
      After converting (\ref{hat-G-E-LSE-implicit}) into position representation\\
      one obtains the corresponding implicit LSE for $G_{\m E}^\pm\n(x,y)=(\ref{G-E-pm-def})$. One has
      \be \label{LSE-implicit-G}
         G_{\m E}^\pm\n(x,y) \; = \; G_{0,E}^\pm\n(x,y) \; + \;
         \int_{a}^{b} \m {\rm d}z \;\, G_{0,E}^\pm\n(x,z) \; V\m(z) \; G_{\m E}^\pm\n(z,y) \mez ;
      \ee
      where $(a,b)$ is the spatial region of nonvanishing potential.\\
      Now, recall the boundary conditions for $G_{0,E}^\pm\n(x,y)=(\ref{Green-take-4})$\\
      $[\,$which are spelled out explicitly at the end of the previous paragraph$\,]$.\\
      The LSE (\ref{LSE-implicit-G}) implies that\\
      $G_{\m E}^\pm\n(x,y)$ must satisfy exactly the same kind of boundary conditions as $G_{0,E}^\pm\n(x,y)$ does.\\
      Indeed, by inspecting (\ref{LSE-implicit-G}) one observes that
      \be \label{G-E-+-asympt}
         G_{\m E}^+\n(x\,\m_{<\,\min(a,y)}^{>\,\max(b,y)}\m,y) \; \simeq \; e^{\pm i K x} \mez ;
      \ee
      while
      \be \label{G-E---asympt}
         G_{\m E}^-\n(x\,\m_{<\,\min(a,y)}^{>\,\max(b,y)}\m,y) \; \simeq \; e^{\mp i K x} \mez .
      \ee
      Thus\\
      $G_{\m E}^+\n(x,y)$ satisfies the outgoing boundary conditions for $x \to \pm\infty$ $[\,y\;{\rm constant}\,]$, whereas\\
      $G_{\m E}^-\n(x,y)$ satisfies the incoming boundary conditions for $x \to \pm\infty$ $[\,y\;{\rm constant}\,]$.\\
      In passing we note that the LSE (\ref{hat-G-E-LSE-implicit}) or (\ref{LSE-implicit-G}) is solvable by iterations.\\
      $[\,$Valid at least for a substantial class of problems where the potential $V\m(x)$ is weak enough\\
      \phantom{$[\,$}as to justify an iterative treatment.$\,]$ One arrives in this way towards\\
      a perturbative expansion for the Green operator/function in increasing powers of the potential.\\
      Written down explicitly, one gets an expansion
      \begin{eqnarray} \label{hat-G-E-LSE-Born}
         \frac{1}{E-\hat{\sf H} \pm i\,0_+} & = & \frac{1}{E-\hat{\sf H}_0 \pm i\,0_+} \; + \;
         \frac{1}{E-\hat{\sf H}_0 \pm i\,0_+} \, \hat{\sf V} \, \frac{1}{E-\hat{\sf H}_0 \pm i\,0_+} \\
         & + & \frac{1}{E-\hat{\sf H}_0 \pm i\,0_+} \, \hat{\sf V} \, \frac{1}{E-\hat{\sf H}_0 \pm i\,0_+} \, \hat{\sf V} \, \frac{1}{E-\hat{\sf H}_0 \pm i\,0_+}
         \;\, + \;\, \bm\cdots \mez ; \nonumber
      \end{eqnarray}
      or, equivalently,
      \begin{eqnarray} \label{LSE-Born-G}
         G_{\m E}^\pm\n(x,y) & = & G_{0,E}^\pm\n(x,y) \; + \;
         \int_{a}^{b} \m {\rm d}z \;\, G_{0,E}^\pm\n(x,z) \; V\m(z) \; G_{0,E}^\pm\n(z,y) \\
         & + & \int_{a}^{b} \m {\rm d}z \int_{a}^{b} \m {\rm d}z' \;\, G_{0,E}^\pm\n(x,z) \; V\m(z) \; G_{0,E}^\pm\n(z,z'\n) \; V\m(z'\n) \; G_{0,E}^\pm\n(z'\n,y)
         \;\, + \;\, \bm\cdots \mez . \nonumber
      \end{eqnarray}
      Equations (\ref{hat-G-E-LSE-Born})-(\ref{LSE-Born-G}) represent the so called Born series for $(E-\hat{\sf H} \pm i\,0_+)^{-1}$ or $G_{\m E}^\pm\n(x,y)$.
      
\newpage

{\color{magenta}
\dotfill\\
\begin{centering}
{\sl Appendix IV}\\
\vspace*{-0.20cm}
\end{centering}
\dotfill}

\item Our analysis of $G_{\m E}^\pm\n(x,y)$ approaches an epic final.\\
      In this paragraph we shall express $G_{\m E}^+\n(x,y)$ explicitly in terms of the eigenfunctions $\psi_{E(\pm 1)}^{+}\n(x)$.\\
      $[\,$We have decided to focus just on the retarded Green function $G_{\m E}^+\n(x,y)$,\\
      \phantom{$[\,$}not only for the sake of definiteness, but also due to\\
      \phantom{$[\,$}the prominent role of $(E-\hat{\sf H} + i\,0_+)^{-1}$ and $G_{\m E}^+\n(x,y)$ in quantum scattering theory,\\
      \phantom{$[\,$}which should be apparent from the main text.\\
      \phantom{$[\,$}In any case, the workout of the advanced Green function $G_{\m E}^-\n(x,y)$ is analogous,\\
      \phantom{$[\,$}one uses $\psi_{E(\pm 1)}^{-}\n(x)=\psi_{E(\mp 1)}^{+*}\n(x)$ instead.$\,]$\\
      The just advertised final treatment of $G_{\m E}^+\n(x,y)$ is based upon key properties (\ref{G-E-pm-SCHE-ihg}), (\ref{G-E-+-asympt})\\
      combined with the boundary conditions (\ref{tilde-psi-E-+-bc-b})-(\ref{tilde-psi-E-+-bc-a}), (\ref{tilde-psi-E---bc-b})-(\ref{tilde-psi-E---bc-a})
      possessed by $\tilde{\psi}_{E(\pm 1)}^{+}\n(x)$.\\ After taking into account all the just mentioned relations, a moment of reflection reveals\\
      that, inevitably,
      \begin{eqnarray}
         \label{get-G-E-+-right} G_{\m E}^+\n(x>y,y) & = & \lambda_+\n(E,y) \; \psi_{E(+1)}^{+}\n(x) \mez ;\\
         \label{get-G-E-+-left} G_{\m E}^+\n(x<y,y) & = & \lambda_-\n(E,y) \; \psi_{E(-1)}^{+}\n(x) \mez ;
      \end{eqnarray}
      where $\lambda_\pm\n(E,y)$ are as yet unknown factors independent of $x$.\\
      Recall that our previous considerations have established the continuity of $G_{\m E}^+\n(x,y)$ at $x=y$.\\
      $[\,$See our discussion pursued immediately after equation (\ref{G-E-pm-spectral}) above.$\,]$\\
      If so, then necessarily
      \be \label{get-G-E-+-equation-1}
         \lambda_+\n(E,y) \; \psi_{E(+1)}^{+}\n(y) \; = \; \lambda_-\n(E,y) \; \psi_{E(-1)}^{+}\n(y) \mez .
      \ee
      In order to see even more clearly what happens at $x=y$,\\
      we shall integrate the inhomogeneous ODE (\ref{G-E-pm-SCHE-ihg}) over $x \in (y-(\Delta\,x),y+(\Delta\,x))$.\\
      This yields the following intermediate outcome:
      \be \label{get-G-E-+-equation-2-preprelim}
         \hspace*{-0.50cm}
         \frac{\hbar^2}{2\,m} \; \partial_{x} \, G_{\m E}^+\n(x,y) \, \Bigr|_{x=y+(\Delta\,x)} \; - \;
         \frac{\hbar^2}{2\,m} \, \partial_{x} \, G_{\m E}^+\n(x,y) \, \Bigr|_{x=y-(\Delta\,x)} \; + \;
         2 \, (\Delta\,x) \Bigl( E - V\m(y) \Bigr) \, G_{\m E}^+\n(y,y) \; = \; 1 \mz . \mz
      \ee
      Let us subsequently push $(\Delta\,x) \to +0$.\\
      Then the last l.h.s.~term of (\ref{get-G-E-+-equation-2-preprelim}), proportional to $(\Delta\,x)$, vanishes.\\
      Hence the $x$-derivative of $G_{\m E}^+\n(x,y)$ is discontinuous at $x=y$. Stated mathematically, one has
      \be \label{get-G-E-+-equation-2-prelim}
         \partial_{x} \, G_{\m E}^+\n(x,y) \, \Bigr|_{x=y+0_+} \; - \; \partial_{x} \, G_{\m E}^+\n(x,y) \, \Bigr|_{x=y-0_+} = \; \frac{2\,m}{\hbar^2} \mez .
      \ee
      After plugging (\ref{get-G-E-+-right}) and (\ref{get-G-E-+-left}) into (\ref{get-G-E-+-equation-2-prelim}) one finds that
      \be \label{get-G-E-+-equation-2}
         \lambda_+\n(E,y) \; \partial_y \, \psi_{E(+1)}^{+}\n(y) \; - \; \lambda_-\n(E,y) \; \partial_y \, \psi_{E(-1)}^{+}\n(y) \; = \; \frac{2\,m}{\hbar^2} \mez .
      \ee
      This formula is important, since the as yet unspecified coefficients $\lambda_\pm\n(E,y)$\\
      can be now uniquely determined via solving equations (\ref{get-G-E-+-equation-1}) and (\ref{get-G-E-+-equation-2}).\\
      
\newpage

{\color{magenta}
\dotfill\\
\begin{centering}
{\sl Appendix IV}\\
\vspace*{-0.20cm}
\end{centering}
\dotfill}
      
      Direct calculation yields
      \be \label{lambda-pm-E-y-prelim}
         \lambda_\pm\n(E,y) \; = \; \frac{2\,m}{\hbar^2} \, \frac{\psi_{E(\mp 1)}^{+}\n(y)}{w(E,y)} \mez ;
      \ee
      where $w(E,y)$ stands for the Wronskian of the two involved eigenfunctions $\psi_{E(\pm1)}^{+}\n(y)$,\\
      which was defined by equation (\ref{w-E-x-def}) of {\sl Appendix II}.\\
      Subsequent equation (\ref{w-E-x-explicit}) of {\sl Appendix II} evaluates $w(E,y)$ explicitly, and implies in turn
      \be \label{lambda-pm-E-y}
         \lambda_\pm\n(E,y) \; = \; -\;\frac{2\,\pi\,i}{T_{E}^{+}} \; \psi_{E(\mp 1)}^{+}\n(y) \mez .
      \ee
      Substitution of (\ref{lambda-pm-E-y}) into (\ref{get-G-E-+-right})-(\ref{get-G-E-+-left}) provides now the sought\\
      finalized, explicit and inspirative expressions for the retarded Green function $G_{\m E}^+\n(x,y)$.\\
      Namely, we have found that
      \begin{eqnarray}
         \label{final-G-E-+-right} G_{\m E}^+\n(x \ge y,y) & = & -\;\frac{2\,\pi\,i}{T_{E}^{+}} \; \psi_{E(-1)}^{+}\n(y) \; \psi_{E(+1)}^{+}\n(x) \mez ;\\
         \label{final-G-E-+-left} G_{\m E}^+\n(x \le y,y) & = & -\;\frac{2\,\pi\,i}{T_{E}^{+}} \; \psi_{E(+1)}^{+}\n(y) \; \psi_{E(-1)}^{+}\n(x) \mez .
      \end{eqnarray}
      According to formulas (\ref{final-G-E-+-right})-(\ref{final-G-E-+-left}),
      the entity $G_{\m E}^+\n(x,y)$ can be constructed\\
      just by solving an eigenvalue problem of $\hat{\sf H}$ for $\psi_{E(\pm 1)}^{+}\n(x)$.\\
      After setting $x=a$, $y=b$ in (\ref{final-G-E-+-left}) and exploiting (\ref{tilde-psi-E-+-bc-b}), (\ref{tilde-psi-E---bc-a}), (\ref{psi-E-eta-+-x-def})\\
      one obtains another interesting and important insight, namely,
      \be \label{G-E-+-a-b-final}
         G_{\m E}^+\n(a,b) \; = \; -\;\frac{2\,\pi\,i}{T_{E}^{+}} \; \psi_{E(+1)}^{+}\n(b) \; \psi_{E(-1)}^{+}\n(a) \; = \;
         \frac{(-i)\,m}{\hbar^2 K} \; T_{E}^{+} \; e^{+iK(b-a)} \mez .
      \ee
      In an absence of the potential we have $T_{E}^{+}=1$, and
      $\psi_{E\eta}^{+}\n(x)=\phi_{E\eta}\m(x)=(\ref{phi-E-eta-def})$.\\
      Thereby $G_{\m E}^+\n(x,y) \, = \, (\ref{final-G-E-+-right}) \; \& \; (\ref{final-G-E-+-left})$
      reduces correctly to $G_{0,E}^+\n(x,y)$ given by equation (\ref{Green-take-4}).
\item Summarizing the contents of the present {\sl Appendix IV},\\
      we have introduced in a tutorial step-by-step fashion\\
      the Green operators $(E-\hat{\sf H} \pm i\,0_+)^{-1}$ and their associated Green functions $G_{\m E}^\pm\n(x,y)$.\\
      Well definedness of these entities was demonstrated along the way.\\
      Subsequently we have written down an ODE satisfied by $G_{\m E}^\pm\n(x,y)$,\\
      and highlighted the fundamental difference\\
      between the retarded Green function $G_{\m E}^+\n(x,y)$ and its advanced counterpart $G_{\m E}^-\n(x,y)$.\\
      We have shown that the mentioned difference consists in picking up\\
      the outgoing boundary conditions for $G_{\m E}^+\n(x,y)$, and\\
      the incoming boundary conditions for $G_{\m E}^-\n(x,y)$.\\
      An explicit form of the free Green functions $G_{0,E}^\pm\n(x,y)$ was determined via contour integration.\\
      The retarded Green function $G_{\m E}^+\n(x,y)$ was expressed explicitly and simply in terms\\
      of the pertinent continuum eigenstates $\psi_{E(\pm 1)}^{+}\n(x)$ of the corresponding Hamiltonian $\hat{\sf H}$.
\end{itemize}

\newpage

{\color{magenta}
\dotfill\\
\begin{centering}
{\sl Appendix V}\\
\vspace*{-0.20cm}
\end{centering}
\dotfill}

\vspace*{+0.50cm}

{\bf Appendix V: Numerical treatment of one-dimensional scattering problems}
\vspace*{+0.10cm}
\begin{itemize}
\item We aim at solving numerically the boundary value problem
      \be \label{TISCHE-tilde-psi-E-eta-pm}
         \left\{ \, -\,\frac{\hbar^2}{2\,m}\,\partial_{xx} \, + \, V\m(x) \, \right\} \tilde{\psi}_{E(+1)}^+\n(x) \; = \; E \, \tilde{\psi}_{E(+1)}^+\n(x) \mez ;
      \ee
      and 
      \begin{eqnarray}
         \label{tilde-psi-bc-1-again} \tilde{\psi}_{E(+1)}^{+}\n(x > b) & = & \hspace*{+1.77cm} T_{E(+1)}^{+} \; e^{+i K x} \mez ;\\
         \label{tilde-psi-bc-2-again} \tilde{\psi}_{E(+1)}^{+}\n(x < a) & = & e^{+i K x} \; + \; R_{E(+1)}^{+} \; e^{-i K x} \n\mez .
      \end{eqnarray}
      Here $E>0$ is a given prescribed continuum energy level, and $K=\sqrt{2\,m\,E}/\hbar$.\\
      The unknown entities to be determined are of course $\tilde{\psi}_{E(+1)}^+\n(x)$, $T_{E(+1)}^{+}$, $R_{E(+1)}^{+}$.\\
      The underlying physical context is explained by equations (\ref{TISCHE-psi-E-eta-pm}), (\ref{tilde-psi-bc-1})-(\ref{tilde-psi-bc-2}) from the main text.
\item Instead of dealing with the problem (\ref{TISCHE-tilde-psi-E-eta-pm}), (\ref{tilde-psi-bc-1-again})-(\ref{tilde-psi-bc-2-again}) directly,\\
      we shall take advantage of the following simple trick. $[\,$"Neuhauser trick"\m, see Ref.~\cite{Neuhauser}$\,]$\\
      Assume for a moment that we have succeeded in solving numerically\\ another slightly modified boundary value problem, namely,
      \be \label{TISCHE-overline-psi-E-eta-pm}
         \left\{ \, -\,\frac{\hbar^2}{2\,m}\,\partial_{xx} \, + \, V\m(x) \, \right\} \overline{\psi}_{E(+1)}^{\,+}\n(x) \; = \; E \, \overline{\psi}_{E(+1)}^{\,+}\n(x) \mez ;
      \ee
      and 
      \begin{eqnarray}
         \label{overline-psi-bc-1-again} \overline{\psi}_{E(+1)}^{\,+}\n(x > b) & = & \hspace*{+0.70cm} e^{+i K x} \hspace*{+2.87cm} ;\\
         \label{overline-psi-bc-2-again} \overline{\psi}_{E(+1)}^{\,+}\n(x < a) & = & A_E \; e^{+i K x} \; + \; B_E \; e^{-i K x} \mez .
      \end{eqnarray}
      Then our originally sought entities $\tilde{\psi}_{E(+1)}^+\n(x)$, $T_{E(+1)}^{+}$, $R_{E(+1)}^{+}$ are obviously obtainable via taking
      \be \label{tilde-from-overline}
         \tilde{\psi}_{E(+1)}^+\n(x) \; = \; A_E^{-1} \; \overline{\psi}_{E(+1)}^{\,+}\n(x) \mez , \mez
         T_{E(+1)}^{+} \; = \; A_E^{-1} \mez , \mez R_{E(+1)}^{+} \; = \; A_E^{-1} \; B_E \mez .
      \ee
      Recall that the coefficient $A_E$ is never zero,\\
      as demonstrated in the second paragraph of {\sl Appendix II}.

\newpage

{\color{magenta}
\dotfill\\
\begin{centering}
{\sl Appendix V}\\
\vspace*{-0.20cm}
\end{centering}
\dotfill}

\item What remains to be done\\
      is to resolve the boundary value problem (\ref{TISCHE-overline-psi-E-eta-pm}), (\ref{overline-psi-bc-1-again})-(\ref{overline-psi-bc-2-again})
      for a given prescribed energy $E > 0$.\\ For this purpose, we shall replace the continuous spatial coordinate $x \in (-\infty,+\infty)$\\
      by a discrete equidistant grid $\{ x_n \}_{n \in {\mathbb Z}}$ of increment $(\Delta\,x)>0$.\\ The second order ODE (\ref{TISCHE-overline-psi-E-eta-pm})
      is thus converted into a difference equation
      \be \label{recursive}
         \hspace*{-0.50cm}
         -\,\frac{\hbar^2}{2\,m} \, \frac{\left(\overline{\psi}_{E(+1)}^{\,+}\n(x_{n+1})\,-\,2\,\overline{\psi}_{E(+1)}^{\,+}\n(x_{n})\,+\,\overline{\psi}_{E(+1)}^{\,+}\n(x_{n-1})\right)}
         {(\Delta\,x)^2} \; + \; V\n(x_n) \; \overline{\psi}_{E(+1)}^{\,+}\n(x_{n}) \; = \; E \; \overline{\psi}_{E(+1)}^{\,+}\n(x_{n}) \mz .
      \ee
      Formula (\ref{recursive}) enables us to determine the value of $\overline{\psi}_{E(+1)}^{\,+}\n(x_{n-1})$\\
      provided that the values of $\overline{\psi}_{E(+1)}^{\,+}\n(x_{n})$ and $\overline{\psi}_{E(+1)}^{\,+}\n(x_{n+1})$ are known.\\
      In other words,\\
      we are able to propagate the wavefunction $\Bigl\{\overline{\psi}_{E(+1)}^{\,+}\n(x_{n})\Bigr\}_{n \in {\mathbb Z}}$ recursively on our $x$-grid,\\
      moving step by step from the right towards the left.\\
      Such a propagation may start in the asymptotic region of $x \ge b$,\\ where the simply looking boundary condition (\ref{overline-psi-bc-1-again}) applies,\\
      and where the values of $\overline{\psi}_{E(+1)}^{\,+}\n(x_{n})$ are thus known.\\
      Suppose that we have propagated $\Bigl\{\overline{\psi}_{E(+1)}^{\,+}\n(x_{n})\Bigr\}_{n \in {\mathbb Z}}$ up to the left asymptotic region of $x \le a$,\\
      where the boundary condition (\ref{overline-psi-bc-2-again}) holds.\\
      Then we have in hand numerical values of $\overline{\psi}_{E(+1)}^{\,+}\n(x_n)$ and $\overline{\psi}_{E(+1)}^{\,+}\n(x_{n-1})$ for some $x_{n-1} < x_n \leq a$.\\
      Yet formula (\ref{overline-psi-bc-2-again}) tells us that
      \begin{eqnarray}
         \label{numerics-AB-1} \overline{\psi}_{E(+1)}^{\,+}\n(x_{n\phantom{-1}}) & = & A_E \; e^{+i K x_{n\phantom{-1}}} \; + \; B_E \; e^{-i K x_{n\phantom{-1}}} \mez ;\\
         \label{numerics-AB-2} \overline{\psi}_{E(+1)}^{\,+}\n(x_{n-1}) & = & A_E \; e^{+i K x_{n-1}} \; + \; B_E \; e^{-i K x_{n-1}} \mez .
      \end{eqnarray}
      Equations (\ref{numerics-AB-1})-(\ref{numerics-AB-2}) can be easily solved for $A_E$ and $B_E$, one gets explicitly
      \begin{eqnarray}
         \label{A-E-explicit} + \, A_E \; e^{+iKx_n} & = & \frac{e^{+iK(\Delta\,x)}\,\overline{\psi}_{E(+1)}^{\,+}\n(x_{n})\,-\,\overline{\psi}_{E(+1)}^{\,+}\n(x_{n-1})}
         {e^{+iK(\Delta\,x)}\,-\,e^{-iK(\Delta\,x)}} \mez ;\\
         \label{B-E-explicit} - \, B_E \; e^{-iKx_n} & = & \frac{e^{-iK(\Delta\,x)}\,\overline{\psi}_{E(+1)}^{\,+}\n(x_{n})\,-\,\overline{\psi}_{E(+1)}^{\,+}\n(x_{n-1})}
         {e^{+iK(\Delta\,x)}\,-\,e^{-iK(\Delta\,x)}} \mez .
      \end{eqnarray}
      In summary, we are able to evaluate numerically both\\
      the sought wavefunction $\Bigl\{\overline{\psi}_{E(+1)}^{\,+}\n(x_{n})\Bigr\}_{n \in {\mathbb Z}}$ and the associated coefficients
      $A_E = (\ref{A-E-explicit})$, $B_E = (\ref{B-E-explicit})$.\\ Thereby our boundary value problem (\ref{TISCHE-overline-psi-E-eta-pm}),
      (\ref{overline-psi-bc-1-again})-(\ref{overline-psi-bc-2-again}) has been successfully worked out.
\item According to (\ref{tilde-from-overline}) we have $T_{E(+1)}^{+} = A_E^{-1}$ and $R_{E(+1)}^{+} = A_E^{-1} \, B_E$.\\
      As an useful check of the whole calculation,\\ one may verify validity of the property (\ref{T-R-1}) from {\sl Appendix II}.\\
      One may look also at the property (\ref{T-R-0}),\\
      provided only that the above given algorithm is slightly adapted to calculate also $\tilde{\psi}_{E(-1)}^+\n(x)$.
\end{itemize}

\newpage

\vspace*{+3.00cm}

\end{document}